\documentclass[aps,pra,onecolumn,showpacs,superscriptaddress,longbibliography]{revtex4-2}

\usepackage[Symbol]{upgreek}
\usepackage{mathrsfs}
\usepackage{graphicx} 
\usepackage{dcolumn}  
\usepackage{booktabs}
\usepackage{scalerel}
\usepackage[english]{babel}
\usepackage{hyperref}
\usepackage{times}
\usepackage{amsmath,amssymb,amsfonts,bm}
\usepackage{bbold}
\usepackage{subfig}
\usepackage{mathtools}
\usepackage{textcomp}
\usepackage{braket}
\usepackage{empheq}
\usepackage{amsthm}
\usepackage[babel]{csquotes}
\usepackage{pifont}
\usepackage[labelfont=bf,labelsep=quad,justification=raggedright,font=small]{caption} 
\usepackage{tabularx}
\usepackage{mdframed}
\usepackage{soul}

\newcommand{\sss}{\scriptscriptstyle}

\DeclareSymbolFontAlphabet{\mathbb}{AMSb}

\begin{document}
\preprint{APS/123-QED}

\title{Visualized Geometric Phase of Caustic Geometric Beams}

\author{Haiyang Li$^{1}$ and Yijie Shen$^{2,3,*}$}
\affiliation{
	{$^{1}$Department of Chemical Engineering, Pohang University of Science and Technology, Pohang 37673, Korea}\\
	{$^{2}$Centre for Disruptive Photonic Technologies, School of Physical and Mathematical Sciences \& The Photonics Institute, Nanyang Technological University, Singapore 637371, Singapore}\\
	{$^{3}$School of Electrical and Electronic Engineering, Nanyang Technological University, Singapore 639798, Singapore} 
}

\date{\today}

\vspace{10cm}
\begin{abstract}
\noindent
    Detecting Pancharatnam-Berry (PB) or geometric phases of light usually requires complex interferometry or diffraction through a specially designed truncated aperture. In this work, we present a simpler method that enables direct, visual measurement of geometric phases in certain complex structured light fields, without additional interferometry or beam truncation.
Our approach exploits the PB phase present in SU(2) structured light. In these SU(2) modes, spatial wave packets naturally align with underlying caustic trajectories. By observing wavepacket evolution—i.e., the morphology of coupled caustics as the beam propagates—we can visually reveal both the geometric phase and the Gouy phase of the beam.
This visual detection technique offers new insights into the geometric phase of structured light and broadens the possibilities for designing optical devices that harness geometric phases.
\end{abstract}

\maketitle

\newpage
\section{Introduction}

The evolution of a quantum system acquires both a dynamic phase, arising from the time-dependent evolution of its state, and a geometric phase, which emerges when system parameters undergo cyclic adiabatic changes in parameter space~\cite{anandan1992geometric,cohen2019geometric}. Similarly, in optics, a paraxial light beam acquires an additional phase shift as it propagates. This effect, known as the Gouy phase, was first observed by Gouy~\cite{gouy1890propriete} and later recognized as having a geometric nature~\cite{simon1993bargmann}. Parallel concepts in polarization optics were pioneered by Pancharatnam, who introduced a geometric phase for polarized states~\cite{pancharatnam1956generalized}, later generalized by Berry to establish its universality across physical systems — now termed the Pancharatnam-Berry (PB) phase~\cite{berry1984quantal,berry1987adiabatic}. Over the decades, both Gouy and PB phases have been rigorously extended and harnessed across disciplines. In ultrafast optics, they enable precise spatiotemporal control of light~\cite{lindner2004gouy,hoff2017tracing,sakakura2020ultralow,drevinskas2017ultrafast}, while metamaterials leverage these phases to engineer anomalous refraction, flat optics, and arbitrary wavefront shaping~\cite{zhang2019direct,mueller2017metasurface,yu2011light,yu2014flat,devlin2017arbitrary,zhong2018controlling}. In quantum informatics, they underpin photonic qubit manipulation and entanglement protocols~\cite{yale2016optical,stav2018quantum,cho2019emergence,shen2022nonseparable}, and in structured light, they govern the dynamics of optical vortices, vector beams, and complex topological fields~\cite{wang2018recent,shen2019optical,slussarenko2016guiding,rosales2018review,gu2018gouy,bliokh2023roadmap}. These geometric phases now stand as foundational tools in modern optics and photonics, bridging abstract quantum mechanics with practical wave engineering.

\begin{figure}
    \centering
    \includegraphics[width=\linewidth]{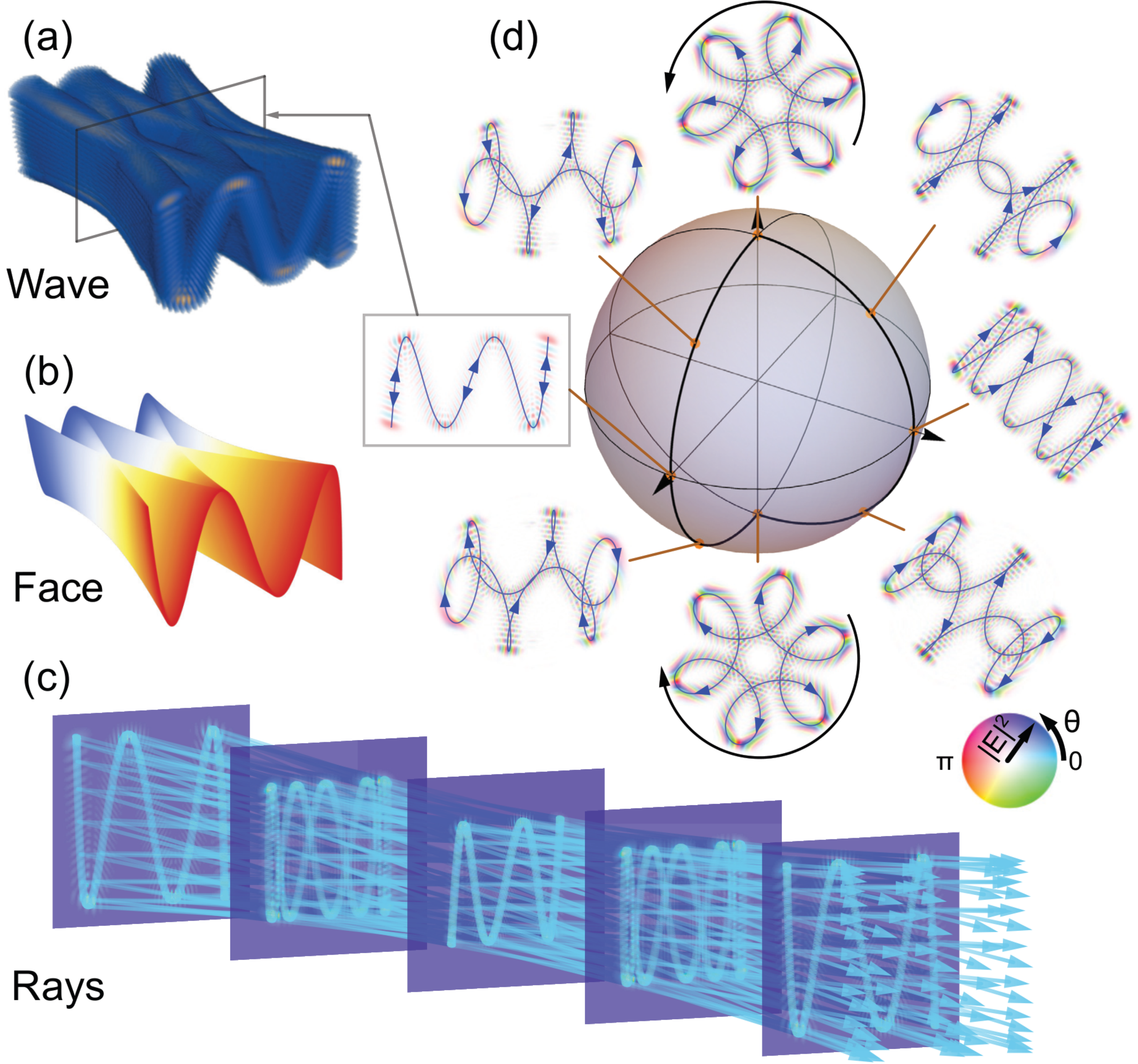}
    \caption{Schematic of caustic light modes mapped on the SU(2) Poincar\'e sphere: (a) 3D wave packet of a caustic-coupled geometric mode. (b) This wave packet is spatially coupled to a Lissajous parametric surface. (c) The Lissajous surface is a ruled surface generated by a family of rays forming caustics; the coupled transverse Lissajous curve evolves upon propagation. (d) The generalized evolution of the Lissajous--trochoidal trajectories and their caustic-coupled modes, with the modes mapped onto the mode Poincar\'e sphere (MPS), illustrates the complex behavior and interactions of these light modes. A phase–intensity color disk serves as a legend indicating phase and intensity, where the hue represents the phase \(\theta\) of the complex field and the brightness corresponds to the field intensity \(|E|^2\).}
    \label{fig:su2_caustic_schematic}
\end{figure}

The PB phase appears in various optical scenarios, including polarization manipulation~\cite{pancharatnam1956generalized}, cylindrical vector beams~\cite{milione2011higher,milione2012higher,cisowski2022colloquium}, and the transverse eigenmodes (i.e., solutions of the paraxial wave equation forming a complete orthonormal basis of spatial modes) of an optical system~\cite{cisowski2022colloquium,galvez2003geometric,padgett1999poincare,habraken2010universal}. A key objective of our work is to develop a noninterferometric method for measuring these geometric phases. To visualize how the PB phase changes as a beam’s mode parameters vary, researchers use the modal Poincaré sphere (MPS)~\cite{galvez2003geometric,padgett1999poincare}. Initially introduced for simple, first-order Gaussian beams carrying orbital angular momentum (OAM) of a single topological charge, the MPS concept has since been extended to a broader class of higher-order two-dimensional modes—Hermite Gaussian (HG), Laguerre Gaussian (LG), and generalized Hermite–Laguerre Gaussian (GG). These modes exhibit more universal patterns in their OAM evolution, meaning they share a common and generalizable framework describing how their orbital angular momentum changes as their parameters vary~\cite{dennis2017swings,alonso2017ray,dennis2019gaussian,gutierrez2019generalized}. On the MPS, the poles correspond to LG modes with opposite OAM values, the equator corresponds to HG modes (these modes rotate within the transverse plane as their position on the equator varies), and areas of the sphere other than the poles and equator represent intermediate GG modes that smoothly bridge these families of modes.

In addition to the MPS representation, the so-called Dennis–Alonso ray picture provides a geometric way to understand complex modes by considering them as superpositions of rays arranged with specific spatial and angular distributions~\cite{alonso2017ray,dennis2019gaussian}. Within this framework, the brightness patterns known as caustics arise naturally~\cite{zannotti2020shaping}. Caustics are formed where rays concentrate, creating bright, curved features in the beam’s intensity profile. These features encode information about the beam’s phase structure. By examining how caustics evolve as a beam propagates, one can infer phase changes, including PB and Gouy phases, without resorting to direct interferometry~\cite{malhotra2018measuring}. However, existing noninterferometric approaches still require introducing a truncated aperture or edge to induce diffraction patterns that make the caustic structures (and thus the underlying phase information) visible~\cite{malhotra2018measuring,hamazaki2006direct}. Because the naturally occurring caustic morphology differs from the stable, propagation-invariant patterns of eigenmodes, it becomes seemingly impossible to measure the phase without altering the beam itself. As a result, beam truncation remains essential for unveiling these caustic effects.

We propose that the PB phase can be detected by analyzing the intrinsic diffraction patterns, which correspond to the three-dimensional geometric structure of the beam's wave packet as it propagates, without relying on interferometric setups or truncation. This approach utilizes a class of ray-wave coupled geometric modes, which form quantum-like coherent states and follow mode evolution on a generalized MPS. In these states, the 3D wave packet is strictly aligned with a caustic ray trajectory, ensuring that the beam’s intrinsic diffraction structure encodes the PB phase. By closely examining how these intrinsic diffraction features (i.e., the naturally occurring intensity distributions as the beam propagates) evolve, we establish a direct relationship between the PB phase and the underlying ray-wave pattern. We further present an experimental method to measure both PB and Gouy phases by tracking these intrinsic diffraction distributions, thus circumventing the need for interferometry or beam truncation.

The family of caustic geometric modes we exploit is constructed by superposing a class of eigenmodes with different mode indices, and the superposed wave packet fulfills the formation of an SU(2) coherent state. The mathematical property of the coherent state determines that the intensity pattern of the wave packet is always coupled with a cluster of caustic ray trajectories bouncing back and forth between two confocal surfaces, usually called ray-wave duality. An example of the caustic geometric modes is demonstrated in Figs.~\ref{fig:su2_caustic_schematic}(a-c), whose 3D wave packet is spatially coupled with a Lissajous parametric ruled surface where the transverse Lissajous curve evolves along the longitudinal direction.

On the other hand, the formation of the SU(2) coherent state also provides many parameters to tune the morphology of the geometric patterns, which can be adjusted into other general geometries, e.g., the trochoidal caustics-coupled modes~\cite{shen2021rays}. In the formation of the SU(2) coherent mode, each decomposed eigenmode can be seen as the GG modes mapped on the MPS, thus the caustic geometric mode can also be mapped on the MPS, named the SU(2) Poincar\'e sphere~\cite{shen20202}. For instance, a class of geometric modes coupled with Lissajous and trochoidal curves represented on an MPS is shown in Fig.~\ref{fig:su2_caustic_schematic}(d). The modal evolution of these is analogous to the OAM evolution on MPS, which is based on the SU(2) structure. The points on the equator represent a set of geometric modes coupled with Lissajous curves, which can gradually evolve into a trochoid-coupled geometric mode carrying OAM when the point moves to the pole, along with the Lissajous-to-trochoidal geometric curve (surface) evolution. The order of the complexity of the Lissajous or trochoidal curve can be controlled by the parameters in the SU(2) coherent state. The north and south poles refer to the right- and left-handed rotating trochoids, respectively.

Akin to the mechanism of the prior eigenmode MPS, the SU(2) MPS representing caustic ray-wave coupled modes also possesses the PB phase effect but with hidden symmetries and properties unveiled. The geometric phase is generated when a point traverses a closed path on the sphere and is related to the spatial solid angle of the route. When the point returns to the initial position, the represented mode evolves back to the initial state but cannot be exactly the same, which results in an additional phase — the PB phase.

Here we show a fully noninterferometric methodology for measuring the PB phase in SU(2) beams. We begin by introducing SU(2) coherent states and their representation on the MPS, establishing the theoretical foundation. We then describe how to determine the geometric phase in both non-superposition and superposition states, demonstrating that our approach does not rely on external interferometric setups or beam truncation. Finally, we detail the implementation of our measurement technique that does not require an interferometer, illustrating how the PB phase influences the 3D structure of SU(2) coherent wave packets and validating our approach experimentally. This approach provides a direct and accessible route to revealing geometric phases that were previously challenging to detect without modifying the beam.

\section{SU(2) Coherent States}

In quantum mechanics, coherent states are eigenstates of the annihilation operator, \(\hat{a} |\alpha\rangle = \alpha |\alpha\rangle\), where \(\alpha = |\alpha| \mathrm{e}^{\mathrm{i}\theta}\) is a complex number, with \( |\alpha| \) and \(\theta\) representing amplitude and phase, respectively. Coherent states can also be expressed as superpositions of harmonic oscillator number states, see Eq.~\eqref{coherent_state_fock}:

\begin{equation}
    |\alpha\rangle = \mathrm{e}^{-\frac{|\alpha|^2}{2}} \sum_{n=0}^{\infty} \frac{\alpha^n}{\sqrt{n!}} |n\rangle.
    \label{coherent_state_fock}
\end{equation}

Here, $|n\rangle$ denotes the Fock (number) state with $n$ quanta. The SU(2) coherent state, defined in Eq.~\eqref{SU(2)_coordinate}, provides an intuitive understanding in laser optics, where a stable optical cavity acts as a macroscopic quantum system supporting frequency-degenerate transverse modes. These states are constructed by coherently superposing orthogonal eigenstates of the optical cavity.

\begin{figure*}[!htbp]
    \centering
    \includegraphics[width=\linewidth]{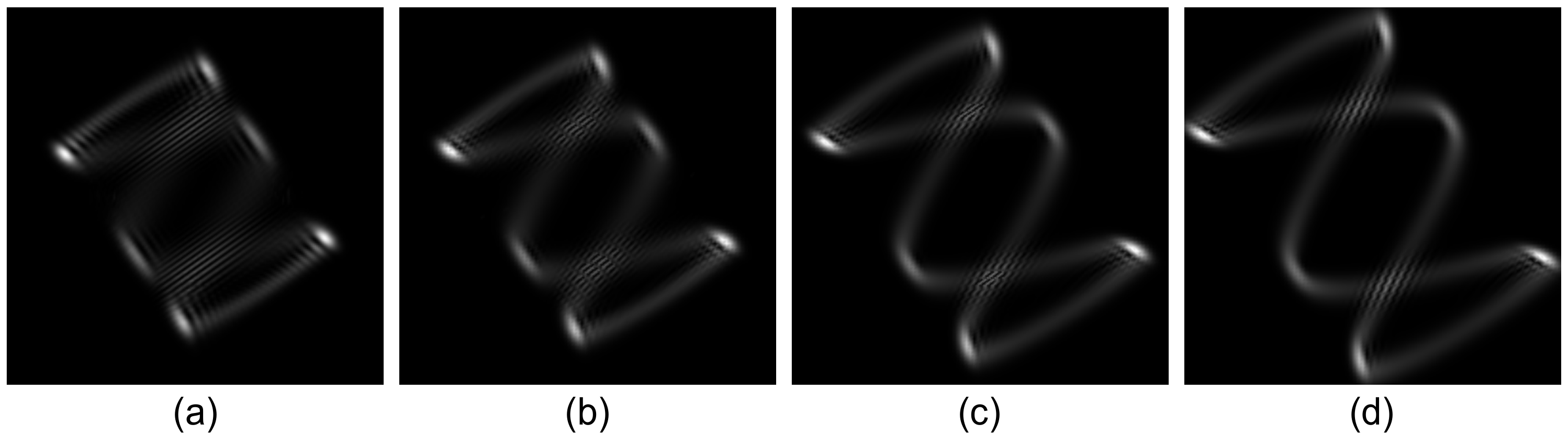}
    \caption{
    Transverse intensity profiles for different \( M \) values.  
    (a)--(d) correspond to \( M = 6, 10, 20, \) and \( 30 \), respectively.  
    As \( M \) increases, the mode profiles become sharper and the ray structures more distinct.
    }
    \label{fig:intensity_vs_M}
\end{figure*}

The resonance frequency of a mode with transverse indices \((m,n)\) and longitudinal index \(l\) is \(f_{m,n,l} = l\,\Delta f_L + (m+n+1)\,\Delta f_T\), where \(\Delta f_L\) and \(\Delta f_T\) denote the longitudinal and transverse mode spacings, respectively. Frequency degeneracy occurs when the ratio is rational, \(\frac{\Delta f_T}{\Delta f_L} = \frac{P}{Q} \in \mathbb{Q},\) with coprime integers \(P\) and \(Q\) (usually \(P=1\)), where the harmonic oscillator parameters \(s\), \(t\), \(P\), and \(Q\) satisfy \(P=1\) and \(Q=s+t\), such that \(Q\) directly relates the degeneracy order to the transverse index steps \(s\) and \(t\)~\cite{tuan2018realizing,chen2017symmetry,chen2011geometry}. Under this condition, a family of Hermite-Gaussian modes becomes degenerate and can interfere coherently to form structured light fields.  

The eigenstate basis for such an optical cavity is expressed as \(\left|\psi_{\sss{m+sK,n+tK,l-K}}\right\rangle\), where \(m\), \(n\), and \(l\) denote the initial transverse and longitudinal indices, \(s\) and \(t\) define the increments in the transverse indices, and \(K\) serves as an integer index labeling the component states in the SU(2) representation, indicating the degree of modal mixing between two orthogonal Gaussian modes. Each basis state corresponds to a GG mode with indices \((m+sK,n+tK)\) and longitudinal order \(l-K\). Collectively, these states form a complete orthonormal set, suitable for constructing SU(2) coherent-state superpositions.

The SU(2) coherent state \(\left|\Psi_{\sss{m,n,l}}^{\sss{M,s,t}}\right\rangle\) is constructed from the basis states \(\left|\psi_{\sss{m+sK,n+tK,\,l-K}}\right\rangle\). 
Each basis state corresponds to a Generalized Gaussian (GG) cavity mode with transverse indices \((m+sK,n+tK)\) and longitudinal order \(l-K\), and its coordinate representation is given in Eq.~\eqref{GG_expression}. 
The coherent state is then written as

\begin{widetext}
    \begin{equation}
        \left|\Psi_{\sss{m,n,l}}^{\sss{M,s,t}}\right\rangle = \frac{1}{2^{\sss{M/2}}} \sum_{K=0}^{M} \binom{M}{K}^{\sss{1/2}} \mathrm{e}^{\sss{\mathrm{i} K (\phi-\frac{\pi}{2})}} \underbrace{\left|\psi_{\sss{m+sK,n+tK, l-K}}\right\rangle}_{\mathclap{\substack{\text{GG mode with transverse indices }(m+sK,n+tK)\\ \text{and longitudinal order }l-K}}},
    \label{SU(2)_coordinate}
    \end{equation}
\end{widetext}

Here, $\phi$ denotes the coherent phase, and $K$ serves as an integer index labeling the component states in the SU(2) representation introduced earlier.
The term $\mathrm{e}^{\mathrm{i}K(\phi - \tfrac{\pi}{2})}$ represents the corresponding coherent phase factor introducing a $K$-dependent $\tfrac{\pi}{2}$ phase shift.  The binomial coefficient $\binom{M}{K}^{\sss{1/2}}$ assigns weights to each mode, and the normalization factor $2^{-M/2}$ ensures proper normalization.  \(M\) determines the number of modes (\(M+1\)) in the superposition, where larger \(M\) yields sharper and more structured wave packets. 
For experimental and simulation purposes, the transverse indices are set to \(m=18\), \(n=30\), with \(M=6\), balancing trajectory clarity and computational feasibility (see Fig.~\ref{fig:intensity_vs_M}). Higher \(M\) values require increased numerical precision to ensure stable convergence of the computed wave packets.

The orbital angular momentum (OAM) Poincaré sphere (PS) representation of GG modes naturally extends to the SU(2) PS for coherent state visualization~\cite{shen20202} (see Fig.~\ref{fig:su2_caustic_schematic}). 
When \(M\ge 1\), multiple eigenmodes undergo coherent superposition, generating wave packets that exhibit ray--wave duality with distributions aligned along parametric curves (2D) or surfaces (3D)~\cite{chen2013exploring}.
Larger \(M\) emphasizes classical trajectories, whereas smaller \(M\) enhances wave characteristics; \(M=0\) corresponds to a single eigenmode (no coherent superposition), i.e., a purely wave-like case.
The PS mapping follows: polar points correspond to vortex LG eigenmodes or trochoidal SU(2) modes with opposite OAM; equatorial points to planar HG eigenmodes or Lissajous SU(2) modes; and intermediate points to elliptical GG eigenmodes or topological transition states between Lissajous and trochoidal modes - all maintaining identical total mode order \( N \)~\cite{shen2019optical}.

\begin{figure}[!htbp]
    \centering
    {\includegraphics[width=1\textwidth]{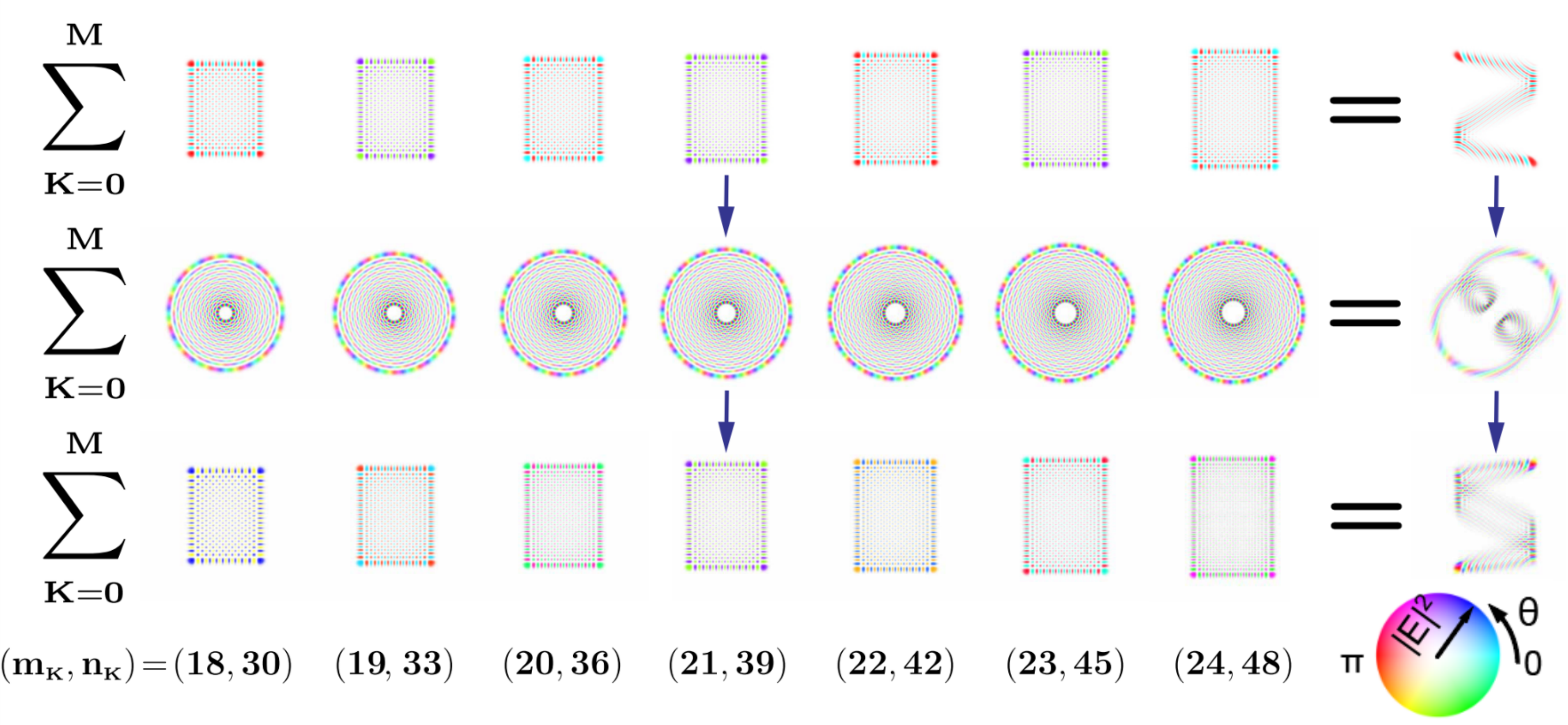}}
    \caption{
    Schematic illustration of the generation and transformation of an SU(2) wave packet through coherent superposition:
    The schematic depicts the creation of an SU(2) wave packet via the coherent superposition of its eigenstate wave packets. The summation symbol \(\sum\) denotes coherent superposition, and the downward-pointing arrow signifies the transformation of upper modes into the specified lower modes. 
    Starting from the top, the first row illustrates an SU(2) equatorial mode, termed the initial mode, generated by superposing HG modes. The mode indices in the two transverse directions, \((m_{\sss{K}}, n_{\sss{K}})\), are \((18,30)\), \((19,33)\), \((20,36)\), \((21,39)\), \((22,42)\), \((23,45)\), and \((24,48)\). These modes correspond to the equatorial points on the OAM PS with orders \(N_{\sss{K}} = 48\), \(52\), \(56\), \(60\), \(64\), \(68\), and \(72\), respectively.
    In the second row, an SU(2) polar mode is created from \(M+1\) LG modes, sharing the same transverse indices and mode orders as the eigenstate modes of the SU(2) equatorial mode. 
    The third row presents an SU(2) mode that emerges from the final HG modes, which also share the same transverse indices and mode orders as the previous rows. These HG modes acquire a PB phase through a complete cycle of mode transformation. Unlike the initial SU(2) mode in the first row, the SU(2) mode in the third row, termed the final mode, includes an additional PB phase. 
    Importantly, any modes from the first and third rows that correspond to the same equatorial point on the OAM PS will also result in coherently superposed SU(2) modes corresponding to the same equatorial point on the SU(2) PS.
    A phase–intensity color disk serves as a legend indicating phase and intensity, where the hue represents the phase \(\theta\) of the complex field and the brightness corresponds to the field intensity \(|E|^2\).
    }
    \label{mode_expansion}
\end{figure}

GG modes, characterized by total mode number \(N\), provide a unified framework encompassing LG, HG, and GG modes~\cite{dennis2019gaussian}. These modes serve as the eigenmode basis for SU(2) coherent states, with their mode order given by \(N = m + n = |\ell| + 2p_{\text{r}}\), where \(m\) and \(n\) are the transverse indices of HG modes, \(\ell\) is the topological charge, and \(p_{\text{r}}\) is the radial index. The GG modes are constructed via angular momentum superposition weighted by Wigner \(d\)-matrix elements:

\begin{widetext}
\begin{equation}
    \mathrm{GG}_{\sss{N, \ell}}(r, \varphi; \alpha, \beta)
    = \sum_{\substack{\ell'=-N \\ \text{steps of } 2}}^{N}
    \underbrace{d^{\sss{\frac{N}{2}}}_{\sss{\frac{\ell'}{2}\frac{\ell}{2}}}(\beta)}_{\substack{\text{Wigner $d$-matrix element} \\ \text{(angular momentum rotation)}}}
    \cdot \mathrm{e}^{-\mathrm{i}\frac{\ell'\alpha}{2}}
    \mathrm{i}^{\sss{\ell'-|\ell'|+\ell-N}}
    \cdot \mathrm{LG}_{\sss{N,\ell'}}(r, \varphi),
    \label{GG_expression}
\end{equation}
\begin{equation}
    \mathrm{LG}_{\sss{N, \ell}}(r, \varphi)
    = \sqrt{\frac{2^{\sss{|\ell|+1}}\left(\frac{N-|\ell|}{2}\right)!}{\left(\frac{N+|\ell|}{2}\right)!}}
    \cdot g_{\sss{0}}(r)
    \cdot \mathrm{e}^{\mathrm{i}\ell\varphi}
    \frac{r^{\sss{|\ell|}}}{w_{\sss{0}}^{\sss{|\ell|}}}
    \cdot \underbrace{L_{\sss{\frac{N-|\ell|}{2}}}^{\sss{|\ell|}}\!\left(\frac{2r^2}{w_{\sss{0}}^{\sss2}}\right)}_{\text{Associated Laguerre polynomial}}.
    \label{normalized_LG}
\end{equation}
\end{widetext}

where the fundamental Gaussian mode is \(g_{\sss 0}(r)=\big( w_{\sss 0}\sqrt{\pi}\big)^{-1}\exp\!\big(-r^{2}/w_{\sss 0}^{\sss 2}\big)\), with \(w_{\sss 0}\) the beam waist radius.
We use polar coordinates \((r,\varphi)\) in the focal plane; \(r\) is the radial coordinate and \(\varphi\) the transverse azimuthal coordinate.
At the focal plane, the normalized GG mode in Eq.~\eqref{GG_expression} represents the coordinate-space eigenstate wave packets~\cite{dennis2019gaussian}, constructed from LG modes as basis states (Eq.~\eqref{normalized_LG}).
Here \(m\) and \(n\) denote the transverse mode indices along the \(x\)- and \(y\)-directions, respectively, while \(l\) is the longitudinal (\(z\)-axis) index; \(\ell\) denotes the topological charge (distinct from \(l\)).
The harmonic-oscillator parameters \(s,t,P,Q\) satisfy \(P=1\) and \(Q=s+t\).
\(L_k^{\alpha}\) is the associated Laguerre polynomial of degree \(k\) and order \(\alpha\), characteristic of LG modes.
Rotations between angular-momentum states \(\lvert j,m'\rangle\) and \(\lvert j,m\rangle\) are described by the Wigner \(d\)-matrix element \(d^{j}_{m'm}(\beta)\), where \(\beta\in[0,\pi]\) is the colatitude angle and \(\alpha\in[0,2\pi)\) is the longitude on the Poincare sphere; in our notation the indices match as \(j=N/2\), \(m'=\ell'/2\), and \(m=\ell/2\).

\section{Geometric Phase in Mode Transformations}

As is well known, the PB geometric phase depends solely on the geometry of the evolution path in parameter space and is independent of dynamical parameters~\cite{shen2019optical,jisha2021geometric,galvez2003geometric,zhang2023geometric,cohen2019geometric,bliokh2019geometric,zhou2020experimental,samuel1988general,van2010geometric,milione2011higher,hayashi2005asymptotic,alonso2017ray}. Under adiabatic conditions, a mode undergoing cyclic evolution in parameter space acquires this geometric phase. While the Poincaré sphere visualization helps understand this relationship, it introduces phase singularities at the poles that require careful treatment. If these phase singularities are properly handled, allowing continuous phase evolution on the Poincaré sphere, the geometric phase may then be quantitatively determined on the PS.

\begin{figure*}[!htbp]
    \centering
    \includegraphics[width=\linewidth]{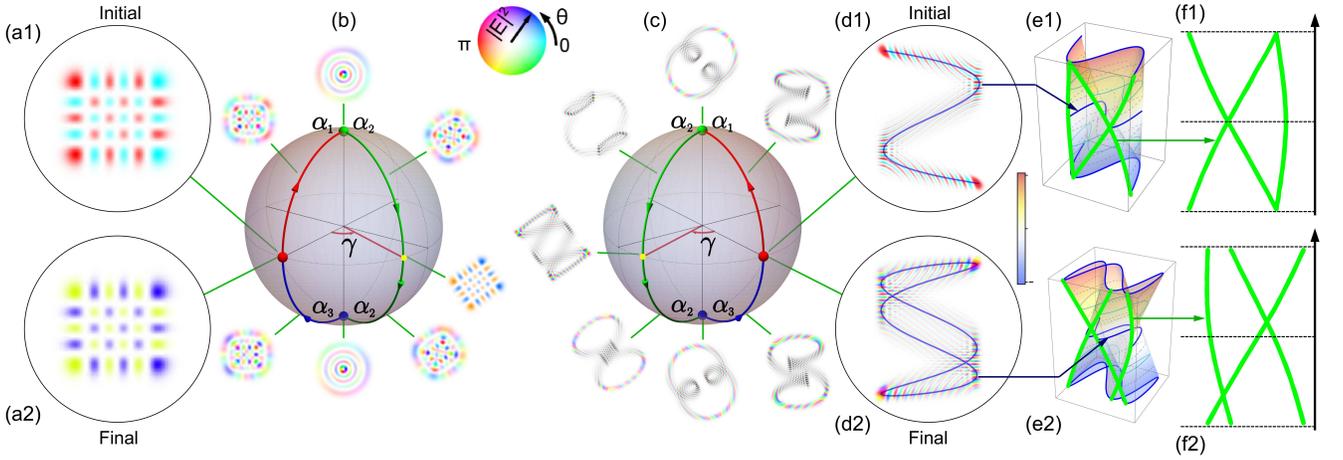}
    \caption{
    Illustration of the effect of the PB phase on the initial and final modes of HG and SU(2) under a complete cycle of mode transformation:
    This diagram represents the mode profiles and transformations. (a1) and (a2) depict the initial and final mode profiles of the HG mode, respectively, showing identical intensity distributions but differing phase distributions. This indicates that interferometric approaches are viable for determining the generated PB phase. (d1) and (d2) present the initial and final mode profiles of the Lissajous geometric mode, showcasing varying intensity and phase distributions, suggesting that non-interferometric methods are applicable to inferring the generated PB phase. (e1) and (e2) display the 3D parametric surfaces of the Lissajous geometric modes, while (f1) and (f2) illustrate the feature lines on these surfaces. The intersections of these feature lines serve as indicators for measuring the PB phase. (b) and (c) show the MPS with several modes along a closed path. Here, \(\alpha\) denotes the azimuthal coordinate; \(\alpha_{1}\), \(\alpha_{2}\), and \(\alpha_{3}\) are the azimuthal coordinates of the red, green, and blue segments, respectively; and \(\gamma\) is the enclosed spherical angle of the closed path. 
    A phase–intensity color disk serves as a legend indicating phase and intensity, where the hue represents the phase \(\theta\) of the complex field and the brightness corresponds to the field intensity \(|E|^2\).
    Collectively, these panels elucidate the different impacts of the PB phase on eigenstates versus SU(2) coherent state wave packets, highlighting the potential of non-interferometric measurement, whereby the PB phase is inferred from PB-induced changes in the intensity distribution and feature-line evolution.}
    \label{fig:pb-phase-mode-transformation}
\end{figure*}

To illustrate what we mean by ``continuous evolution'', consider a closed path on the Poincar\'{e} sphere along which the mode evolves under mode transformations.
Along this path the intensity profile varies smoothly, establishing a unique correspondence between the path parameter and the spatial pattern.  The phase, however, cannot remain smooth everywhere on the sphere because polar singularities interrupt its continuity; special care is therefore needed at those points. To make the subsequent analysis easier to follow, we must first state the precise meanings of several key terms.
\textit{(i) Mode distribution}: the complete spatial pattern containing both intensity and phase.  
\textit{(ii) Phase distribution}: the spatial variation of the field’s phase alone.  
\textit{(iii) Intensity distribution}: the spatial variation of optical energy, independent of phase. 

\begin{figure}[!htbp]
    \centering
    \includegraphics[width=0.5\linewidth]{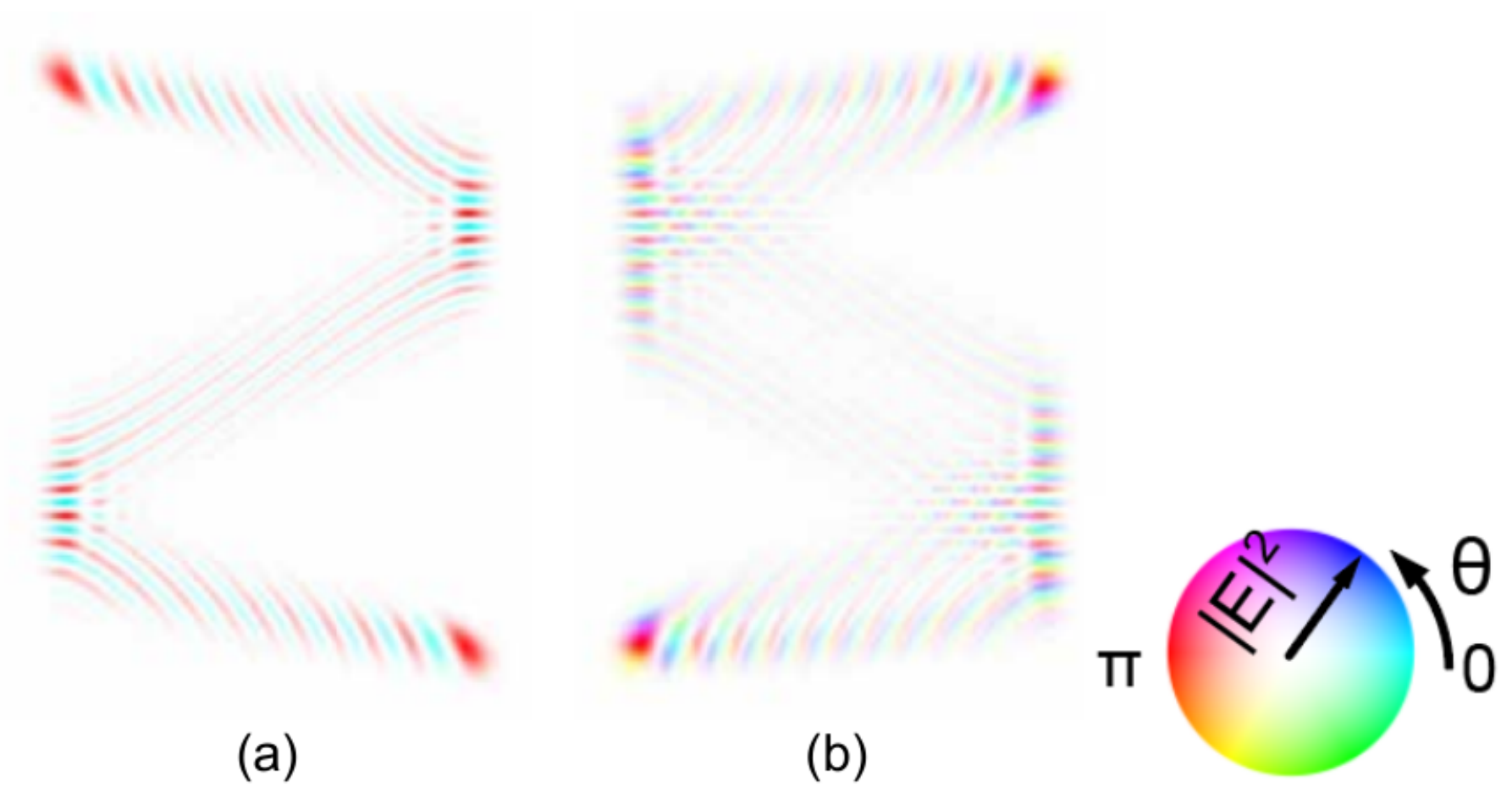}
    \caption{
    Comparison of SU(2) modes for \((s,t)=(1,3)\) before and after continuous evolution.
    (a) Initial mode.
    (b) Final mode after a full continuous evolution where the solid angle \(\Omega = 140^\circ\).
    The fields are presented in the transverse plane \((x,y)\).
    A phase–intensity color disk serves as a legend indicating phase and intensity, where the hue represents the phase \(\theta\) of the complex field and the brightness corresponds to the field intensity \(|E|^2\).
    It can be seen that the PB phase alters both the SU(2) mode’s phase and intensity structure.
    This change is consistent with both the mode profiles obtained through continuous evolution and those predicted by Eq.~\eqref{SU(2)_coordinate_with_PB}.
    }
    \label{initial_and_final_for_SU(2)}
\end{figure}

With these concepts in place, we now clarify what continuous evolution means in this work. Here, continuous evolution refers to a mode transformation in which the phase distribution—once freed from the influence of the polar singularities—changes as smoothly as the intensity distribution while the mode traces a closed path on the PS. Consequently, the entire mode distribution evolves continuously along that path. The process of continuous evolution is presented in Supplemental Material~\cite{SM} Section I.~\textit{Process of Continuous Evolution}.

To make our description clear, we introduce a few definitions. The mode at the starting point of the red path, with coordinates $(\alpha,\beta)=(0,\frac{\pi}{2})$, is called the Initial Mode, referring either to the Initial HG Eigenmode on OAM PS (an HG mode) or the Initial SU(2) Coherent Mode on SU(2) PS. The mode at the end of the red path (which is also the beginning of the green path), at coordinates $(\alpha,\beta)=(0,0)$, is named the North Polar Mode. It refers either to the North Polar Eigenmode on OAM PS (an LG mode) or the North Polar SU(2) Coherent Mode on SU(2) PS. Similarly, the mode located at $(\alpha,\beta)=(0,\pi)$ is termed the South Polar Mode, referring either to the South Polar Eigenmode on OAM PS (an LG mode) or the South Polar SU(2) Coherent Mode on SU(2) PS. Finally, the mode at the end of the blue path, again located at $(\alpha,\beta)=(0,\frac{\pi}{2})$, is called the Final Mode, meaning either the Final Eigenmode on OAM PS (an HG mode) or the Final SU(2) Coherent Mode on SU(2) PS (see Fig.~\ref{fig:pb-phase-mode-transformation}). The initial mode has a PB phase of \(0^{\circ}\), indicating the starting point of mode transformation. In contrast, the final mode carries a non-zero PB phase. 
Denoting the initial and final modes as \(\Psi^{\sss{(\text{initial})}}|_{\sss{\text{GG/SU(2)}}}\) and \(\Psi^{\sss{(\text{final})}}|_{\sss{\text{GG/SU(2)}}}\), respectively, their relationship is given by Eq.~\eqref{mode_relation_2}. 
Here, \(\exp[\text{i}g(\gamma)]\) represents the PB phase factor, and \(g(\gamma)\) denotes the PB phase as a function of the spherical angle \(\gamma\), defined as the spherical angle between the two geodesic arcs corresponding to azimuthal coordinates.

\begin{equation}
    \Psi^{\sss{(\text{final})}}|_{\sss{\text{GG/SU(2)}}}=\exp[\text{i}\;g(\gamma)]\cdot\Psi^{\sss{(\text{initial})}}|_{\sss{\text{GG/SU(2)}}}
    \label{mode_relation_2}
\end{equation}

We employ two methods to determine the PB phase accumulated during mode transformation: (1) the ``method of continuous evolution``, which tracks the phase gained as the mode moves adiabatically along a closed path on the PS, and (2) the ``method of phase-factor'', which evaluates the phase directly from an established relation—e.g., Eq.~\eqref{PB_phase_eigen} for GG modes. Both methods give the same result: the first shows how the phase builds up dynamically, whereas the second yields a direct analytical expression.

Unlike GG modes, the geometric phase for arbitrary SU(2) modes involves more complex considerations. An SU(2) mode’s PB phase depends on both the PB phases of its constituent eigenmodes and their interference. Thus, determining this phase requires understanding the superposition mechanism that generates an SU(2) mode from its eigenmodes. This involves identifying which specific GG eigenmode combinations produce a given SU(2) state and how their geometric phases combine. With this in mind, we discuss the construction of SU(2) coherent states through eigenmode superposition in the next section.

\subsection{SU(2) Coherent States from Eigenmode Superposition}

Each intensity distribution of GG modes uniquely corresponds to points on closed paths of the OAM-Poincar\'{e} sphere, just as SU(2) modes map completely to closed trajectories on their SU(2)-Poincar\'{e} sphere. The construction of an SU(2) mode of order $N_{\sss 0}$ involves a coherent superposition of multiple GG modes with orders $N_{\sss K}$. Here, each constituent eigenmode associates with corresponding points on distinct OAM PSs of order $N_{\sss K}$, where $K$ ranges from $0$ to $M$ ($K \in \{0,\ldots,M\} \subset \mathbb{Z}$). Expressing $N_{\sss 0} = m_{\sss 0} + n_{\sss 0}$ in terms of transverse indices $m_{\sss 0}$ and $n_{\sss 0}$, the precise relationship between these mode orders is governed by Eq.~\eqref{Nk_N0_relation}, which determines the complete set of $M+1$ participating eigenmodes in the superposition.

\begin{equation}
    N_{\sss{K}} = m_{\sss{K}} + n_{\sss{K}}, 
    \label{Nk_N0_relation}
\end{equation}
where \( m_{\sss{K}} = m_{\sss{0}} + s \cdot K \) and \( n_{\sss{K}} = n_{\sss{0}} + t \cdot K \).
Upon substituting \( s+t = Q \) into Eq.~\eqref{Nk_N0_relation}, we deduce Eq.~\eqref{index_relation}.

\begin{equation}
    N_{\sss K} = N_{\sss 0} + Q \cdot K \quad (0 \leq K \leq M).
    \label{index_relation}
\end{equation}

For the case of $(s,t) = (1, 3)$, $(m_{\sss{0}}, n_{\sss{0}}) = (18, 30)$, $M = 6$, and $Q = 4$, the SU(2) mode construction requires superposition of seven GG eigenmodes with transverse indices:

\begin{equation*}
(m_{\sss{K}}, n_{\sss{K}}) = 
\begin{cases}
(18, 30) & K=0 \\
(19, 33) & K=1 \\
(20, 36) & K=2 \\
(21, 39) & K=3 \\
(22, 42) & K=4 \\
(23, 45) & K=5 \\
(24, 48) & K=6
\end{cases}
\end{equation*}

This progression exhibits three key characteristics: First, both transverse indices increase linearly with fixed steps ($\Delta m_{\sss{K}} = s = 1$, $\Delta n_{\sss{K}} = t = 3$). Second, the mode order $N_{\sss{K}}$ advances with the constant $Q = 4$. Third, the sequence spans continuously from base ($K=0$) to maximum ($K=M=6$). Fig.~\ref{mode_expansion} illustrates the resulting superposition.

\textit{Compensation of Phase Factors.} To ensure continuous evolution along the closed path, two compensating phase factors are required—one for the green segment and one for the blue segment on the OAM PS. The phase factor for the green segment is given by Eq.~\eqref{PB_phase for eigenmode_north_1}, whereas the one for the blue segment is given by Eq.~\eqref{PB_phase for eigenmode_south_1}.

\begin{align}
    \Phi_{\mathrm{g|GG}} &= \exp\left[\text{i}\frac{\alpha-\alpha_{1}}{2}(m-n)\right]
    \label{PB_phase for eigenmode_north_1} \\[1ex]
    \Phi_{\mathrm{b|GG}} &= \exp\left[-\text{i}(\alpha-\alpha_{2})(m-n)\right]
    \label{PB_phase for eigenmode_south_1}
\end{align}

Here, $\alpha$ is the azimuthal coordinate on the MPS; $\alpha_{1}$, $\alpha_{2}$, and $\alpha_{3}$ denote the azimuthal coordinates of the red, green, and blue segments, respectively; and $\gamma$ is the enclosed spherical angle of the closed path. 
The evolution starts at $(\alpha,\beta)=(0,\pi/2)$, so $\alpha=\alpha_{1}=0$. 
Since the red and blue segments lie on the same geodesic, $\alpha_{3}=0$, and the point on the green segment has azimuth coordinate $\alpha_{2}=\gamma$. 
Therefore, the PB phase factor of the eigenstate mode is given in Eq.~\eqref{PB_phase_eigen} as the product of the two phase factors, where in the last line we set $\alpha=\alpha_{3}=0$ because the mode with PB phase is located at the end of the red segment.

\begin{equation}
    \begin{aligned}
        \Phi_{\sss{\text{GG}}}^{\sss{\text{PB}}} = ~~~\!&\Phi_{\mathrm{g|GG}} \cdot \Phi_{\mathrm{b|GG}}\\
        =~~~\!&\exp\left[-\mathrm{i} \frac{\alpha - 2\gamma}{2}(m - n)\right] \\
        \quad\stackrel{\alpha=\alpha_{\sss 3}}{=}&\exp\left[\mathrm{i}(m - n)\gamma\right] \\
    \end{aligned}
    \label{PB_phase_eigen}
\end{equation}

The PB phase of GG modes is encoded in the phase distribution of the eigenstate wave packets (see Fig.~\ref{fig:pb-phase-mode-transformation}(a1)-(a2)) and is therefore accessible via interference measurements. Note that, by incorporating the relevant phase factors into the GG mode expression [Eq.~\eqref{GG_with_PB_phase}], two intermediate forms are obtained that describe the green and blue evolution segments when \(\Phi_{\text{GG}} = \Phi_{\mathrm{g|GG}}\) or \(\Phi_{\mathrm{b|GG}}\). Note that, when the relevant phase factors are incorporated into the GG mode expression (see Eq.~\eqref{GG_with_PB_phase}), three forms are obtained: two intermediate ones corresponding to the green and blue evolution segments, where $\Phi_{\text{GG}} = \Phi_{\mathrm{g|GG}}$ and $\Phi_{\mathrm{b|GG}}$, respectively, and the final GG mode carrying the PB phase after a complete cycle, where $\Phi_{\text{GG}} = \Phi_{\sss{\text{GG}}}^{\sss{\text{PB}}}$.

\begin{widetext}
    \begin{equation}
            \text{GG}_{N, \ell}(r, \varphi ; \alpha, \beta)=\Phi_{\text{GG}}
            \sum_{\ell^{\prime}=-N \atop \text{ steps of } 2}^{N} d_{\sss{ \frac{\ell^{\prime}}{2} \frac{\ell}{2} }}^{\sss{\frac{N}{2} }}(\beta) \;\;\text{e}^{\sss{-\text{i} \frac{\ell^{\prime} \alpha}{2} }}\;\text{i}^{\sss{\ell^{\prime}-\left|\ell^{\prime}\right|+\ell-N}}\;\text{LG}_{\sss{N,\ell^{\prime}}}(r,\varphi)
        \label{GG_with_PB_phase}
    \end{equation}
\end{widetext}

Our focus is on the PB phase of coherent wave packets rather than that of GG mode wave packets. 
For eigenstates, there is a one-to-one correspondence between the PB phase and the evolution of state vectors in Hilbert space. 
Because a superposition state is built from GG modes, its PB phase can be inferred by applying analogous operations to the PB phases of the corresponding eigenstates. 
In what follows, we outline the calculation of the PB phase for the SU(2) mode.

\subsection{Continuous Evolution of SU(2) Mode}

The preceding analysis indicates that if each eigenstate wave packet of an SU(2) coherent wave packet undergoes continuous evolution, then the SU(2) mode also evolves continuously and acquires a PB phase. The accumulation of this phase is illustrated in Fig.~\ref{mode_expansion} and leads to Eqs.~\eqref{SU(2)_coordinate_with_PB}, where \(\Phi_{K}\) is the phase factor that ensures continuous evolution of the SU(2) mode along the closed path on the SU(2) PS. When $\Phi_{K}=\Phi_{\mathrm{g|}K}$ or $\Phi_{\mathrm{b|}K}$, the SU(2) modes are the intermediate modes corresponding to the green or blue segments, respectively. And when $\Phi_{K}=\Phi_{\sss{K}}^{\sss{\text{PB}}}$, the SU(2) modes are the final modes carrying the PB phase after a complete cycle. Here the subscript SU(2) indicates that these phase factors are associated with the eigenstates of the SU(2) mode. Then let's derive the closed-form expressions for the 3 phase factors that ensure continuous evolution.

\begin{widetext}
    \begin{equation}
	\left|\Psi_{\sss{m,n,l}}^{\sss{M,s,t}}\right\rangle=\frac{1}{2^{\sss{M/2}}}\sum_{K=0}^{M}\binom{M}{K}^{\sss{1/2}}\cdot\Phi_{K} \cdot\mathrm{e}^{\sss{\text{i} K (\phi-\frac{\pi}{2})}}\left|\psi_{\sss{m+sK,n+tK, l-K}}\right\rangle
    \label{SU(2)_coordinate_with_PB}
    \end{equation}
\end{widetext}

By solving Eqs.~\eqref{PB_phase for eigenmode_north_1}, \eqref{PB_phase for eigenmode_south_1}, \eqref{PB_phase_eigen} and \eqref{GG_with_PB_phase}, together with Eq.~\eqref{SU(2)_coordinate_with_PB}, we obtain the compensated phase factors for the green and blue segments, given by Eqs.~\eqref{PB_SU(2)_green} and \eqref{PB_SU(2)_blue}.

\begin{align}
    \Phi_{\mathrm{g|}K} &= \exp\left\{\text{i}\;[m-n+K(s-t)]\;\frac{\alpha}{2}\right\} \label{PB_SU(2)_green}, \\[1ex]
    \Phi_{\mathrm{b|}K} &= \exp\left\{-\text{i}\;[m-n+K(s-t)]\;(\alpha-\gamma)\right\} \label{PB_SU(2)_blue}.
\end{align}

Determination of the PB phase for the SU(2) mode under continuous evolution proceeds identically to that for the GG mode.  Two steps are involved: (i) multiply Eqs.~\eqref{PB_SU(2)_green} and \eqref{PB_SU(2)_blue}; (ii) set $\alpha=\alpha_{3}=0$ in the resulting expression. The final result is given by Eq.~\eqref{SU(2)_PB_compensation}.

\begin{equation}
    \begin{aligned}
            \Phi_{\sss{K}}^{\sss{\text{PB}}}&\;=\;\Phi_{\mathrm{g|}K} \cdot\Phi_{\mathrm{b|}K} \\[1ex]
            &\;=\;\exp\left\{-\text{i}\;\frac{\alpha-2\gamma}{2}\;[m-n+K(s-t)]\right\}\\[1ex]
            &\stackrel{\alpha=\alpha_{3}}{=}\exp\left\{\text{i}\;\gamma\;[m-n+K(s-t)]\right\}.
    \end{aligned}
    \label{SU(2)_PB_compensation}
\end{equation}

When Eqs.\eqref{PB_SU(2)_green}, \eqref{PB_SU(2)_blue}, and \eqref{SU(2)_PB_compensation} are combined with Eq.\eqref{SU(2)_coordinate_with_PB}, continuous evolution of the SU(2) mode along the closed path is established, as confirmed in the Supplemental Material~\cite{SM} Figs.~S2-S3, where modes \ding{172} and \ding{173} are identical, as are \ding{174} and \ding{175}. Additionally, the final mode, including the PB phase, is obtained by applying Eq.~\eqref{SU(2)_PB_compensation} to the initial mode.

Overall, we obtained three final SU(2) modes to verify our results: 
(i) the mode generated by coherently superposing the eigenstate final modes, see the third line of Fig.~\ref{mode_expansion}; 
(ii) the mode obtained from ``continuous evolution'' on the PS, see Fig.~\ref{fig:pb-phase-mode-transformation}~(d2); 
and (iii) the mode derived by combining Eqs.~\eqref{SU(2)_PB_compensation} and \eqref{SU(2)_coordinate_with_PB}, or equivalently by inserting the SU(2) PB phase into its expression, see Fig.~\ref{initial_and_final_for_SU(2)}~(b). 
These three modes are identical, confirming the validity of Eq.~\eqref{SU(2)_PB_compensation}. 
To maintain generality, we avoid using special values such as $75^\circ$ and set the spherical angle to $\gamma = 70^\circ$ throughout this work.

Note that, the final accumulated PB phase for the SU(2) mode is the sum of the PB phases of all its constituent eigenmodes. Thus, by summing Eq.~\eqref{SU(2)_PB_compensation} over $K$ from $0$ to $M$, we obtain the total PB phase for the SU(2) mode, as given in Eq.~\eqref{SU(2)_PB_phase}, see details in the section III. \textit{Derivation of PB and Gouy Phases in SU(2) Beams} of the Supplemental Material~\cite{SM}. Additionally, the Gouy phase for the SU(2) mode is derived by summing the Gouy phases of its constituent eigenmodes, leading to Eq.~\eqref{SU(2)_Gouy_phase}.

\begin{align}
    \Theta_{\sss{\text{SU(2)}}}^{\sss{\text{PB}}} &=\sum_{\sss{K=0}}^{\sss M}\gamma[m-n+K(s-t)]
    \label{SU(2)_PB_phase}\\
    \Theta_{\sss{\text{SU(2)}}}^{\sss{\text{Gouy}}} &= -\sum_{\sss{K=0}}^{\sss{M}} \tan^{\!\sss-1}\!\left(\frac{z}{z_{\sss R}}\right)\;[K(s\!+\!t)\!+\!m\!+\!n\!+\!1]
    \label{SU(2)_Gouy_phase}
\end{align}

\textit{Summary.} We explore the PB phase in SU(2) modes. 
By realizing continuous evolution on the PS, we derive the compensated phase factors and then the PB phase for SU(2) modes, see Eq.\eqref{SU(2)_PB_phase}. The consistency between the mode profiles obtained through continuous evolution, coherent superposition, and phase compensation validates this framework. In addition, the Gouy phase of the coherent mode, given by Eq.~\eqref{SU(2)_Gouy_phase}, plays a complementary role in the overall phase dynamics. For a geometric viewpoint, a heuristic interpretation based on fiber-bundle 
structure is provided in Section~VII, 
\textit{A Heuristic Fiber-Bundle Interpretation of the PB Phase in SU(2) Beams}, 
of the Supplemental Material~\cite{SM}.

\section{Non-Interferometric Measurement of Geometric Phase}

\subsection{Physical basis of non-interferometric measurement}

\begin{figure*}[!htbp]
    \centering
    \includegraphics[width=\textwidth]{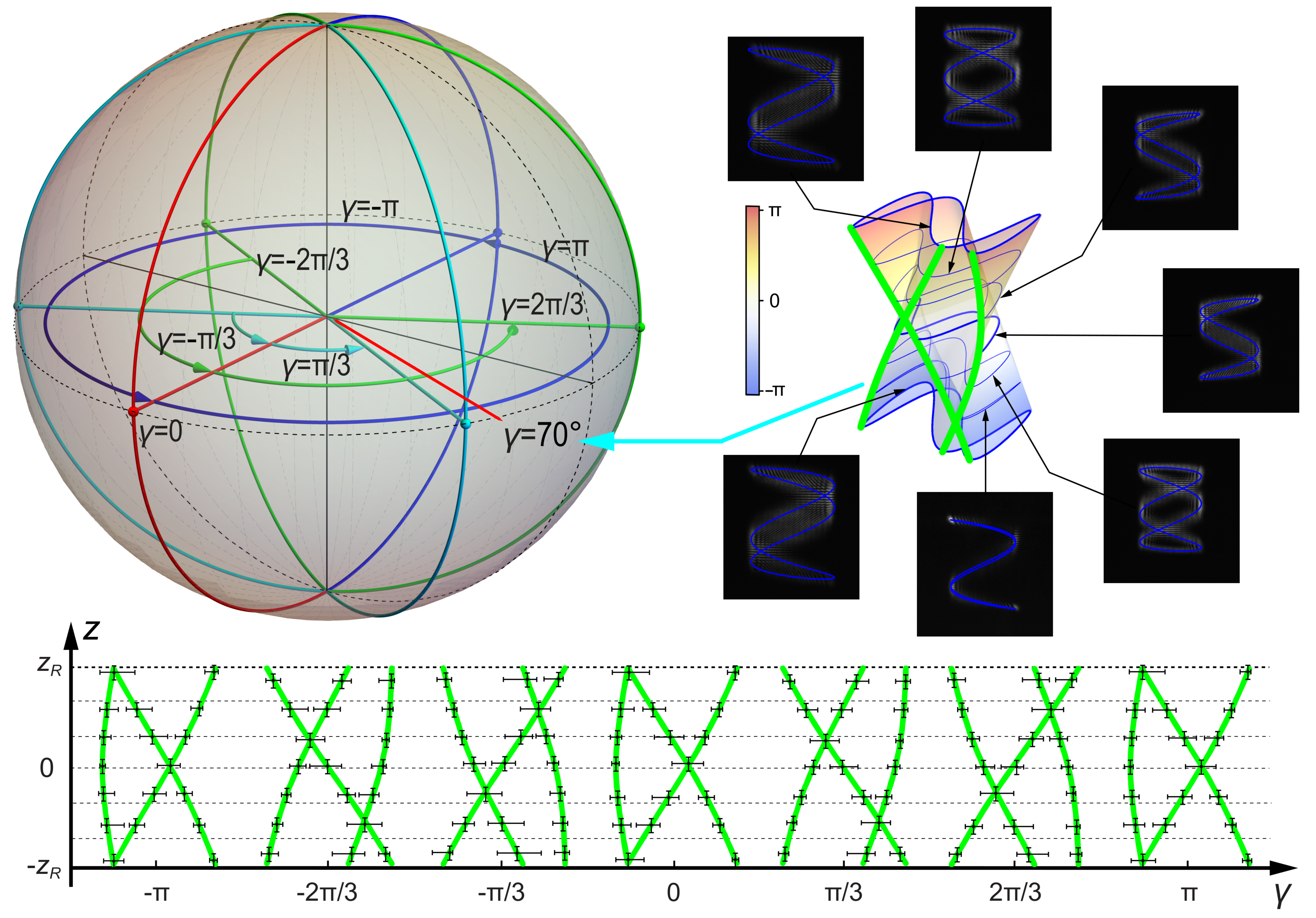}
    \caption{
    Experimental visualization of geometric phase effects in SU(2) modes.  
    Top-Left: the Poincaré sphere shows multiple geodesic loops enclosing different solid angles \(\Omega=2\gamma\), each corresponding to a closed-path transformation of the SU(2) mode. These \(\gamma\) values match the horizontal axis in the feature line plot at the bottom.  
    Top-Right: the 3D wave packet surface at \(\gamma = 70^\circ\), a representative case selected to visualize the spatial structure of the feature line on the parametric surface. The transverse intensity distributions along the propagation direction (\(z\)) span the Rayleigh range \(z_{\sss{R}}\) and can be used to extract Gouy phase characteristics.  
    Bottom: evolution of feature lines versus \(\gamma\), reflecting the PB phase accumulation during the mode transformation. Experimental data are shown as error bars for PB analyses.}
    \label{fig:wave_packet_surface_SU2_gamma_70}
\end{figure*}

This section examines how the PB phase influences the spatial structure of SU(2) wave packets. An analogy is drawn between the formation of SU(2) coherent states through superposition and multibeam interference. As shown in Supplemental Material~\cite{SM} Section VI. \textit{Multibeam Interference Patterns} (Fig. S12), interference patterns generated by Gaussian beams resemble the spatial profiles of SU(2) states~\cite{liu2020investigation,kumar2015digitally,vala2016multiple}. It is well established that the PB phase causes a measurable shift in two-beam interference fringes~\cite{hannonen2020measurement,pancharatnam1956generalized,anandan1992geometric,PhysRevA.99.053826}. Interferometric methods then extract the PB phase by analyzing these fringe shifts in intensity. Moreover, because an SU(2) coherent state beam is formed by superposing multiple Gaussian modes, its spatial profile resembles interference fringes.
Expectedly, variations in the PB phase reshape the spatial profile of the SU(2) coherent state beam. As a result, the PB phase can be determined directly from the beam’s spatial profile rather than only from interference intensity—this forms the basis of our measurement technique that does not require an interferometer. In our work, we use the feature line on the 3D wave-packet surface as a spatial reference (see Fig.~\ref{fig:wave_packet_surface_SU2_gamma_70}). Table~\ref{Law_of_mode} lists feature-line periodic law concerning the PB phase. Here, \( P(\gamma) \) represents the periods of transverse packet intensity distribution changes with respect to \(\gamma\), calculated using \( 2\pi / |s - t| \). For a detailed derivation of this relationship, readers may consult Supplemental Material~\cite{SM} Section IV. \textit{Periodicity of the PB Phase}.

\begin{table}[!htbp]
\centering
\caption{
Feature-line periods of SU(2) beam intensity as a function of the spherical angle $\gamma$.
Each column corresponds to a distinct SU(2) mode configuration $(s,t)$, and $P(\gamma)$ denotes
the corresponding feature-line period associated with the geometric evolution on the modal Poincaré sphere.}
\label{Law_of_mode}
\renewcommand{\arraystretch}{1.25}
\setlength{\tabcolsep}{4.5pt}
\begin{tabular}{c|ccccccccc}
\hline\hline
$(s,t)$ 
& $(1,3)$ & $(3,1)$ & $(0,4)$ & $(4,0)$ 
& $(-1,5)$ & $(5,-1)$ & $(-2,6)$ & $(6,-2)$ & $(2,2)$ \\ \hline
$P(\gamma)$ 
& $\pi$ & $\pi$ & $\tfrac{\pi}{2}$ & $\tfrac{\pi}{2}$
& $\tfrac{\pi}{3}$ & $\tfrac{\pi}{3}$ & $\tfrac{\pi}{4}$ 
& $\tfrac{\pi}{4}$ & $0$ \\ 
\hline\hline
\end{tabular}
\end{table}

\subsection{PB Phase Dynamics in SU(2) Modes}

\begin{figure*}[!htbp]
    \centering
    \includegraphics[width=\linewidth]{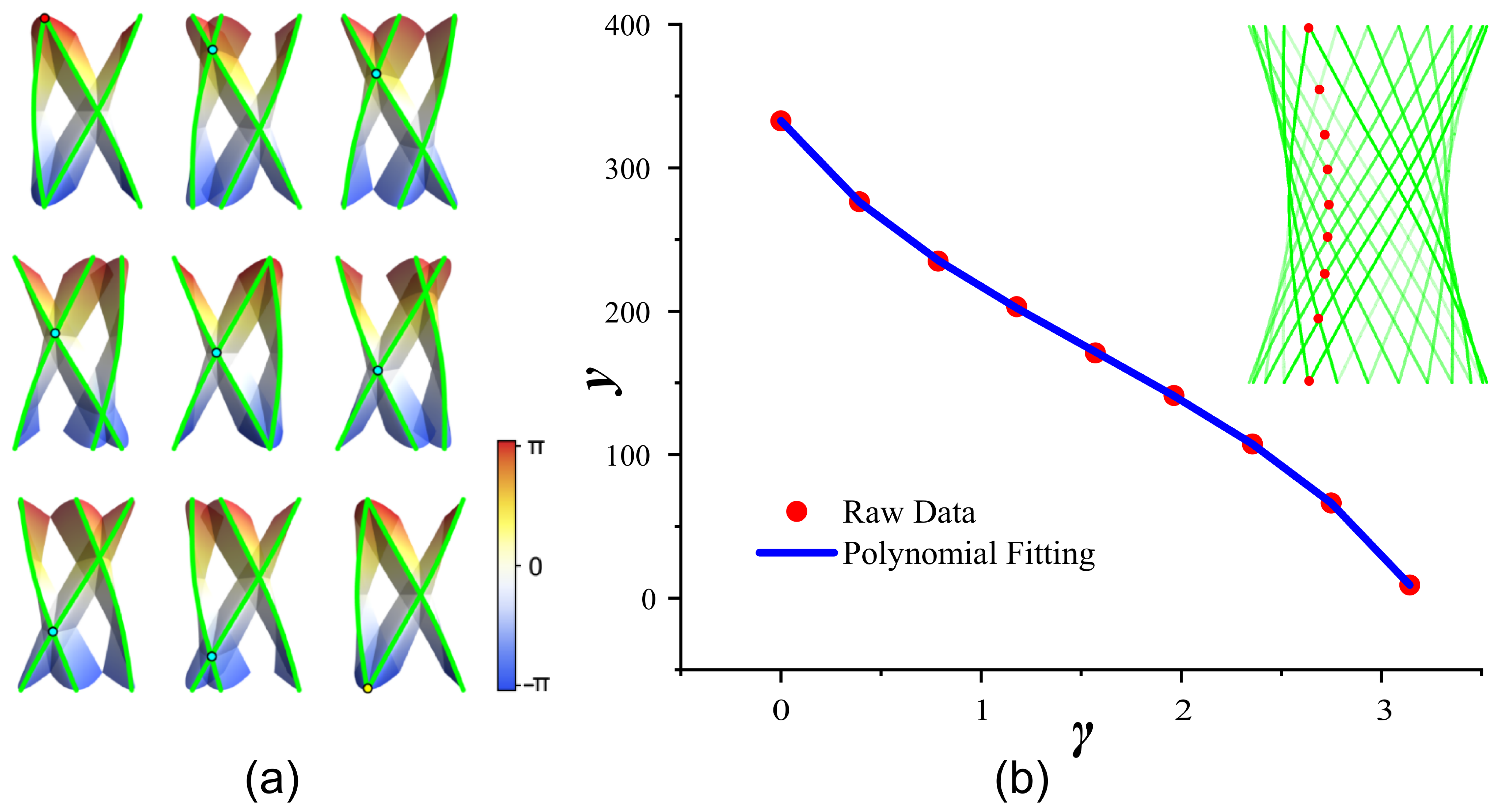}
    \caption{
    Illustration of PB phase determination using polynomial curve fitting.
    (a) Intersection points of two feature lines serve as indicators of the PB phase.
    The red- and yellow-faced circles mark the initial (\(\gamma = 0^{\circ}\)) and final (\(\gamma = \pi\)) intersections, respectively, corresponding to the period of the 3D wave packet for \((s,t) = (1,3)\).
    The associated \(\gamma\) values and PB phases are summarized in Table~\ref{table_for_SU2_parameters}.
    (b) Fifth-order polynomial fitting of the obtained data. The red dots denote the intersection points extracted from the superposed feature lines (inset), which serve as discrete \(\gamma\)–\(y\) data pairs for the fit. The polynomial curve smoothly interpolates these points, enabling continuous estimation of the PB phase from the analyzed intersections.}
    \label{fig:PB_phase_determination_and_fitting}
\end{figure*}

Fig. ~\ref{fig:pb-phase-mode-transformation} shows how the PB phase affects HG and SU(2) modes. For SU(2) modes, both the phase (shown by hue) and the intensity pattern change with the PB phase. For non-superposition modes, the PB phase affects only the phase distribution (hue), while the intensity remains the same. This leads to visible differences in the final output patterns. In SU(2) modes, because the PB phase alters both the phase and intensity, the change in the spatial pattern alone can be used to identify the PB phase. In contrast, for eigenmodes, the PB phase does not change the intensity profile and cannot be directly observed. Instead, it must be measured indirectly through interference-based methods, such as analyzing shifts in interference fringes. The non-interferometric measurement of the geometric phase is based on how the PB phase changes the three-dimensional structure of an SU(2) coherent wave packet. This is similar to how interference affects Gaussian beams, with both leading to new spatial patterns. To measure the PB phase in an SU(2) beam, we use specific features of the wave packet as references. In particular, the “feature line” on the 3D wave packet surface (see Fig.~\ref{fig:PB_phase_determination_and_fitting}(a)), serves as a key reference. Understanding how this feature line changes with the PB phase is crucial, and the detailed results are provided in the Supplemental Material~\cite{SM} Section V. \textit{Intensity and Mode Distribution Dynamics}.

\subsection{Non-interferometric Measurement Implementation}

\begin{figure*}[!htbp]
    \centering
    \includegraphics[width=\linewidth]{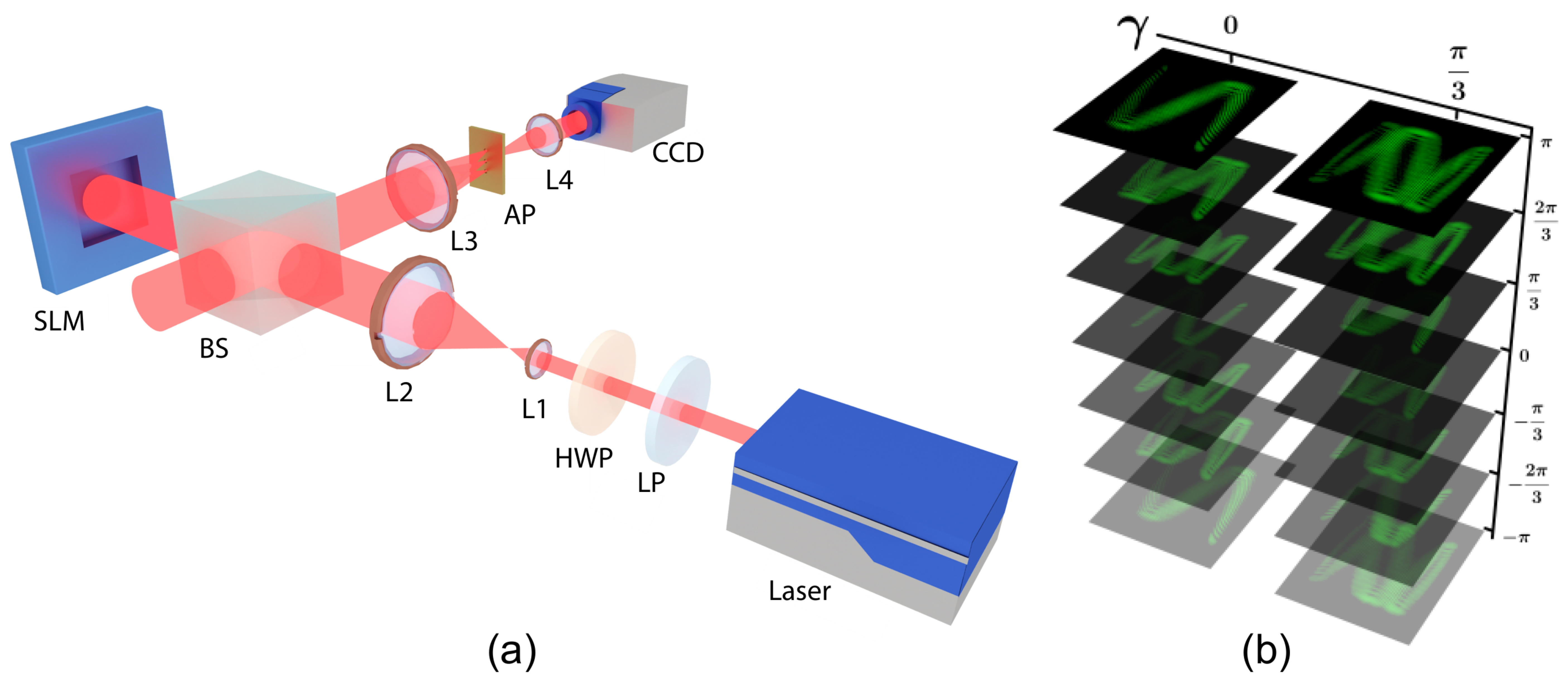}
    \caption{
    Experimental setup for observation of SU(2) mode transformation and PB phase accumulation:
    (a) Experimental setup for generating and analyzing SU(2) geometric modes. HWP: half-wave plate; LP: linear polarizer; L1-L4: lenses; BS: beam splitter; SLM: spatial light modulator; AP: aperture; CCD: charge-coupled device. 
    (b) Experimental observation of SU(2) mode transformation and PB phase accumulation obtained using the setup in (a). Left: Initial SU(2) mode; Right: Final mode after completing a closed loop in the parameter space (\(\gamma=\pi/3\)), with the accumulated PB phase determined by analyzing the feature lines.}
    \label{fig:experimental_setup}
\end{figure*}

This section describes the implementation and validation of a non-interferometric method for measuring the PB phase in SU(2) coherent state beams, as illustrated in Fig.~\ref{fig:experimental_setup}(a). A continuous-wave laser provides the light source, followed by a linear polarizer (LP) and a half-wave plate (HWP) to prepare the desired polarization state. 
The beam is expanded into a near-plane wave using a telescope (L1 and L2) and directed onto a phase-only spatial light modulator (SLM), which displays a sequence of computer-generated holograms (CGHs)~\cite{arrizon2007pixelated,rosales2017shape}.
These CGHs encode the phase profiles required for SU(2) mode evolution along closed geodesic paths on the Poincar\'e sphere.
Following the method developed by Wan~\cite{wan2018quadrant}, this approach enables high-fidelity, cavity-free generation and transformation of SU(2) modes via spatial modulation.
The modulated beam passes through a 4f imaging system (L3 and L4) with a spatial filter (SF) at the Fourier plane to isolate the desired diffraction order, and is finally imaged onto a CCD camera. After completing a full transformation cycle, the beam returns to its original spatial configuration with an accumulated PB phase.

Overlaid dark error bars indicate experimental uncertainties in PB phase extraction. The top-right panel illustrates the simulated 3D wave packet surface at \(\gamma = 70^\circ\), selected as a representative case to visualize the spatial structure of the feature line on the parametric surface. For illustration, the transverse intensity profiles along the propagation direction (\(z\)), normalized by the Rayleigh range, are also shown. Although not used for quantitative analysis, these profiles offer additional insight into the beam’s internal structure. The consistency between experimental and theoretical feature lines—both in their spatial configuration and in their \(\gamma\)-dependent evolution—demonstrates the accuracy and robustness of our spatial-structure-based, non-interferometric approach for PB phase characterization.

Fig.~\ref{fig:wave_packet_surface_SU2_gamma_70} presents a comprehensive visualization of the geometric phase characterization.
The top-left panel shows multiple closed geodesic loops on the PS, each enclosing a distinct solid angle \(\Omega = 2\gamma\); these loops represent the transformation paths of SU(2) modes.
These \(\gamma\) values match the horizontal axis of the bottom panel, which shows how the feature lines evolve with \(\gamma\). And the dark error bars indicate the experimental uncertainties. The top-right panel shows the 3D wave packet surface for \(\gamma = 70^\circ\), which is selected as a representative case to visualize the spatial configuration of the feature line on the parametric surface. And the transverse intensity profiles of this 3D wave packet surface are also shown. Although not used for quantitative analysis, these profiles offer additional insight into the beam’s internal structure. We compare simulated and measured SU(2) modes as they evolve along a closed geodesic loop.  
The PB phase is extracted by analyzing the evolution of feature lines on the 3D wave packet surface, which enables a quantitative, non-interferometric measurement of the PB phase.  
The strong agreement confirms the validity of this spatial-structure-based method. Note that for the selected case \((s,t,M,\gamma,m_0,n_0) = (1,3,6,7\pi/18,18,30)\), the PB phase calculated by Eq.~\eqref{SU(2)_PB_phase} is
\(\Phi_{\sss{\text{SU(2)}}} = \pi\), whose equivalent in \([0^\circ, 360^\circ)\) is
\(\Phi'_{\sss{\text{SU(2)}}} = 180^\circ\).

The PB phase can be obtained directly from experimental results by identifying the intersection points of two feature lines, as shown in Fig.~\ref{fig:PB_phase_determination_and_fitting}(b).
The highest intersection corresponds to $\gamma=0$, and the next one below corresponds to $\gamma=\pi/8$, with subsequent intersections continuing in this pattern.
Substituting these $\gamma$ values into Eq.~\eqref{SU(2)_PB_phase} yields the corresponding PB phases $\Phi{\sss{\text{SU(2)}}}$.
The summarized results are presented in Table~\ref{table_for_SU2_parameters}, where the first row lists the spherical angle $\gamma$, the second row provides the calculated PB phases $\Phi_{\sss{\text{SU(2)}}}$, and the third row shows the modulo-$2\pi$–normalized phases $\Phi'_{\sss{\text{SU(2)}}}$.
The last row gives the pixel coordinate $y$ extracted from the processed experimental images.
These $y$ values correspond to the positions of the feature-line intersections for each $\gamma$ and serve as input for subsequent modeling.
Although slight variations may occur depending on the image-processing method, the final PB phase extraction remains unaffected.
In this work, we directly record the pixel coordinates of the intersection points, providing a straightforward and reproducible measurement approach.

\begin{table}[!htbp]
\centering
\caption{
Summary of SU(2) mode evolution parameters.
The first row lists the spherical angle $\gamma$, followed by the associated PB phase factors 
$\Phi_{\sss{\text{SU(2)}}}$ and $\Phi'_{\sss{\text{SU(2)}}}$, and the corresponding $y$ positions (in pixels) extracted from image processing.
Here, $\Phi_{\sss{\text{SU(2)}}}$ represents the PB phase calculated from Eq.~\eqref{SU(2)_PB_phase}, while 
$\Phi'_{\sss{\text{SU(2)}}}$ denotes the same phase values after applying a modulo $2\pi$ normalization to confine them within one full phase cycle ($0$--$2\pi$).
}
\label{table_for_SU2_parameters}
\renewcommand{\arraystretch}{1.25}
\setlength{\tabcolsep}{4.5pt}
\begin{tabular}{c|ccccccccc}
\hline\hline
$\gamma$ 
& $0$ & $\tfrac{\pi}{8}$ & $\tfrac{\pi}{4}$ & $\tfrac{3}{8}\pi$ 
& $\tfrac{\pi}{2}$ & $\tfrac{5}{8}\pi$ & $\tfrac{3}{4}\pi$ 
& $\tfrac{7}{8}\pi$ & $\pi$ \\ \hline
$\Phi_{\sss{\text{SU(2)}}}$ 
& $0$ & $-\tfrac{63}{4}\pi$ & $-\tfrac{63}{2}\pi$ & $-\tfrac{189}{4}\pi$ & $-63\pi$ 
& $-\tfrac{315}{4}\pi$ & $-\tfrac{189}{2}\pi$ & $-\tfrac{441}{4}\pi$ & $-126\pi$ \\[4pt]
$\Phi'_{\sss{\text{SU(2)}}}$ 
& $0$ & $\tfrac{\pi}{4}$ & $\tfrac{\pi}{2}$ & $\tfrac{3}{4}\pi$ & $\pi$ 
& $\tfrac{5}{4}\pi$ & $\tfrac{3}{2}\pi$ & $\tfrac{7}{4}\pi$ & $0$ \\[4pt]
$y$ 
& $332.7$ & $276.3$ & $234.9$ & $203.1$ & $171$ & $141.3$ & $107.4$ & $66.3$ & $9.2$ \\ 
\hline\hline
\end{tabular}
\end{table}

Next, we apply a fifth-order polynomial fitting to the extracted intersection points to characterize the relationship between the PB phase and the spherical angle \(\gamma\).
This approach provides a simple yet accurate means of estimating the geometric phase, as the fitted curve captures the smooth dependence of the PB phase on \(\gamma\) while remaining robust against minor variations in the analyzed data.

\subsection{Determining PB Phase by Extrapolation}

We apply polynomial curve fitting to model the relationship between \(y\) and \(\gamma\). To accurately capture the underlying trend, a fifth-order polynomial is used for fitting. The resulting relationship is given in Eq.~\eqref{curve_fitting}. For example, using this equation with \(y = 100\) gives a predicted value of \(\gamma = 2.435\) (\(139.5^\circ\)), which corresponds to a PB phase of \(\Phi^{\prime}_{\text{SU(2)}} = 243^\circ\) after modulo adjustment. As shown in Fig.~\ref{fig:PB_phase_determination_and_fitting}b, the fitted curve follows all experimental data points closely, confirming the numerical consistency of the extrapolation. The high-order polynomial provides sufficient flexibility to describe the nonlinear relationship between \(\gamma\) and \(y\) while maintaining smooth continuity across the full phase range. 

\begin{widetext}
    \begin{equation}
        y(\gamma) = -1.73\gamma^5 + 13.62\gamma^4 - 49.2\gamma^3 + 96.86\gamma^2 - 175.56\gamma + 332.77
    \label{curve_fitting}
    \end{equation}
\end{widetext}

Although the polynomial reproduces all the experimental data, it can, in principle, be sensitive to small measurement variations. To verify the stability of the fitted parameters, we examined multiple subsets of the data and found that the coefficients of the polynomial remained nearly unchanged, suggesting that the model captures the essential geometric dependence rather than noise-induced fluctuations. Accordingly, the polynomial fit was adopted as a reliable and transparent method for PB phase prediction.
The full fitting code, together with the visualization videos illustrating the evolution of feature lines with the geometric phase, is openly available in the accompanying Zenodo repository~\cite{li_2025_su2viz}.

\section{Conclusion}

This study examined the generation and measurement of the PB phase in SU(2) coherent states. To improve conceptual clarity, we utilized the Poincaré sphere for visualization and the PB phase formalism as a theoretical basis. This approach provided insights into the coherent superposition mechanisms of SU(2) wave packets, facilitating understanding of their spatial synthesis and evolution. Our theoretical framework was tested experimentally, incorporating a step-by-step geometric phase extraction procedure. Data analysis employed high-order polynomial extrapolation, which ensured accurate prediction of PB phases while maintaining stability across the measured domain.

We investigated the parameter \( M \)'s role in non-interferometric measurements, demonstrating that larger \( M \) values improve the ``trace'' property for more accurate PB phase extraction. However, current computational constraints limit exploration of the high-\( M \) regime. Future work could address this limitation and explore specific cases like \( \Omega = 1/8 \) to potentially uncover further aspects of PB phase dynamics in structured Gaussian beams. Overall, this work presents a framework for analyzing PB phases within the complex structured light fields of SU(2) coherent states.

\newpage
\section{Acknowledgments}

We thank Dr. Zhensong Wan for his assistance in collecting necessary experimental results for this study.

Y. Shen acknowledges the support from Nanyang Technological University Start Up Grant, Singapore Ministry of Education (MOE) AcRF Tier 1 grant (RG157/23), MoE AcRF Tier 1 Thematic grant (RT11/23), and Imperial-Nanyang Technological University Collaboration Fund (INCF-2024-007).

\nocite{shen2020su2, forbes2021structured,
          chen2006devil, chen2019laser, chen2019origin,
          chen2010spatial, leinonen2023noncyclic}
          
\bibliography{references}
\end{document}


\preprint{APS/123-QED}

\title{Supplementary Materials for\\
\textbf{Visualized Geometric Phase of Caustic Geometric Beams}} 

\author{
Haiyang Li$^{1}$, and Yijie Shen$^{2,3,*}$\\[1ex]
\small $^{1}$Department of Chemical Engineering, Pohang University of Science and Technology, Pohang 37673, Korea\\
\small $^{2}$Centre for Disruptive Photonic Technologies, School of Physical and Mathematical Sciences \& The Photonics Institute,\\
\small Nanyang Technological University, Singapore 637371, Singapore\\
\small $^{3}$School of Electrical and Electronic Engineering, Nanyang Technological University, Singapore 639798, Singapore
}

\maketitle

\begin{center}
\textbf{Corresponding author:} \texttt{yijieshen@ntu.edu.sg}
\end{center}

\vspace{2em}
\textbf{The PDF file includes:}
\vspace{-0.5em}

\begingroup
\setlength{\itemsep}{0.8em}
\setlength{\parskip}{0pt}
\singlespacing
\begin{itemize}[label={}, leftmargin=2.5em]
  \item Supplementary Figures S1 to S13
  \item Process of continuous evolution (Figs.~S1--S3)
  \item Continuous evolution under mode transformation (Figs.~S4--S5)
  \item Derivation of PB and Gouy phases
  \item PB phase periodicity analysis (Fig.~S6)
  \item Intensity and mode distribution dynamics (Figs.~S7--S11)
  \item Multibeam interference patterns (Fig.~S12)
  \item Fiber-bundle interpretation of SU(2) PB phase (Figs.~S13)
\end{itemize}
\endgroup

\newpage
\tableofcontents

\newpage
\section{Process of Continuous Evolution}

This section illustrates how both eigenmodes and SU(2) coherent modes undergo continuous evolution on the Poincaré sphere. Figure~S1 shows that for eigenmodes, the PB phase leads to a global phase shift without altering the intensity profile. In contrast, Figures~S2 and S3 reveal that SU(2) coherent modes exhibit structural changes in both phase and intensity due to the PB phase, indicating a deeper geometrical transformation beyond mere phase redistribution.

\begin{figure*}[!htbp]
    \centering
    \includegraphics[width=0.8\textwidth]{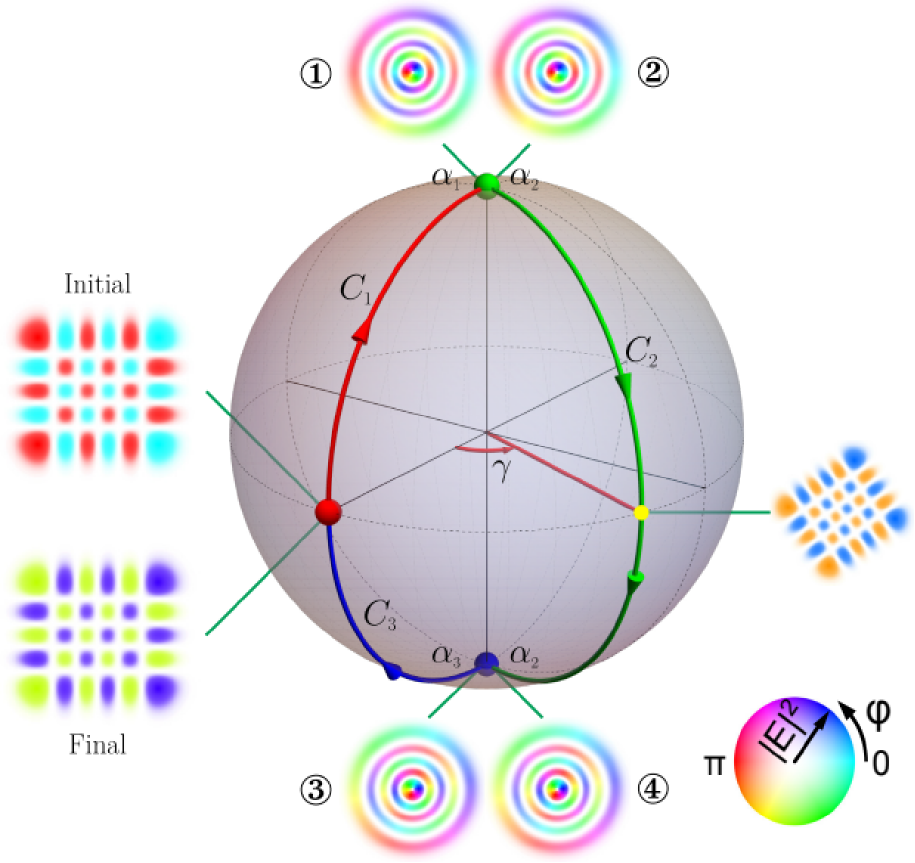}
    \caption{
    Schematic diagram illustrating the continuous evolution in parameter space and the generation of the PB phase for non-superposition modes: 
    The initial mode, an HG\(_{\sss{5,4}}\) mode, starts from an equatorial point \((\alpha,\beta)=(0,\frac{\pi}{2})\). It evolves along the red arc \(C_1\) to point \(\alpha_1\), the green arc \(C_2\) to point \(\alpha_2\), and the blue arc \(C_3\) to point \(\alpha_3\). This transformation follows geodesics, composed of three arc segments in red, green, and blue. The mode first evolves to reach the north pole, a phase singularity. This implies that the mode at the end of the red arc (denoted by \ding{172}) should differ from the mode at the start of the green arc (denoted by \ding{173}) due to the singularity. However, through phase compensation, these two modes are made identical, achieving a purely bijective mapping between modes and points along the closed transformation path on a PS. Similarly, modes \ding{174} and \ding{175} are also made identical by phase compensation. This approach ensures that the requirement for continuous evolution is satisfied throughout the transformation path. Ultimately, the final mode is obtained by attaching a PB phase to the initial mode as a result of the mode transformation. This modification alters the phase distribution of the initial mode. Consequently, the initial and final modes exhibit distinct hues, illustrating the impact of the PB phase on the mode's phase characteristics.}
    \label{OMA_PS_with_numbers}
\end{figure*}

\newpage
\begin{figure*}[!htbp]
    \centering
    \includegraphics[width=0.8\textwidth]{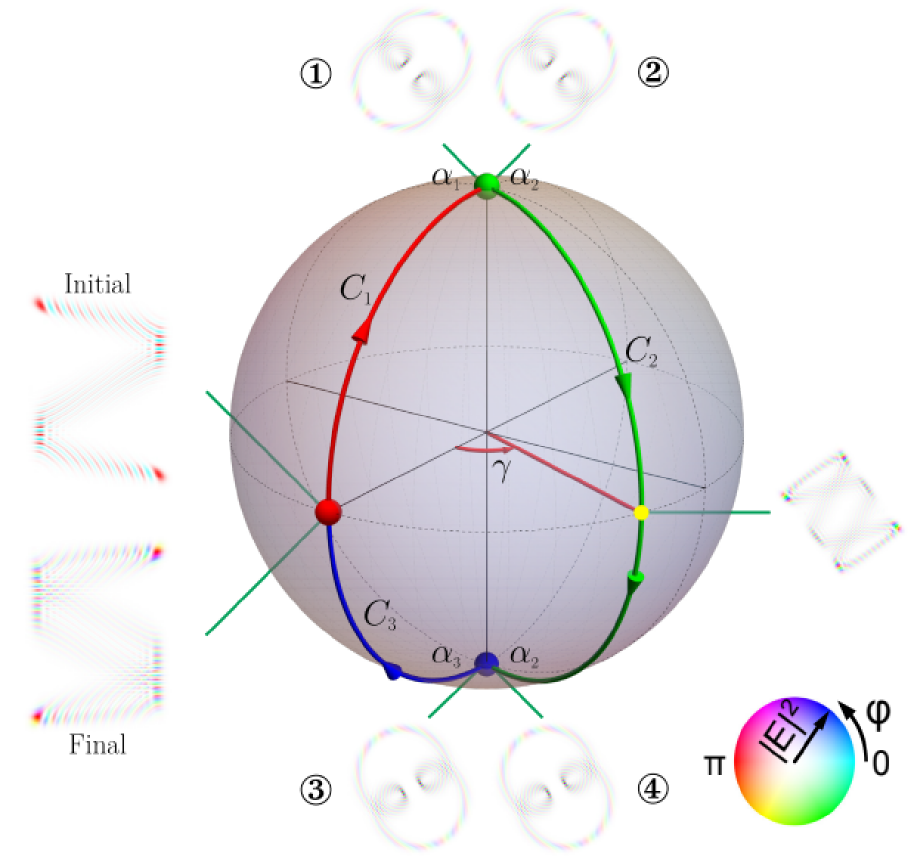}
    \caption{
    Schematic of the continuous evolution of SU(2) wave packets on the PS for \((s,t) = (1, 3)\): 
    Starting from \((\alpha, \beta) = (0, \frac{\pi}{2})\), the initial mode evolves along the red arc \(C_1\) to point \(\alpha_1\), the green arc \(C_2\) to point \(\alpha_2\), and the blue arc \(C_3\) to point \(\alpha_3\). The evolution follows geodesic lines and involves three segments of continuous mode transformation. The initial mode starts from an equatorial point and evolves through these three arc segments to reach a phase singularity at the north pole. Due to the singularity, the mode at the end of the red arc (\ding{172}) should differ from the mode at the start of the green arc (\ding{173}), but phase compensation ensures they are identical. Similarly, phase compensation aligns modes \ding{174} and \ding{175}. This ensures continuous evolution along the transformation path. The final mode, obtained by attaching a PB phase to the initial mode, not only alters the phase distribution but also changes the intensity distribution of the SU(2) coherent state wave packets, making non-interferometric measurements of the PB phase possible.}
    \label{SU(2)_PS_with_numbers_13}
\end{figure*}

\newpage
\begin{figure*}[!htbp]
    \centering
    \includegraphics[width=0.8\textwidth]{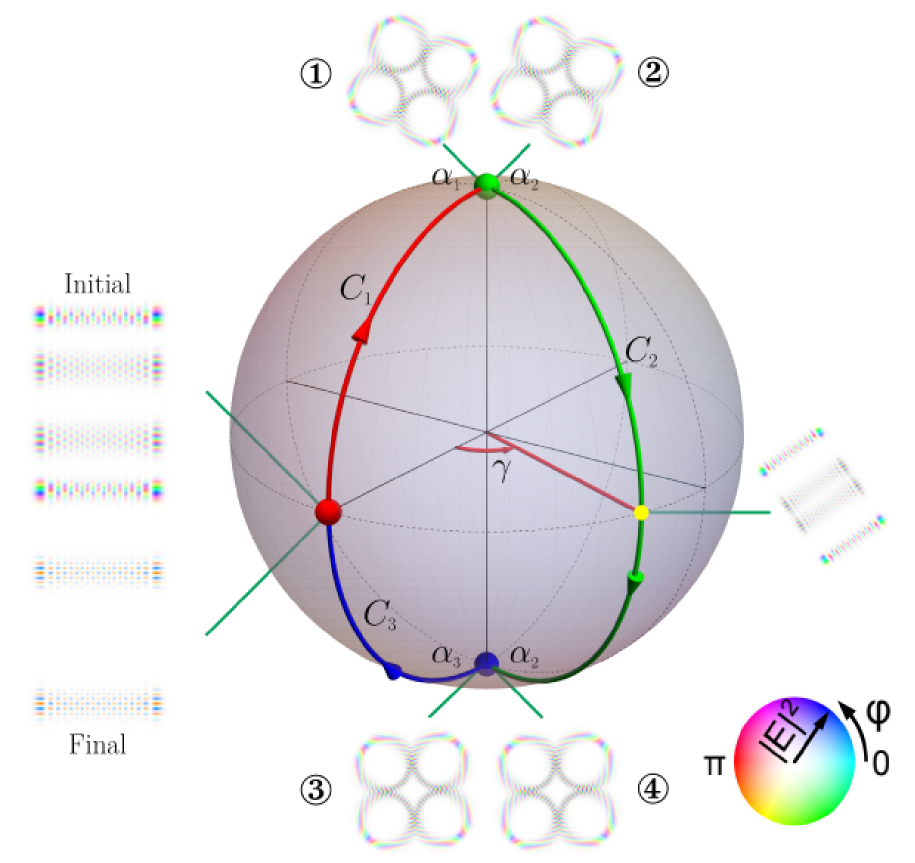}
    \caption{Schematic of the continuous evolution of SU(2) wave packets on the SU(2) PS for \((s,t) = (4, 0)\): Starting from \((\alpha, \beta) = (0, \frac{\pi}{2})\), the initial mode follows a similar transformation path as in the \((s,t) = (1, 3)\) case, evolving through three geodesic arc segments to reach the north pole. Phase compensation ensures continuous mode transformation at the singularity points. The final mode, with an attached PB phase, shows altered phase and intensity distributions, making non-interferometric measurements of the PB phase possible.}
    \label{SU(2)_PS_with_numbers_40}
\end{figure*}

\section{Continuous Evolution under Mode Transformation}

This section gives a more detailed explanation of the concept of continuous evolution.
For simplicity, consider selecting a point on the equator of the OAM PS as the initial point, typically designated by \((\alpha,\beta)=(0,\pi/2)\), which corresponds to the \(\mathrm{HG}_{\mathrm{m,n}}\) mode. This mode then undergoes continuous evolution on the PS. During this process, an additional phase--the PB phase--appears when the mode completes a full cycle and returns to the starting point. This continuous evolution must meet three conditions:

\begin{enumerate}
    \item During evolution, the mode's distribution stays continuous at every point \(q\) on the PS. This includes both the phase distribution \(f_{\sss P}(q)\) and the intensity distribution \(f_{\sss I}(q)\). Here, \(q\) represents any point on the PS.
    
    \item The intensity distribution \(f_{\sss I}(q)\) forms a bijective (one-to-one) mapping at each point \(q\) on the PS. This means every intensity distribution corresponds to exactly one point on the sphere, and vice versa.
    
    \item The phase distribution \(f_{\sss P}(q)\) also forms a bijective mapping, but at each point \(q^{\prime}\) along the closed path traced by continuous mode transformation. Here, \(q^{\prime}\) is any point on this closed path (where the transformation happens), which is different from \(q\) (any point on the PS).\\
    (Note: Before the continuous evolution is made, only the polar points did not satisfy this condition.)
\end{enumerate}

\begin{figure*}[!htbp]
    \centering
    \begin{minipage}{0.85\textwidth}
    \centering
        \subfloat[\((0, 0)\)]{\includegraphics[width=0.15\textwidth]{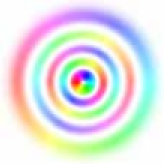}}
        \,
        \subfloat[\(\left(\frac{\pi}{2}, 0\right)\)]{\includegraphics[width=0.15\textwidth]{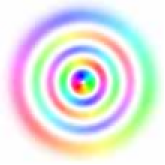}}
        \,
        \subfloat[\((\pi, 0)\)]{\includegraphics[width=0.15\textwidth]{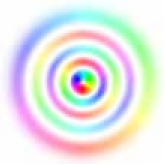}}
        \,
        \subfloat[\(\left(\frac{3\pi}{2}, 0\right)\)]{\includegraphics[width=0.15\textwidth]{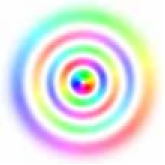}}
        \,
        \subfloat[\((2\pi, 0)\)]{\includegraphics[width=0.15\textwidth]{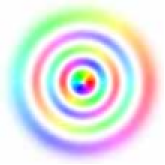}}
        \\
        \subfloat[\((0, \frac{\pi}{2})\)]{\includegraphics[width=0.15\textwidth]{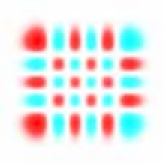}}
        \,
        \subfloat[\(\left(\frac{\pi}{2}, \frac{\pi}{2}\right)\)]{\includegraphics[width=0.15\textwidth]{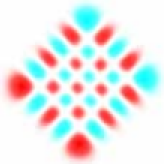}}
        \,
        \subfloat[\((\pi, \frac{\pi}{2})\)]{\includegraphics[width=0.15\textwidth]{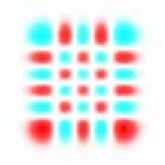}}
        \,
        \subfloat[\(\left(\frac{3\pi}{2}, \frac{\pi}{2}\right)\)]{\includegraphics[width=0.15\textwidth]{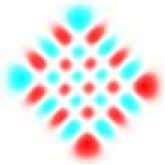}}
        \,
        \subfloat[\((2\pi, \frac{\pi}{2})\)]{\includegraphics[width=0.15\textwidth]{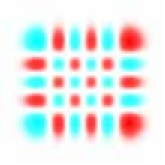}}
        \\
        \subfloat[\((0, \pi)\)]{\includegraphics[width=0.15\textwidth]{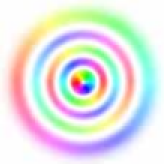}}
        \,
        \subfloat[\(\left(\frac{\pi}{2}, \pi\right)\)]{\includegraphics[width=0.15\textwidth]{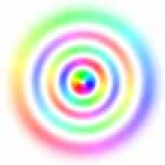}}
        \,
        \subfloat[\((\pi, \pi)\)]{\includegraphics[width=0.15\textwidth]{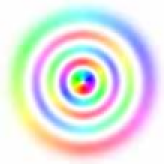}}
        \,
        \subfloat[\(\left(\frac{3\pi}{2}, \pi\right)\)]{\includegraphics[width=0.15\textwidth]{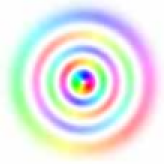}}
        \,
        \subfloat[\((2\pi, \pi)\)]{\includegraphics[width=0.15\textwidth]{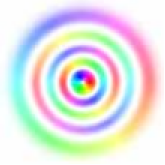}}
    \end{minipage}
    \begin{minipage}{0.13\textwidth}
        \centering
        \includegraphics[width=1\textwidth]{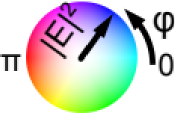}
    \end{minipage}
    \caption{
    Schematic of intensity-phase distributions for HLG modes at various points on the OAM PS: 
    (a-e) Intensity-phase distributions at the north pole (\(\beta = 0\)) for \(\alpha = 0, \; \frac{\pi}{2}, \; \pi, \; \frac{3\pi}{2}, \; 2\pi\). 
    (f-j) Intensity-phase distributions at equatorial points (\(\beta = \frac{\pi}{2}\)) for \(\alpha = 0, \; \frac{\pi}{2}, \; \pi, \; \frac{3\pi}{2}, \; 2\pi\). 
    (k-o) Intensity-phase distributions at the south pole (\(\beta = \pi\)) for \(\alpha = 0, \; \frac{\pi}{2}, \; \pi, \; \frac{3\pi}{2}, \; 2\pi\). 
    Each row corresponds to a different \(\beta\) value, showcasing the intensity-phase distributions for different \(\alpha\) values. The colorbar on the right indicates the intensity and phase. These patterns represent LG modes. Despite involving only two polar points, the phase distribution varies with changing \(\alpha\). 
    This variation leads to a scenario where the phase distribution does not coincide. Consequently, the bijective mapping requirement, which is crucial for achieving continuous mode evolution, is not met. This lack of congruence signifies a break in continuity. Therefore, corrective measures must be taken to restore continuity in the mode transformation process.
    }
    \label{modes_on_OAM_PS}
\end{figure*}

Eq.~\eqref{GG_expression} is continuous and bijective along the closed path during the mode transformation, except at the poles. Hence, the mode can evolve smoothly along any path on the PS, except at the north and south poles. These poles are singular: they include all \((\alpha,\beta)\) with \(\alpha \in [0,2\pi]\) and \(\beta = 0\) (north pole) or \(\beta = \pi\) (south pole).
To better understand the polar singularities, Fig.~\ref{modes_on_OAM_PS} shows the modes on the PS obtained from Eq.~\eqref{GG_expression}.  
The top, middle, and bottom rows display the north-pole, equatorial, and south-pole modes, labeled \(\mathrm{LG}_{\mathrm{n}}\), \(\mathrm{HG}\), and \(\mathrm{LG}_{\mathrm{s}}\), respectively.

\begin{figure*}[!htbp]
    \centering  
    \subfloat{\includegraphics[width=0.45\textwidth]{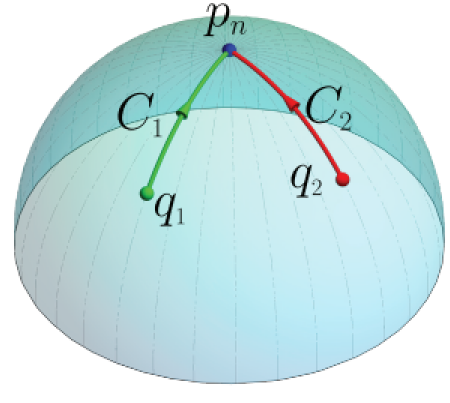}\label{fig:C1_curve}}\hfill
    \subfloat{\includegraphics[width=0.45\textwidth]{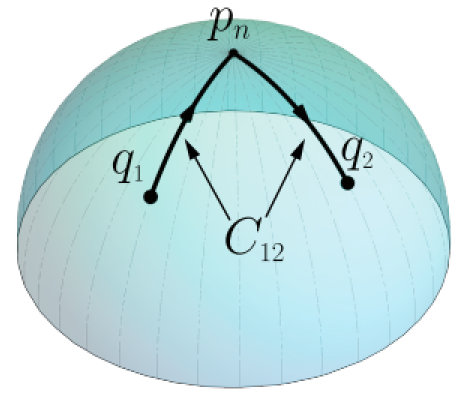}\label{fig:C2_curve}}
    \caption{Schematic of continuous evolution in parameter space: 
    (a) Geodesic \(C_{\sss{1}}\) connects \(q_{\sss{1}}\) to the north pole \(p_{\sss{n}}\), and geodesic \(C_{\sss{2}}\) connects \(q_{\sss{2}}\) to \(p_{\sss{n}}\); the blue dot at \(p_{\sss{n}}\) marks the phase singularity.
    (b) Geodesic \(C_{\sss{12}}\) runs through \(q_{\sss{1}}\), \(p_{\sss{n}}\), and \(q_{\sss{2}}\); phase compensation removes the singularity at \(p_{\sss{n}}\), establishing a one-to-one correspondence between modal states and points along the path.}
    \label{fig:composite_figure_SU2_modes_phase_paths_continuous_evolution_PS}
\end{figure*}

The subscripts \(\mathrm{n}\) and \(\mathrm{s}\) denote the north and south poles.  
The polar modes have a phase singularity, which impedes continuous evolution.
For clarity, we focus on the north pole, labelled \(p_{\sss n}\) (see Fig.~\ref{fig:composite_figure_SU2_modes_phase_paths_continuous_evolution_PS}); the same reasoning applies to the south pole. The intensity distribution \(f_{\sss I}(p_{\sss n})\) is the same for every \(\alpha \in [0,2\pi]\). However, the phase distribution \(f_{\sss P}(p_{\sss n})\) changes with \(\alpha\), so \(p_{\sss n}\) is a phase singularity. 

\newpage

\begin{widetext}
    \begin{equation}
        \mathrm{GG}_{\sss{N, \ell}}(r, \varphi ; \alpha, \beta)=\sum_{\substack{\ell'=-N \\ \text{steps of } 2}}^{N} \underbrace{d^{\frac{N}{2}}_{\frac{\ell'}{2}\frac{\ell}{2}}(\beta)}_{\substack{\text{Wigner d-matrix element} \\ \text{(angular momentum rotation)}}} 
        \cdot \mathrm{e}^{-\mathrm{i}\frac{\ell'\alpha}{2}} \mathrm{i}^{\ell'-|\ell'|+\ell-N} \cdot \mathrm{LG}_{\sss{N,\ell'}}(r,\varphi),
        \label{GG_expression}
    \end{equation}
\end{widetext}

Consider any two distinct points on the PS (see Figs.~\ref{fig:composite_figure_SU2_modes_phase_paths_continuous_evolution_PS}), excluding the poles. Let these points be denoted by \(q_{\sss 1}\) and \(q_{\sss 2}\). The corresponding modes for these points are denoted by \(\mathrm{GG}_1\) and \(\mathrm{GG}_2\), respectively. When these points are connected to the north pole \(p_{\sss n}\) along their respective geodesics, we obtain the curves \(C_1\) and \(C_2\). Furthermore, by connecting \(q_{\sss 1}\), \(p_{\sss n}\), and \(q_{\sss 2}\), we form the curve \(C_{12}\). The polar point corresponds to the \(\mathrm{LG}\) mode. For mode transformation along curve \(C_1\), the north pole is denoted by \(\mathrm{LG}_1\), while for curve \(C_2\), it is denoted by \(\mathrm{LG}_2\). If the mode \(\mathrm{LG}_1\) is identical with \(\mathrm{LG}_2\), then the mode transformation along the curve \(C_{12}\) remains continuous. However, these two \(\mathrm{LG}\) modes have different phase distributions \(f_{\sss P}(\alpha,0)\) due to different \(\alpha\) values. Thus, Eq.~\eqref{mode_relation} gives their relationship

\begin{equation}
    \mathrm{LG}_2 = \exp[\mathrm{i}f(\alpha)]\mathrm{LG}_1,
    \label{mode_relation}
\end{equation}

where \(f(\alpha)\) is a phase function that depends on \(\alpha\), and \(\exp(\mathrm{i}f(\alpha))\) is the corresponding phase factor. Multiplying the modes on curve \(C_2\) by \(\exp(-\mathrm{i}f(\alpha))\) makes the mode transformation along the path \(C_{12}\) continuous; see Eq.~\eqref{add_phase_factor_to_C2}.

\begin{equation}
    \begin{aligned}
        & \exp[-\mathrm{i}f(\alpha)]\mathrm{LG}_2\\[1ex]
        = & \exp[-\mathrm{i}f(\alpha)]\exp[\mathrm{i}f(\alpha)]\mathrm{LG}_1\\[1ex]
        = & \mathrm{LG}_1
    \end{aligned}
    \label{add_phase_factor_to_C2}
\end{equation}

The previous analysis shows that a phase factor is essential for avoiding singularities during mode transformations, so the transformation remains continuous.  
For the closed path in Fig.~\ref{OMA_PS_with_numbers}, two phase factors are required: one at the north pole and one at the south pole. A detailed derivation of these phase factors follows.


The phase factors for the green and blue paths are derived from the initial conditions of the mode transformation. Starting at \((\alpha, \beta) = (0, \pi/2)\), we have \(\alpha_2 = \gamma\). Additionally, since the red and blue paths lie on the same geodesic, \(\alpha_{3} = \alpha_1 = 0\). Substituting these into Eqs.~\eqref{PB_phase for GGmode_north_1} and \eqref{PB_phase for GGmode_south_1} yields the simplified phase factors in Eqs.~\eqref{PB_phase for GGmode_north_2} and \eqref{PB_phase for GGmode_south_2}:

The phase factors for the green and blue segments are derived from the initial conditions of the mode transformation.
Let \(\alpha_{1}, \alpha_{2},\) and \(\alpha_{3}\) be the \(\alpha\) coordinates of the first (red), second (green), and third (blue) segments, respectively.  
Since the red and blue segments share the same geodesic, we obtain \(\alpha_{3}=\alpha_{1}=0\).  
Starting from \((\alpha,\beta)=(0,\pi/2)\), the green segment satisfies \(\alpha_{2}=\gamma\).

\begin{align}
    \Phi_{\mathrm{g|GG}} &= \exp\left[\text{i}\frac{\alpha-\alpha_{1}}{2}(m-n)\right]
    \label{PB_phase for GGmode_north_1} \\[1ex]
    \Phi_{\mathrm{b|GG}} &= \exp\left[-\text{i}(\alpha-\alpha_{2})(m-n)\right]
    \label{PB_phase for GGmode_south_1}
\end{align}

Substituting these values into Eq.~\eqref{PB_phase for GGmode_north_1} and Eq.~\eqref{PB_phase for GGmode_south_1} yields the simplified phase factors given in Eq.~\eqref{PB_phase for GGmode_north_2} and Eq.~\eqref{PB_phase for GGmode_south_2}.

\begin{align}
    \Phi_{\mathrm{g|GG}} &= \exp\left[\text{i}\frac{\alpha}{2}(m-n)\right]
    \label{PB_phase for GGmode_north_2}, \\[1ex]
    \Phi_{\mathrm{b|GG}} &= \exp\left[-\text{i}(\alpha-\gamma)(m-n)\right]
    \label{PB_phase for GGmode_south_2}.
\end{align}


The PB phase is obtained by multiplying the phase factors for the green and blue paths (see Eq.~\eqref{PB_phase_GG}).  
This gives the PB phase \(\Phi^{\sss{\text{PB}}}_{\text{GG}}\), which depends on the angle \(\gamma\) between the geodesics.

\begin{equation}
    \begin{aligned}
        \Phi_{\mathrm{g|GG}} \cdot \Phi_{\mathrm{b|GG}}
        =~~~\!&\exp\left[-\mathrm{i} \frac{\alpha - 2\gamma}{2}(m - n)\right] \\
        \quad\stackrel{\alpha=\alpha_{3}}{=}&\exp\left[\mathrm{i}(m - n)\gamma\right] \\
        =~~~\!&\Phi^{\sss{\text{PB}}}_{\sss\text{GG}}
    \end{aligned}
    \label{PB_phase_GG}
\end{equation}

\section{Derivation of PB and Gouy Phases in SU(2) Beams}

In this section, we detail the derivation of expressions for calculating both the PB and the Gouy phases for an SU(2) beam, thereby reinforcing the validity of prior analyses.

With the spatial propagation factor added, the LG mode expression is given by Eq.~\eqref{normalized_LG_with_z}, as corroborated by several studies~\cite{shen2020su2, shen2021rays}. Here, \(z_{\sss{\mathrm{R}}}=\pi\,\omega_{\sss{0}}^{\sss{2}}/\lambda\) denotes the Rayleigh distance, \(w(z)=w_{\sss{0}}\sqrt{1+(z/z_{\sss{\mathrm{R}}})^2}\) represents the beam width, \(\tilde{z}(x,y,z)=z\bigl[1+(x^{\sss{2}}+y^{\sss{2}})/\bigl(2(z^{\sss{2}}+z_{\sss{\mathrm{R}}}^{\sss{2}})\bigr)\bigr]\) specifies the wavefront curvature, and \(\vartheta(z)=\tan^{-1}(z/z_{\sss{\mathrm{R}}})\) defines the Gouy phase for the fundamental Gaussian mode. Notably, for a Gaussian mode with total mode number \(N\), the Gouy phase is given by Eq.~\eqref{Gouy_phase_of_N_G}.

\begin{widetext}
    \begin{equation}
    \begin{aligned}
        \mathrm{LG}_{\sss{N, \ell}}(x,y,z) = &
        \sqrt{\frac{2^{\sss{| \ell | +1}} \left(\frac{1}{2} (N-\sss{| \ell |})\right)!}{\left(\frac{1}{2} (N+\sss{| \ell |} )\right)!}}\cdot  \frac{1}{w(z)\sqrt{\pi}}\exp\left(-\frac{x^{\sss{2}}+y^{\sss{2}}}{w(z)^{\sss{2}}}\right)
        \cdot\mathrm{e}^{\mathrm{i}\ell \arg (x+\mathrm{i} y)}
        \cdot\frac{\sqrt{x^{\sss{2}}+y^{\sss{2}}}^{\sss{| \ell | }}}{w(z)^{\sss{| \ell | }}}
        \times \\
        & L_{\sss{\frac{1}{2} (N-\sss{| \ell |} )}}^{\sss{| \ell |}}\left(\frac{2\left(x^{\sss{2}}+y^{\sss{2}}\right)}{w(z)^{\sss{2}}}\right)
        \cdot \exp \left(\mathrm{i}\frac{2 \pi}{\lambda}\tilde{z}(x,y,z)-\mathrm{i} (N+1) \vartheta(z)\right)
    \end{aligned}
    \label{normalized_LG_with_z}
    \end{equation}
\end{widetext}

\begin{equation}
    \begin{aligned}
        \Theta_{\sss{\text{GG}}}^{\sss{\text{Gouy}}} 
        &= -\mathrm{i}\,(N+1) \vartheta(z)\\[1ex]
        &=-\,(m+n+1)\,\tan^{-1}(z/z_{\mathrm{R}})
    \end{aligned}
    \label{Gouy_phase_of_N_G}
\end{equation}

To rigorously investigate potential concealed terms in the coherent superposition, we assume \(w_{\sss 0}=1\) and \(\lambda=1\), so that \(z_{\sss \mathrm{R}}=\pi\), \(\vartheta(z)=\tan^{-1}(z/\pi)\), \(w(z)=\sqrt{1+(z/\pi)^{2}}\), and \(\tilde{z}(x,y,z)=z\bigl[1+(x^{\sss{2}}+y^{\sss{2}})/\bigl(2(z^{\sss{2}}+z_{\sss{\mathrm{R}}}^{\sss{2}})\bigr)\bigr]\); we then substituting Eqs.~\eqref{GG_with_PB_phase} (which includes the PB phase factor) and \eqref{normalized_LG_with_z} into Eq.~\eqref{SU(2)_coordinate} leads to Eq.~\eqref{SU(2)_with_PB_expansion}, which explicitly includes the PB and Gouy phase terms of the SU(2) mode. Comparing this with Eq.~\eqref{SU(2)_PB_compensation} confirms the accuracy of the PB factor derived through the continuous evolution method, thus affirming the validity of the previously determined PB.

\begin{widetext}
    \begin{equation}
            \mathrm{GG}_{N, \ell}(r, \varphi ; \alpha, \beta)={\exp[\mathrm{i}(m-n)\gamma]}
            \sum_{\ell^{\prime}=-N \atop \text{ steps of } 2}^{N} d_{\sss{ \frac{\ell^{\prime}}{2}  \frac{\ell}{2} }}^{\sss{\frac{N}{2}}}(\beta) \;\;\mathrm{e}^{\sss{-\mathrm{i} \frac{\ell^{\prime} \alpha}{2}}}\;\mathrm{i}^{\sss{\ell^{\prime}-\left|\ell^{\prime}\right|+\ell-N}}\;\mathrm{LG}_{\sss{N,\ell^{\prime}}}(r,\varphi)
        \label{GG_with_PB_phase}
    \end{equation}
\end{widetext}

\begin{equation}
	\left|\Psi_{\sss{m,n,l}}^{\sss{M,s,t}}\right\rangle=\frac{1}{2^{\sss{M/2}}} \sum_{K=0}^{M}\left(\begin{array}{l}
M\\
K
\end{array}\right)^{\sss{1/2}}\cdot\mathrm{e}^{\sss{\mathrm{i} K (\phi-\frac{\pi}2)}}\left|\psi_{\sss{m+sK,n+tK, l-K}}\right\rangle
    \label{SU(2)_coordinate}
\end{equation}

\begin{widetext}
    \begin{equation}
        \begin{aligned}
            \left|\Psi(\gamma)\right\rangle &=2^{\sss{-\frac{M}{2}}}\mathlarger{\mathlarger{\mathlarger{\mathlarger\sum}}}_{\sss{K=0}}^{\sss{M}}\sqrt{\binom{M}{K}}
            \;\;\exp\bigl\{\mathrm{i}{\color{black}\gamma\left[m\!-\!n+K(s\!-\!t)\right]}\bigr\}\cdot
            \exp\!\left\{-\mathrm{i}\tan^{\!\sss-1}\!\left(\frac{z}{\pi}\right)\;[K(s\!+\!t)\!+\!m\!+\!n\!+\!1)]\right\}
            \cdot\exp\!\left\{{\mathrm{i}K\left(\phi+\frac{\pi}{2}\right)}\right\}\times\\[10pt]
            &\mathlarger{\mathlarger{\mathlarger\sum}}_{\substack{\sss{\mu=-[K(s+t)+m+n]}\\ \sss{\text{step=2}}}}^{\sss{K(s+t)+m+n}}\frac{1}{\sqrt{\pi}}\Bigg\{\left(\frac{z^{\sss 2}}{\pi^{\sss2}}+1\right)^{\sss{-\frac{|\mu|+1}{2}}}\cdot(x^{\sss2}+y^{\sss2})^{\sss{\frac{|\mu|}{2}}}\cdot\mathrm{i}^{\sss{-2(Kt+n)-|\mu|+\mu}}\cdot\sqrt{\frac{2^{\sss{|\mu|+1}}\left[\frac{1}{2}(K(s+t)+m+n-|\mu|)\right]!}{\left[\frac{1}{2}(K(s+t)+m+n+|\mu|)\right]!}}\times\\[10pt]
            & d_{\sss{\frac{\mu}{2}, \frac{1}{2}[K(s-t)+m-n]}}^{\sss{\frac{1}{2}(K(s+t)+m+n)}}(\beta)\cdot L_{\sss{\frac{1}{2}(Ks+Kt+m+n|\mu|)}}^{\;\sss{\mu}}\!\!\left(\frac{2(x^{\sss2}+y^{\sss2})}{\frac{z^{\sss2}}{\pi^{\sss2}}+1}\right)\cdot\exp\!\!\left[\mathrm{i}\left(2\pi z+\frac{\pi z(x^{\sss 2}+y^{\sss 2})}{z^{\sss 2}+\pi^{\sss 2}}
            -\mu\left[\frac{\alpha}{2}\!-\!\tan^{\sss-1}\!\left(\frac{y}x\right)\right]\right)
            -\frac{x^{\sss2}\!+\!y^{\sss2}}{\frac{z^{\sss2}}{\pi^{\sss2}}+1}\right]\Bigg\}
        \end{aligned}
        \label{SU(2)_with_PB_expansion}
    \end{equation}
\end{widetext}

\begin{equation}
    \Phi_{\sss{K}}^{\sss{\text{PB}}}
    =\exp\left\{\mathrm{i}\;\gamma\;[m-n+K(s-t)]\right\}
    \label{SU(2)_PB_compensation}
\end{equation}


Note that these relations are implicit in Eq.~\eqref{normalized_LG_with_z} and Eq.~\eqref{SU(2)_with_PB_expansion}; here \(\ell\) is the azimuthal mode number, which sets the beam’s orbital angular momentum, while \(p_{\text{r}}\) is the radial mode number, which gives the number of radial intensity nodes in an LG mode~\cite{forbes2021structured}:

\begin{enumerate}
    \item \(m+n=N=|\ell|+2p\)
    \item \(p=\frac{N-|\ell|}{2}=\min(m,n)\)
    \item \(\ell=\mu=m-n\)
    \item \(m=\frac{N+\mu}{2}\)
    \item \(n=\frac{N-\mu}{2}\)
\end{enumerate}

The PB phase factor of an SU(2) beam is the sum of the weighted PB factors of its eigenstate modes, see Eq.~\eqref{SU(2)_weighted_PB_expansion}, with constants \(\omega_{\sss K}\) representing all terms except the PB factor.

\begin{equation}
    \begin{aligned}
        \left|\Psi(\gamma)\right\rangle 
        &=\sum_{K=0}^{M}\Phi_{\sss {K}}^{\sss{\text{PB}}}(\gamma) \cdot \Phi_{\sss {K}}^{\sss{\text{Gouy}}}(\vartheta) \cdot C_{\sss K} \left|\psi_{\sss K}\right\rangle\\[1ex]
        &=\sum_{K=0}^{M}\omega_{\sss K}\cdot \Phi_{\sss {K}}^{\sss{\text{PB}}}(\gamma)
    \end{aligned}
    \label{SU(2)_weighted_PB_expansion}
\end{equation}

The PB of the SU(2) beam is determined by calculating the argument of \(\left|\Psi(\gamma)\right\rangle\), expressed in radians, see Eq.~\eqref{SU(2)_PB_phase}.

\begin{equation}
    \begin{aligned}
        \Theta_{\sss{\text{SU(2)}}}^{\sss{\text{PB}}} 
        & = \arg\Bigl[\left|\Psi(\gamma)\right\rangle\Bigr]\\[1ex]
        & =\arg\Biggl[ \sum_{K=0}^{M}\omega_{\sss K}\cdot \Phi_{\sss {K}}^{\sss{\text{PB}}}(\gamma)\Biggr]\\[1ex]
        &=\sum_{K=0}^{M}\arg\Bigl[\omega_{\sss K}\cdot \Phi_{\sss {K}}^{\sss{\text{PB}}}(\gamma) \Bigr]\\[1ex]
        &=\sum_{K=0}^{M}\arg\Bigl[\Phi_{\sss {K}}^{\sss{\text{PB}}}(\gamma)\Bigr]\\[1ex]
        &=\sum_{K=0}^{M}\gamma\,[m-n+K(s-t)]
    \end{aligned}
    \label{SU(2)_PB_phase}
\end{equation}

Then, when \(M=0\), Eq.~\eqref{SU(2)_PB_phase} degenerates to the eigenstate case, as demonstrated in Eq.~\eqref{SU(2)_PB_phase_eigen}.

\begin{equation}
    \begin{aligned}
        \Theta_{\sss{\text{GG}}}^{\sss{\text{PB}}} &=  
        \Theta_{\sss{\text{SU(2)}}}^{\sss{\text{PB}}}\bigl|_{M=0} \\[1ex]
        &= \sum_{K=0}^{0}\gamma\,[m-n+K(s-t)]\\[1ex]
        &=\gamma(m-n)
    \end{aligned}
    \label{SU(2)_PB_phase_eigen}
\end{equation}

For the same reason, the Gouy phase for the SU(2) beam is elucidated in Eq.~\eqref{SU(2)_Gouy_phase}, with constants \(\sigma_{\sss K}\) representing all terms except the Gouy factor itself. Moreover, \(\vartheta=\tan^{-1}(z/\pi)\) is defined as the Gouy phase for a fundamental mode Gaussian beam.

\begin{equation}
    \begin{aligned}
        \Theta_{\sss{\text{SU(2)}}}^{\sss{\text{Gouy}}} 
        &= \arg\Bigl[\left|\Psi(\vartheta)\right\rangle\Bigr]\\[1ex]
        & = \arg \Biggl[\sum_{K=0}^{M}\sigma_{\sss K}\cdot\Phi_{\sss {K}}^{\sss{\text{Gouy}}}(\vartheta)\Biggr]\\[1ex]
        & = -\sum_{K=0}^{M} \tan^{-1}\!\left(\frac{z}{\pi}\right)\,[K(s+t)+m+n+1]
    \end{aligned}
    \label{SU(2)_Gouy_phase}
\end{equation}

Then, when \(M=0\), Eq.~\eqref{SU(2)_Gouy_phase} degenerates to the eigenstate case, as demonstrated in Eq.~\eqref{SU(2)_Gouy_phase_eigen}.

\begin{equation}
    \begin{aligned}
        \Theta_{\sss{\text{GG}}}^{\sss{\text{Gouy}}} &=  
        \Theta_{\sss{\text{SU(2)}}}^{\sss{\text{Gouy}}}\bigl|_{M=0}\\[1ex]
        &= \sum_{K=0}^{0}\Bigl[-\tan^{-1}\!\left(\frac{z}{\pi}\right)[K(s+t)+m+n+1]\Bigr]\\[1ex]
        &= -\tan^{-1}\!\left(\frac{z}{\pi}\right)(m+n+1)
    \end{aligned}
    \label{SU(2)_Gouy_phase_eigen}
\end{equation}

By comparing Eq.~\eqref{SU(2)_Gouy_phase_eigen} (where the Rayleigh distance term \(z_{\sss R}=\pi\)) with Eq.~\eqref{Gouy_phase_of_N_G}, we validate the methodology once more for calculating the PB of the SU(2) beam. This comparison demonstrates its consistency by showing how the Gouy phase degenerates to the eigenstate scenario. The derived expressions for both the PB and Gouy phases provide a robust framework for analyzing SU(2) beams, confirming the theoretical insights presented in prior studies.

\section{Periodicity of the PB Phase}
\label{Periodicity_of_the_PB_Phase}

In this section, we provide a comprehensive analysis aimed at elucidating how to ascertain the periodicity of the intensity distribution as a function of \(\gamma\). Referring to Eq.~\eqref{SU(2)_with_PB_expansion}, the PB phase factor is articulated as demonstrated in Eq.~\eqref{SU(2)_PB_phase_with_mn}. 

\begin{equation}
    \exp\left[\mathrm{i}\;\Theta_{\sss{\text{SU(2)}}}^{\sss{\text{PB}}}\right] = \sum_{K=0}^{M}\exp\Bigl\{\mathrm{i}\,\gamma\,[m-n+K(s-t)]\Bigr\}
    \label{SU(2)_PB_phase_with_mn}
\end{equation}

Given that \(m\) and \(n\) are part of a shared phase factor that does not influence the periodicity of the SU(2) beam, these variables are excluded, leading to Eq.~\eqref{SU(2)_PB_phase_no_mn}.

\begin{equation}
    \exp\left[\mathrm{i}\;\Theta_{\sss{\text{SU(2)}}}^{\sss{\text{PB}}}\right] = \sum_{K=0}^{M}\exp\Bigl\{\mathrm{i}\,\gamma\,[K(s-t)]\Bigr\}
    \label{SU(2)_PB_phase_no_mn}
\end{equation}

\begin{figure*}
    \centering
    \subfloat[M=0]{\includegraphics[width=0.22\textwidth]{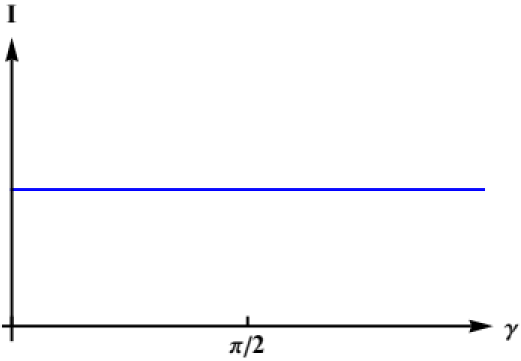}}\;\;\;\;
    \subfloat[M=1]{\includegraphics[width=0.22\textwidth]{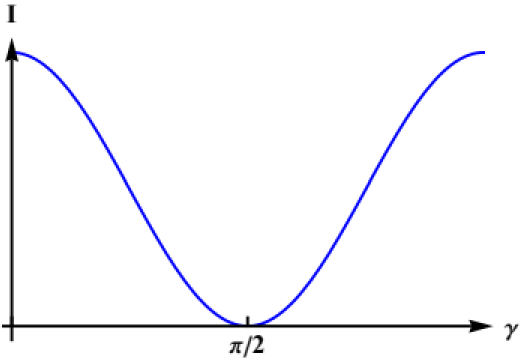}}\;\;\;\;
    \subfloat[M=2]{\includegraphics[width=0.22\textwidth]{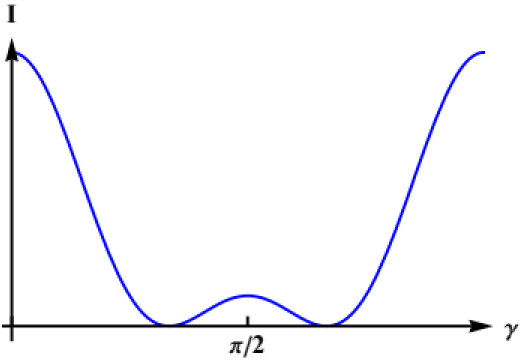}}\;\;\;\;
    \subfloat[M=3]{\includegraphics[width=0.22\textwidth]{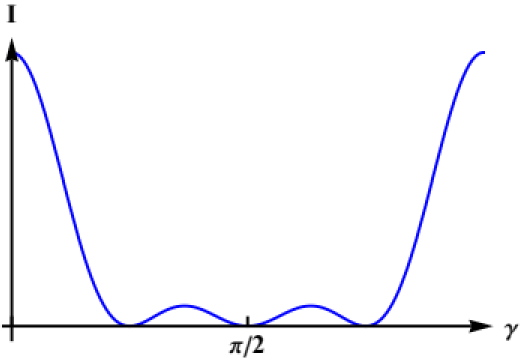}}\\
    \subfloat[M=0]{\includegraphics[width=0.22\textwidth]{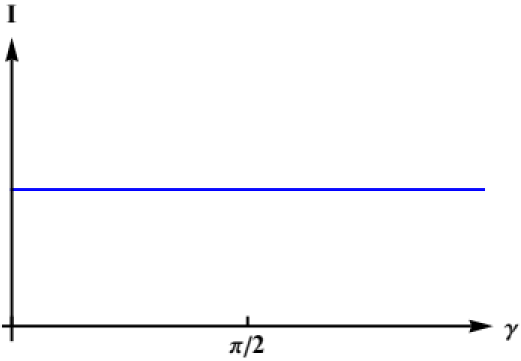}}\;\;\;\;
    \subfloat[M=1]{\includegraphics[width=0.22\textwidth]{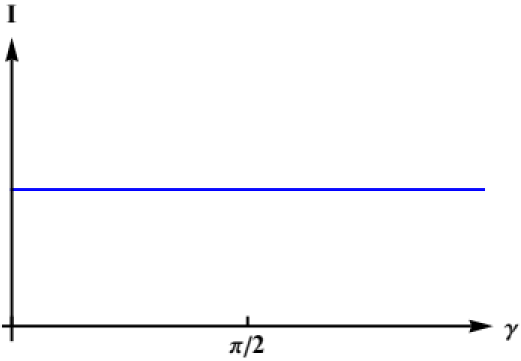}}\;\;\;\;
    \subfloat[M=2]{\includegraphics[width=0.22\textwidth]{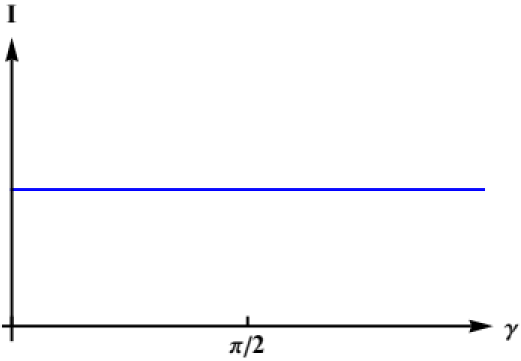}}\;\;\;\;
    \subfloat[M=3]{\includegraphics[width=0.22\textwidth]{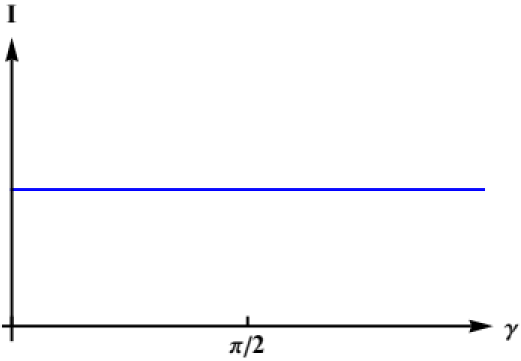}}\\
    \subfloat[M=0]{\includegraphics[width=0.22\textwidth]{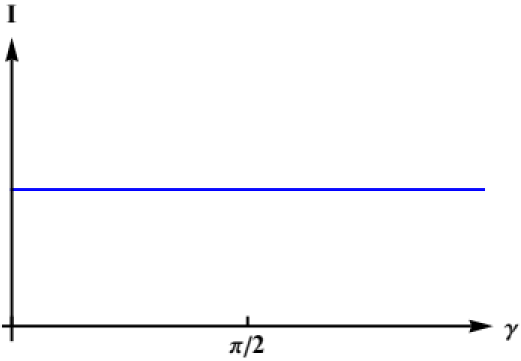}}\;\;\;\;
    \subfloat[M=1]{\includegraphics[width=0.22\textwidth]{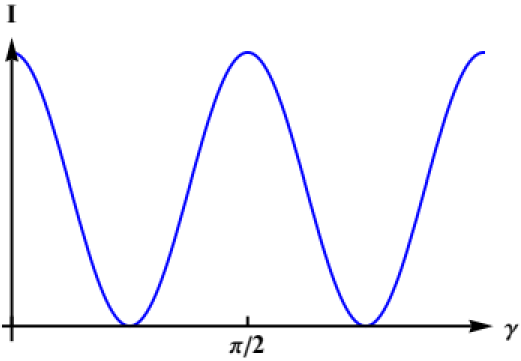}}\;\;\;\;
    \subfloat[M=2]{\includegraphics[width=0.22\textwidth]{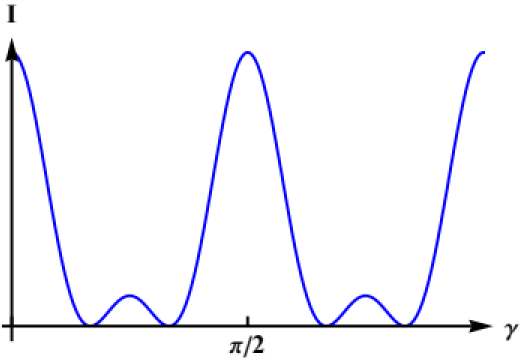}}\;\;\;\;
    \subfloat[M=3]{\includegraphics[width=0.22\textwidth]{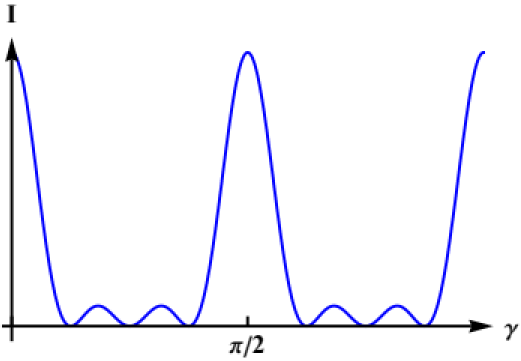}}\\
    \subfloat[M=0]{\includegraphics[width=0.22\textwidth]{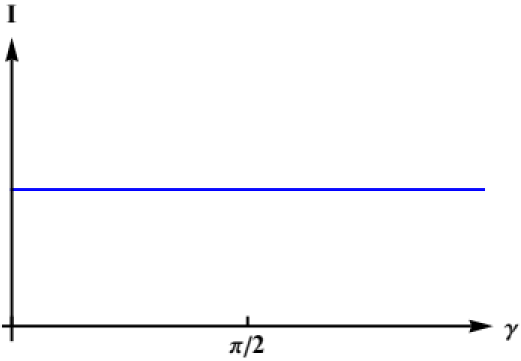}}\;\;\;\;
    \subfloat[M=1]{\includegraphics[width=0.22\textwidth]{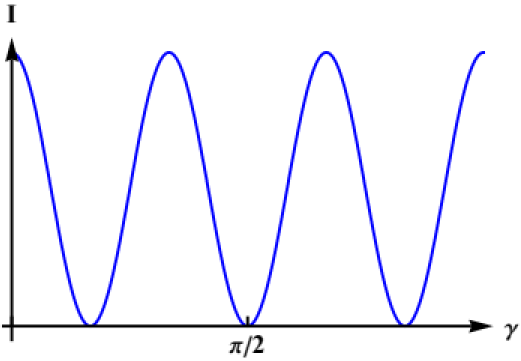}}\;\;\;\;
    \subfloat[M=2]{\includegraphics[width=0.22\textwidth]{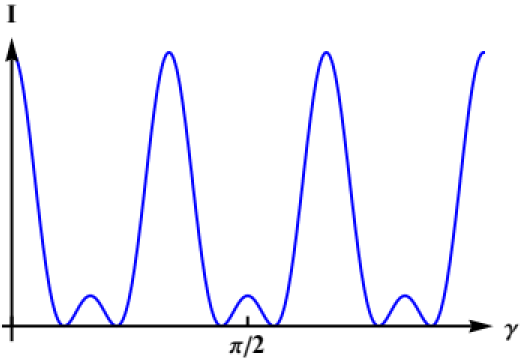}}\;\;\;\;
    \subfloat[M=3]{\includegraphics[width=0.22\textwidth]{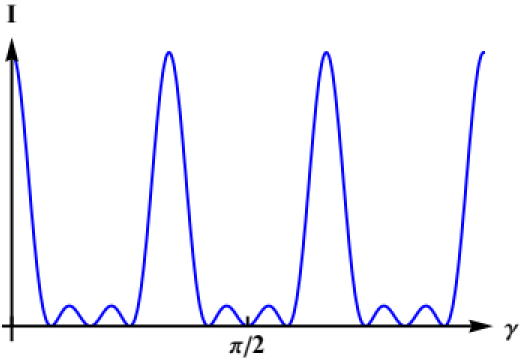}}
    \caption{
    Schematic of wave patterns for varying values of \( M \) across different \((s,t)\) scenarios:
    Top row: \((s,t) = (1,3)\); second row: \((s,t) = (2,2)\); third row: \((s,t) = (0,4)\); and bottom row: \((s,t) = (-1,5)\). The cases for \( M = 0 \) to \( M = 3 \) are depicted. In the scenario where \( M = 0 \), the absence of periodicity is evident, as it represents a state with no superposition effects. Conversely, for \( M \geq 1 \), the periodicity is consistent, with an increase in \( M \) resulting in the emergence of additional sub-wave packets. Notably, the case of \((s,t) = (2,2)\) demonstrates a lack of periodicity, indicating that the non-interference methodology is inapplicable for this particular configuration.
    }
    \label{periodicity_of_PB_phase}
\end{figure*}

When \(M=0\), there is no coherent superposition, and consequently no interference pattern (i.e. no trace property) is observed. When \(M=1\), two eigenstate modes are incorporated into the superposition, manifesting periodicity. However, the inclusion of additional eigenstate modes serves only to make the pattern more intricate without affecting its inherent periodicity, as illustrated in Fig.~\ref{periodicity_of_PB_phase}. Here, the intensity \(\mathbf{I}\) along the \(y\)-axis is computed as the squared norm of Eq.~\eqref{SU(2)_PB_phase_no_mn} (or equivalently Eq.~\eqref{SU(2)_PB_phase_with_mn}). Therefore, the periodicity can be exclusively determined by the case when \(M=1\). For \(M=1\), the expression becomes:
\[
1 + \mathrm{e}^{\mathrm{i}(s-t)\gamma}
\]
To find the norm (or modulus) of this complex number, we calculate:
\[
\left|1 + \mathrm{e}^{\mathrm{i}(s-t)\gamma}\right|
\]
Recall that the norm of a complex number \(a + b\mathrm{i}\) is given by:
\[
\sqrt{a^2 + b^2}.
\]
For the complex number \(1 + \mathrm{e}^{\mathrm{i}(s-t)\gamma}\), we rewrite the exponential term using Euler's formula:
\[
\mathrm{e}^{\mathrm{i}(s-t)\gamma} = \cos[(s-t)\gamma] + \mathrm{i}\sin[(s-t)\gamma].
\]
Thus, the expression becomes:
\[
1 + \cos[(s-t)\gamma] + \mathrm{i}\sin[(s-t)\gamma],
\]
with real part \(1+\cos[(s-t)\gamma]\) and imaginary part \(\sin[(s-t)\gamma]\). Therefore, the norm is:
\[
\left|1 + \mathrm{e}^{\mathrm{i}(s-t)\gamma}\right| = \sqrt{\Bigl[1+\cos((s-t)\gamma)\Bigr]^2 + \sin^2[(s-t)\gamma]}.
\]
Using the Pythagorean identity,
\[
\Bigl[1+\cos((s-t)\gamma)\Bigr]^2 + \sin^2[(s-t)\gamma] = 2+2\cos[(s-t)\gamma],
\]
we obtain
\[
\left|1 + \mathrm{e}^{\mathrm{i}(s-t)\gamma}\right| = \sqrt{2\Bigl[1+\cos((s-t)\gamma)\Bigr]} = 2\left|\cos\left[\frac{(s-t)\gamma}{2}\right]\right|.
\]

To find the periodicity with respect to \(\gamma\), note that the cosine function has period \(2\pi\). For the argument \(\frac{(s-t)\gamma}{2}\) to complete a full cycle, it must increase by \(2\pi\). Thus, the period \(T\) in \(\gamma\) satisfies:
\[
\frac{(s-t)T}{2} = 2\pi \quad\Longrightarrow\quad T = \frac{4\pi}{|s-t|}.
\]
Alternatively, if we consider the periodicity of the complex exponential \(\mathrm{e}^{\mathrm{i}(s-t)\gamma}\), the period is:
\[
T = \frac{2\pi}{|s-t|}.
\]
Since the norm expression involves a cosine of half the angle, the periodicity of the intensity distribution with respect to \(\gamma\) is 
\[
\frac{2\pi}{|s-t|},
\]
assuming \(s\neq t\). If \(s=t\), the expression becomes independent of \(\gamma\) and no periodicity is observed.

\section{Intensity and Mode Distribution Dynamics}

The wave function \(\left|\Psi_{\sss{m,n,l}}^{\sss{M,s,t}}\right\rangle\) from Eq.~\eqref{SU(2)_coordinate_with_PB1} can be deduced to have its intensity distribution concentrated on the parametric surface~\cite{chen2006devil, chen2019laser, chen2019origin, chen2010spatial, chen2011geometry}. From Eq.~\eqref{SU(2)_with_PB_expansion}, it is clear that the PB phase of the SU(2) beam depends on the parameters \(m\), \(n\), \(s\), \(t\), and \(K\). For our analysis, we set the values to \(m_0 = 18\), \(n_0 = 30\), \(s = 1\), \(t = 3\), and \(K \in [0, M]\) with \(M = 6\). Besides the PB phase, the coherent phase also plays a role. Although these phases are distinct in their origins and mechanisms, they both influence the position of the interference pattern~\cite{leinonen2023noncyclic}. This influence enables the indirect determination of the PB phase value by observing the periodic movement of the feature line in the SU(2) wave packet. Since our main interest lies in the PB phase, we will set \(z = 0\) to fix the Gouy phase when analyzing the PB phase. In this section, we will explore the period law of the mode and intensity variations under \(\gamma\). This appendix contains detailed results, where the intensity variations are depicted in grayscale and mode changes are depicted in hue.

\begin{widetext}
    \begin{equation}
	\left|\Psi_{\sss{m,n,l}}^{\sss{M,s,t}}\right\rangle_{\text{g}}=\frac{1}{2^{\sss{M/2}}}\sum_{K=0}^{M}\left(\begin{array}{l}M\\K\end{array}\right)^{\sss{1/2}}\cdot{\Phi_{\mathrm{g|}K}}\cdot\mathrm{e}^{\sss{\mathrm{i} K (\phi-\frac{\pi}2)}}\left|\psi_{\sss{m+sK,n+tK, l-K}}\right\rangle
    \label{SU(2)_coordinate_with_PB1}
    \end{equation}
    \begin{equation}
        \left|\Psi_{\sss{m,n,l}}^{\sss{M,s,t}}\right\rangle_{\text{b}}=\frac{1}{2^{\sss{M/2}}} \sum_{K=0}^{M}\left(\begin{array}{l}M\\K\end{array}\right)^{\sss{1/2}}\cdot{\Phi_{\mathrm{b|}K}}\cdot\mathrm{e}^{\sss{\mathrm{i} K (\phi-\frac{\pi}2)}}\left|\psi_{\sss{m+sK,n+tK, l-K}}\right\rangle
    \label{SU(2)_coordinate_with_PB2}
    \end{equation}
    \begin{equation}
        \left|\Psi_{\sss{m,n,l}}^{\sss{M,s,t}}\right\rangle_{\text{PB}}=\frac{1}{2^{\sss{M/2}}} \sum_{K=0}^{M}\left(\begin{array}{l}M\\
        K\end{array}\right)^{\sss{1/2}}\cdot{\Phi_{\sss{K} }^{\sss{\text{PB}}}}\cdot\mathrm{e}^{\sss{\mathrm{i} K (\phi-\frac{\pi}2)}}\left|\psi_{\sss{m+sK,n+tK, l-K}}\right\rangle
    \label{SU(2)_with_PB_phase}
    \end{equation}
\end{widetext}

\begin{figure*}[!htbp]
    \centering
    \subfloat[\((s,t)=(1,3)\)]{\includegraphics[width=0.4\textwidth]{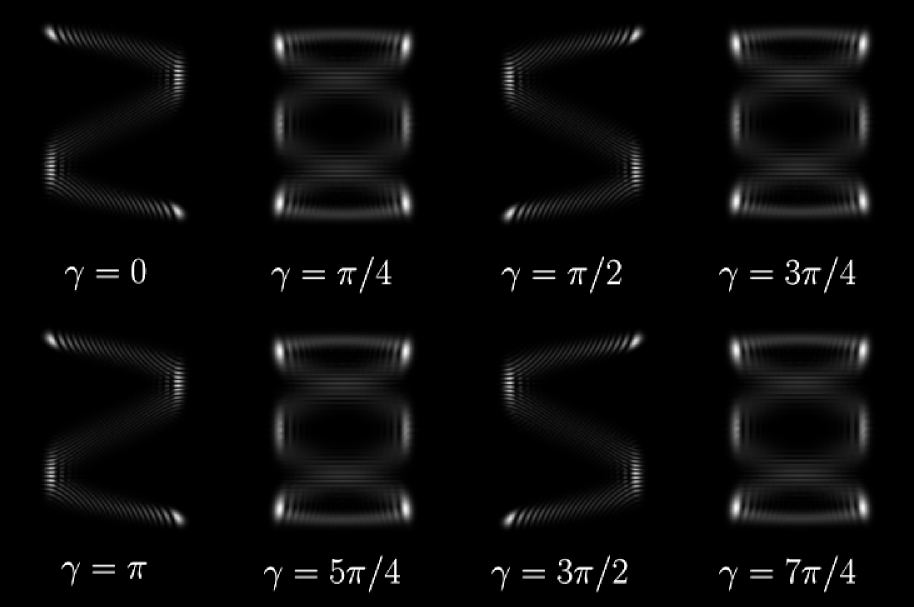}}\;\;\;\;\;
    \subfloat[\((s,t)=(3,1)\)]{\includegraphics[width=0.4\textwidth]{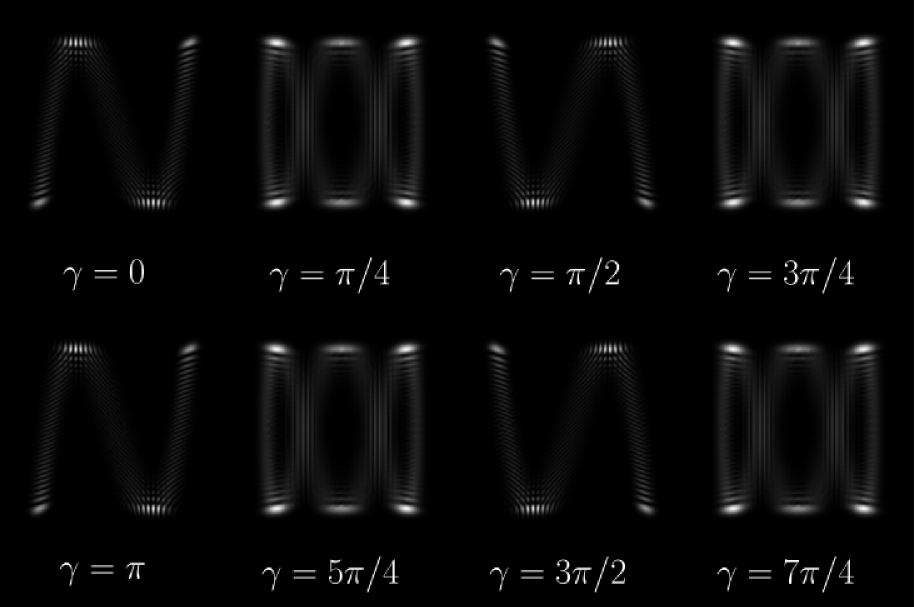}}\\
    \subfloat[\((s,t)=(0,4)\)]{\includegraphics[width=0.4\textwidth]{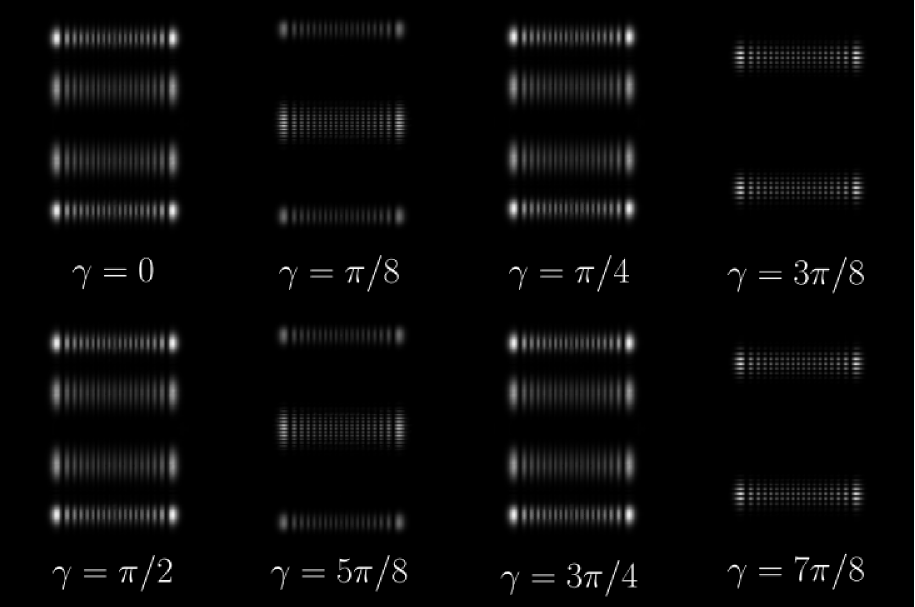}}\;\;\;\;\;
    \subfloat[\((s,t)=(4,0)\)]{\includegraphics[width=0.4\textwidth]{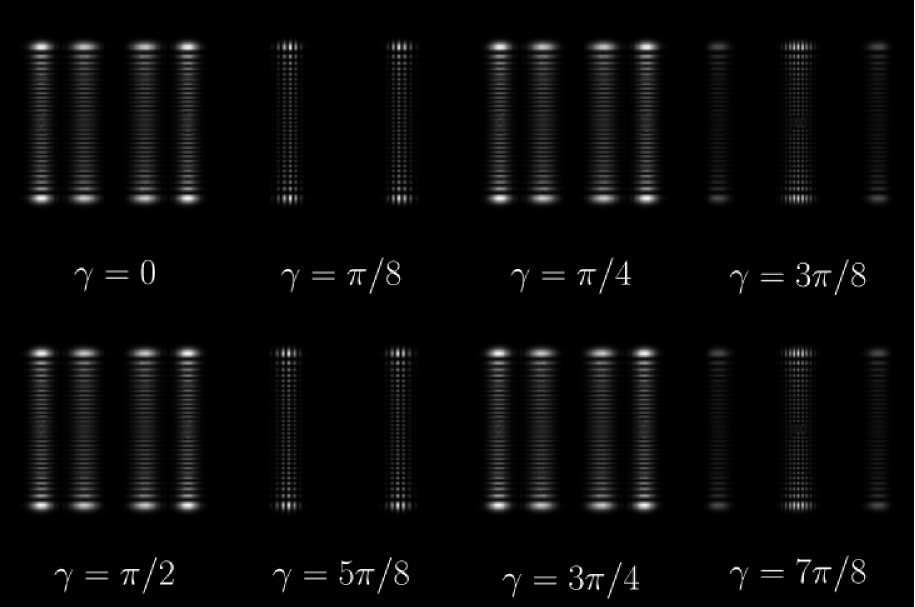}}\\
    \subfloat[\((s,t)=(-1,5)\)]{\includegraphics[width=0.4\textwidth]{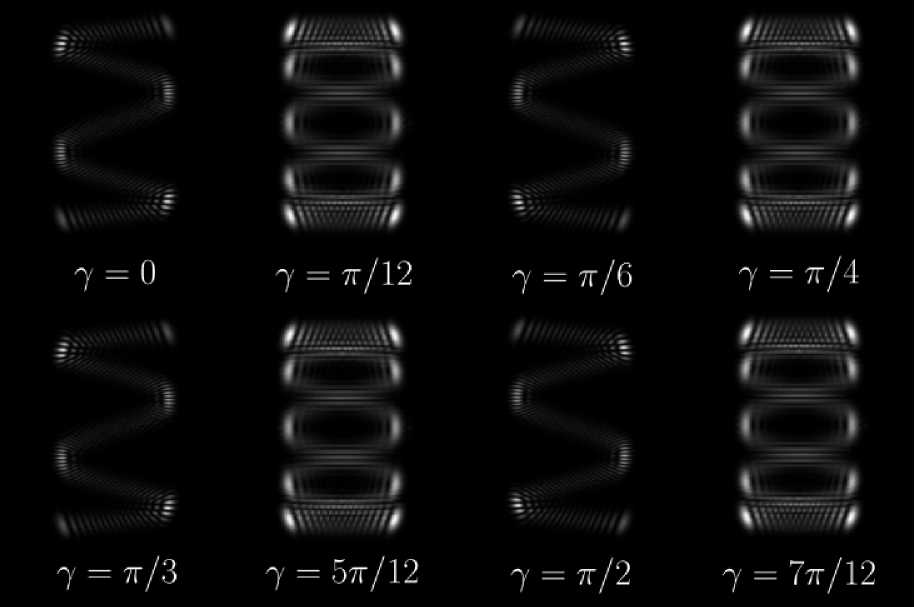}}\;\;\;\;\;
    \subfloat[\((s,t)=(5,-1)\)]{\includegraphics[width=0.4\textwidth]{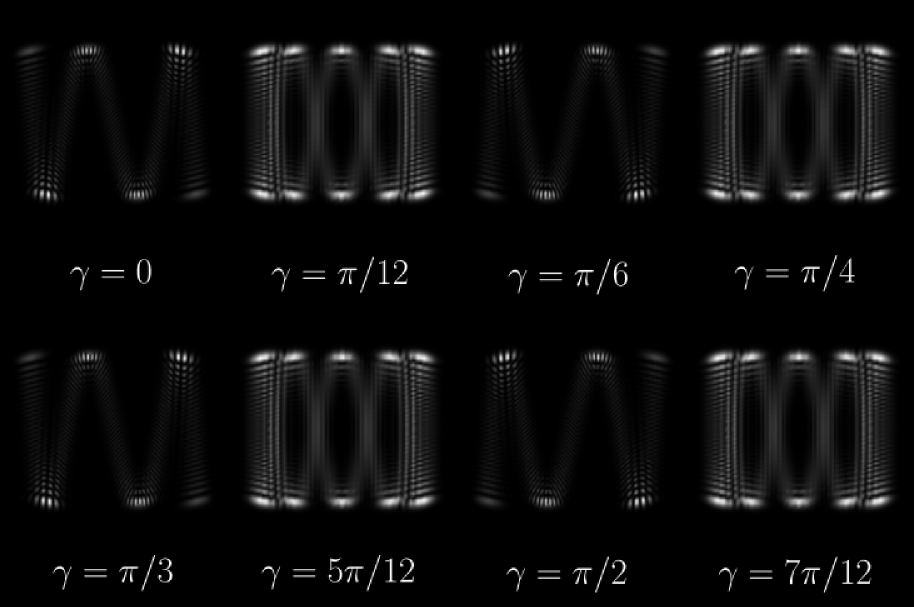}}\\
    \subfloat[\((s,t)=(-2,6)\)]{\includegraphics[width=0.4\textwidth]{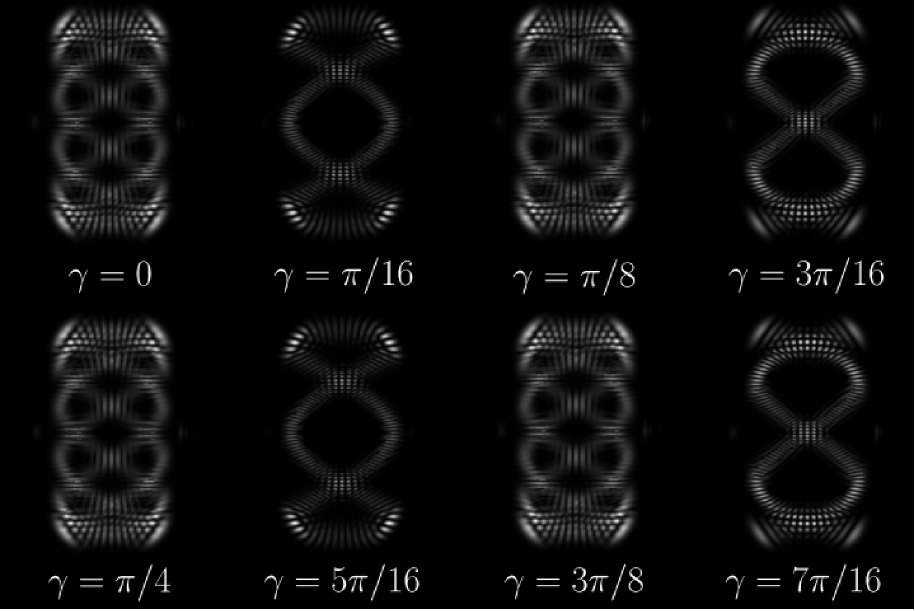}}\;\;\;\;\;
    \subfloat[\((s,t)=(6,-2)\)]{\includegraphics[width=0.4\textwidth]{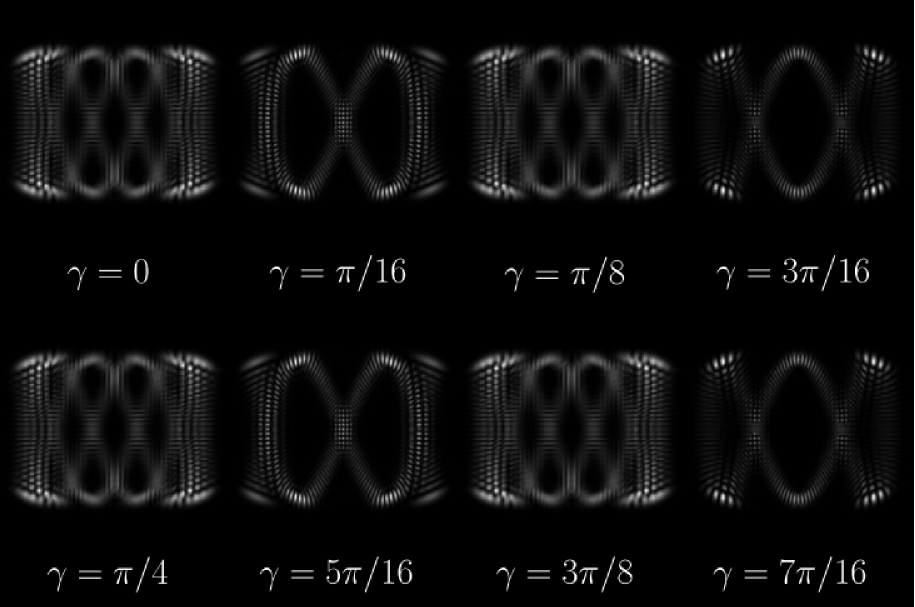}}\\
    \caption{
    Schematic of intensity distribution profiles for SU(2) modes under various \((s,t)\) configurations influenced by the PB phase:
    Top row: \((s,t) = (1,3)\) and \((3,1)\) with a periodicity \(P(\gamma) = \pi\); second row: \((s,t) = (0,4)\) and \((4,0)\) with \(P(\gamma) = \frac{\pi}{2}\); third row: \((s,t) = (-1,5)\) and \((5,-1)\) with \(P(\gamma) = \frac{\pi}{3}\); bottom row: \((s,t) = (-2,6)\) and \((6,-2)\) with \(P(\gamma) = \frac{\pi}{4}\). Each column corresponds to different values of \(\gamma\), showcasing the evolution of the intensity pattern as \(\gamma\) progresses.
    }
    \label{periodic_law_gamma}
\end{figure*}

\begin{figure*}[!htbp]
    \centering
    \subfloat[Intensity vs. \(\gamma\)]{%
        \includegraphics[width=0.4\textwidth]{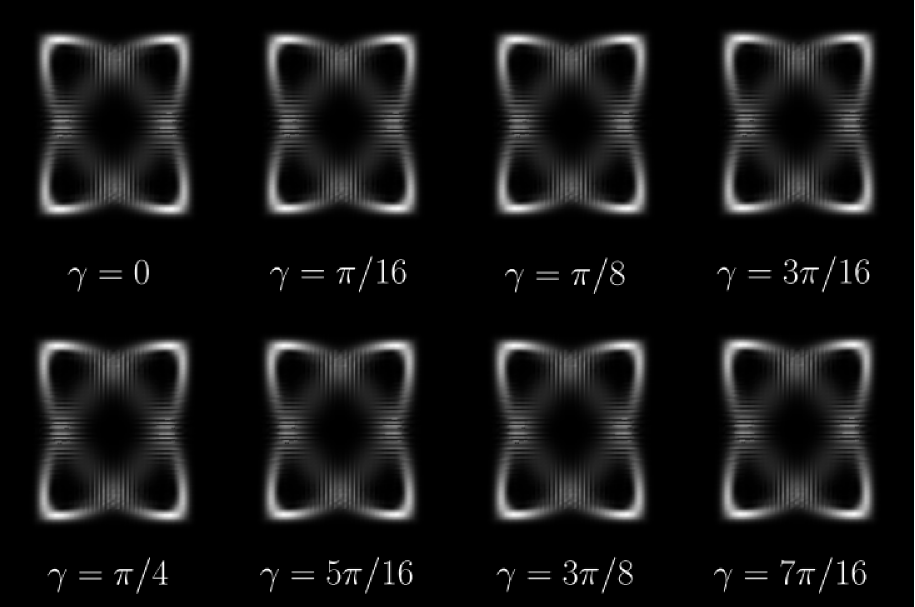}
    }\quad
    \subfloat[Intensity vs. \(\phi\)]{%
        \includegraphics[width=0.4\textwidth]{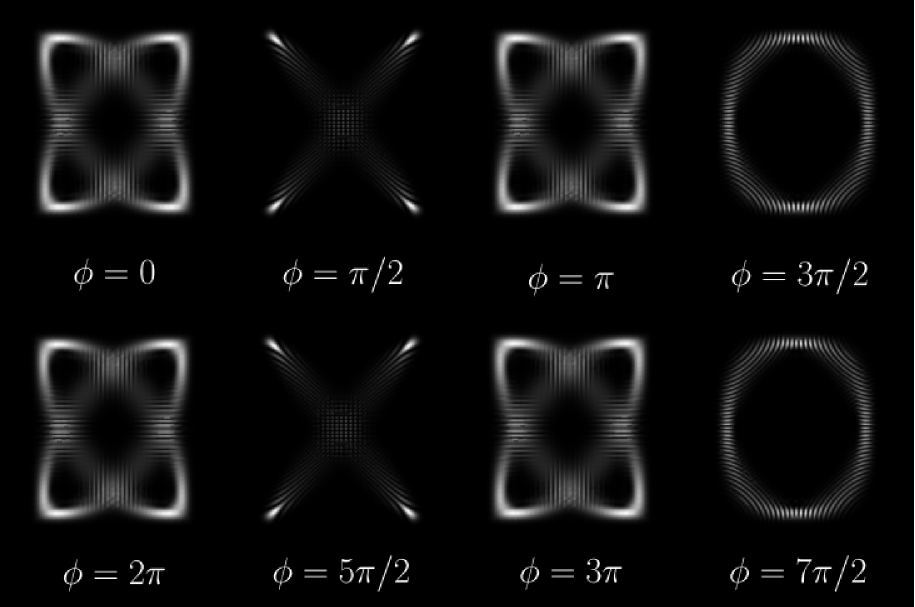}
    }\\[0.5em]
    \subfloat[Phase vs. \(\gamma\)]{%
        \includegraphics[width=0.4\textwidth]{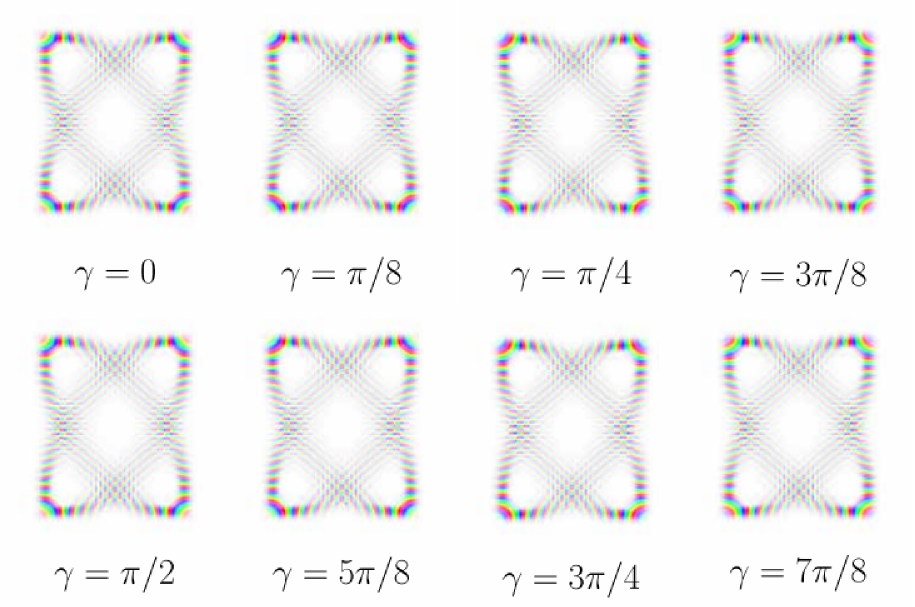}
    }\quad
    \subfloat[Phase vs. \(\phi\)]{%
        \includegraphics[width=0.4\textwidth]{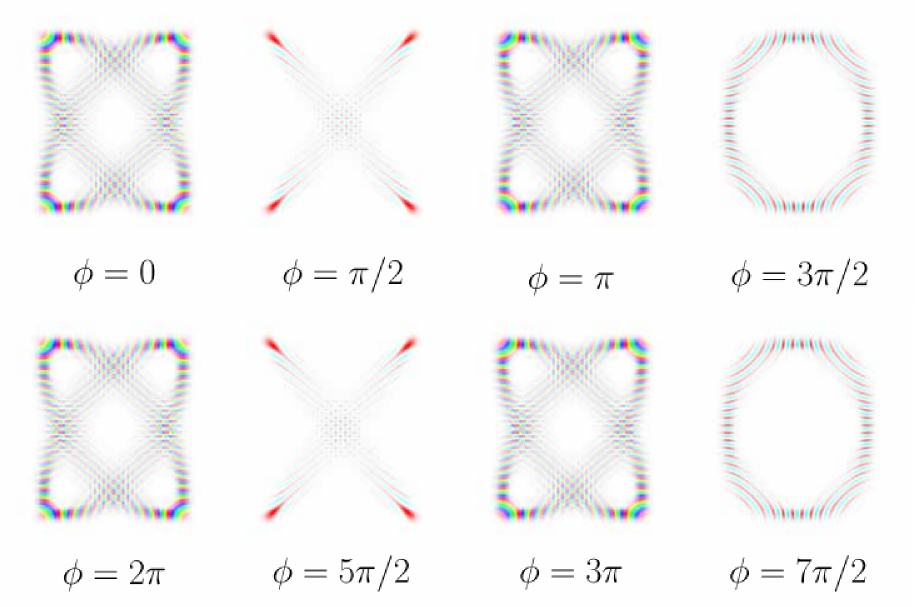}
    }
    \caption{
    Intensity and phase distribution profiles for SU(2) modes with \((s,t) = (2,2)\), demonstrating the effects of the PB phase and the coherent phase: 
    (a) Intensity distribution as a function of \(\gamma\), indicating that the PB phase does not result in periodicity, thus suggesting the non-applicability of the non-interference methodology for this SU(2) mode. 
    (b) Intensity distribution as a function of \(\phi\), exhibiting a periodic evolution of the wave packet.
    (c) Phase distribution as a function of \(\gamma\), showing the absence of periodicity.
    (d) Phase distribution as a function of \(\phi\), exhibiting a periodic evolution of the wave packet.
    }
    \label{periodic_law_gamma_phi}
\end{figure*}

Table I in the main text summarizes the periodic law concerning the PB phase. Here, \( P(\gamma) \) represents the periods of transverse packet intensity distribution changes with respect to \(\gamma\), calculated using the equation \( 2\pi / |s-t| \). For a comprehensive derivation of this equation, readers may consult Section~\ref{Periodicity_of_the_PB_Phase}. The table, along with the intensity distribution profile illustrated in Fig.~\ref{periodic_law_gamma}, conclusively shows that the PB phase modifies both the phase and intensity patterns of the SU(2) beam. It should be noted that for the case of \((s, t) = (2, 2)\), despite the periodicity in \(\phi\) depicted in Fig.~\ref{periodic_law_gamma_phi}b, the absence of periodicity as shown in Fig.~\ref{periodic_law_gamma_phi}a and the second row of Fig.~\ref{periodicity_of_PB_phase} signifies that the non-interference methodology cannot be applied to this specific SU(2) mode. Note that we consistently use \(P = 1\), \(Q = s+t\), and \(\Omega = P / Q\), and only consider the case of \(\Omega = 1/4\).

However, Figs.~\ref{periodic_law_gamma} and \ref{periodic_law_gamma_phi} only show the transverse profile at the beam width. The feature lines for measuring the PB phase are within the range of \((-z_{\sss{R}},z_{\sss{R}})\), where \(z_{\sss{R}}\) is the Rayleigh distance. A dynamic demonstration of how the feature lines evolve with \(\gamma\) is provided in the visualization videos of \textit{Parametric Surface with Feature Lines} and \textit{Pure Feature Lines}, to help understand the rules~\cite{li_2025_su2viz}.

\newpage

\newpage
\begin{figure*}[!htbp]
    \centering
    \subfloat[\((s,t)=(1,3)\)]{\includegraphics[width=0.4\textwidth]{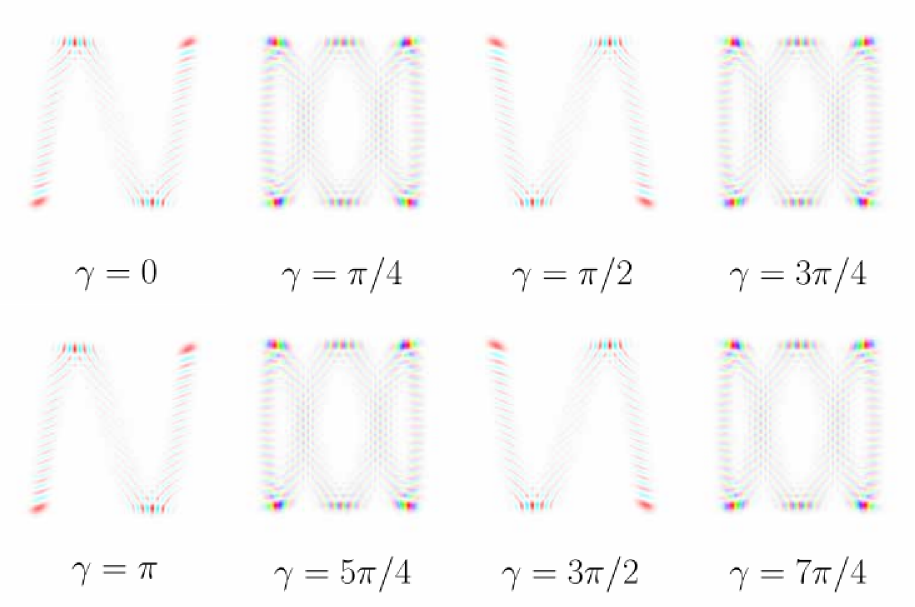}}\;\;\;\;\;
    \subfloat[\((s,t)=(3,1)\)]{\includegraphics[width=0.4\textwidth]{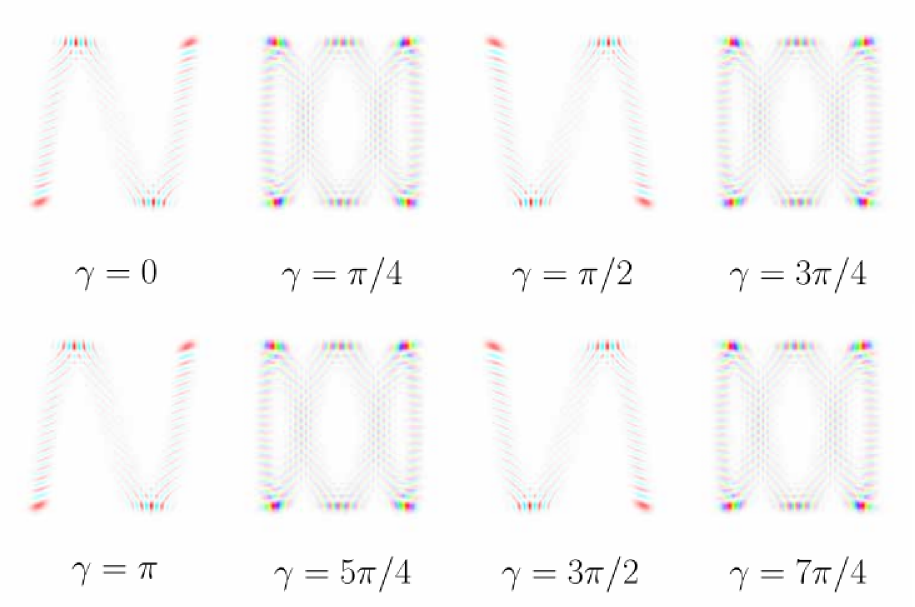}}\\
    \subfloat[\((s,t)=(0,4)\)]{\includegraphics[width=0.4\textwidth]{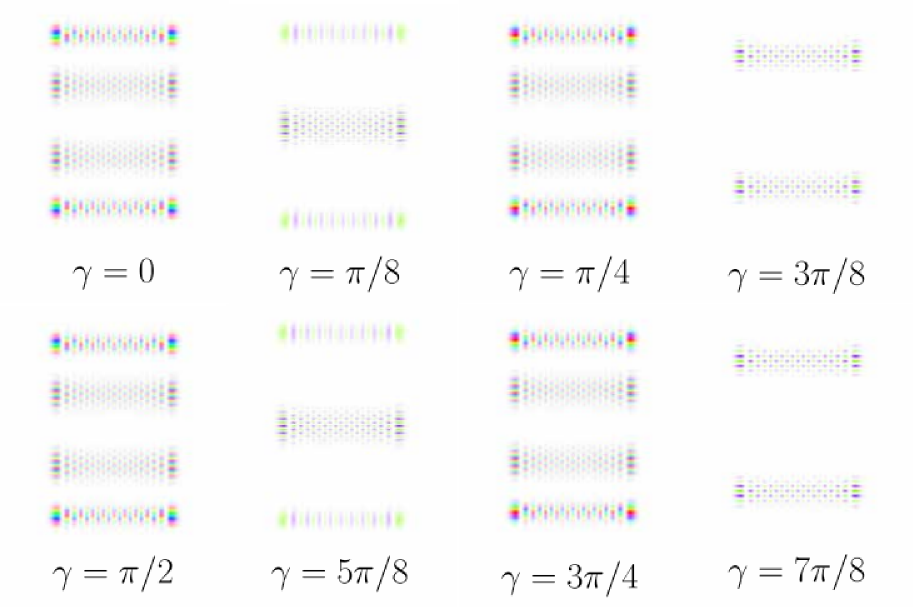}}\;\;\;\;\;
    \subfloat[\((s,t)=(4,0)\)]{\includegraphics[width=0.4\textwidth]{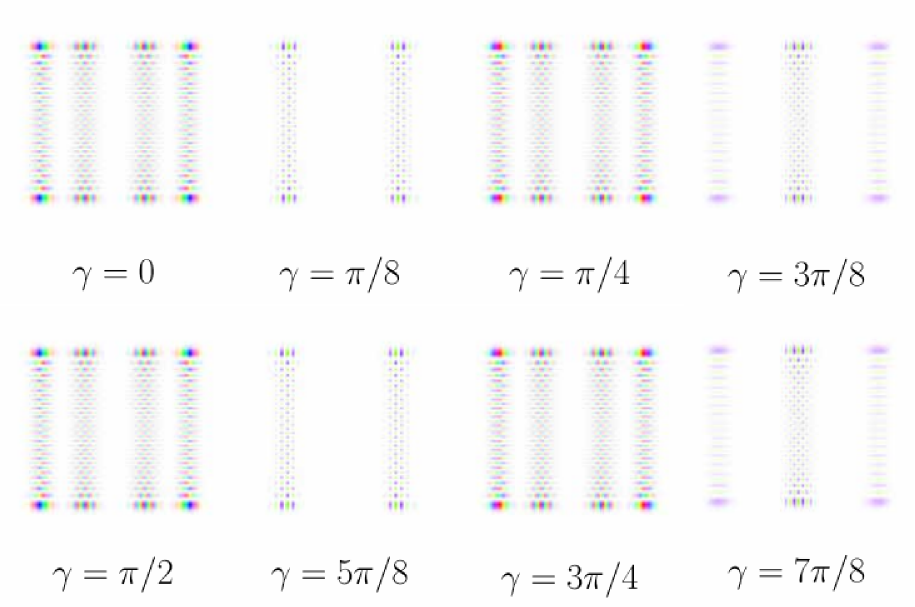}}\\
    \subfloat[\((s,t)=(-1,5)\)]{\includegraphics[width=0.4\textwidth]{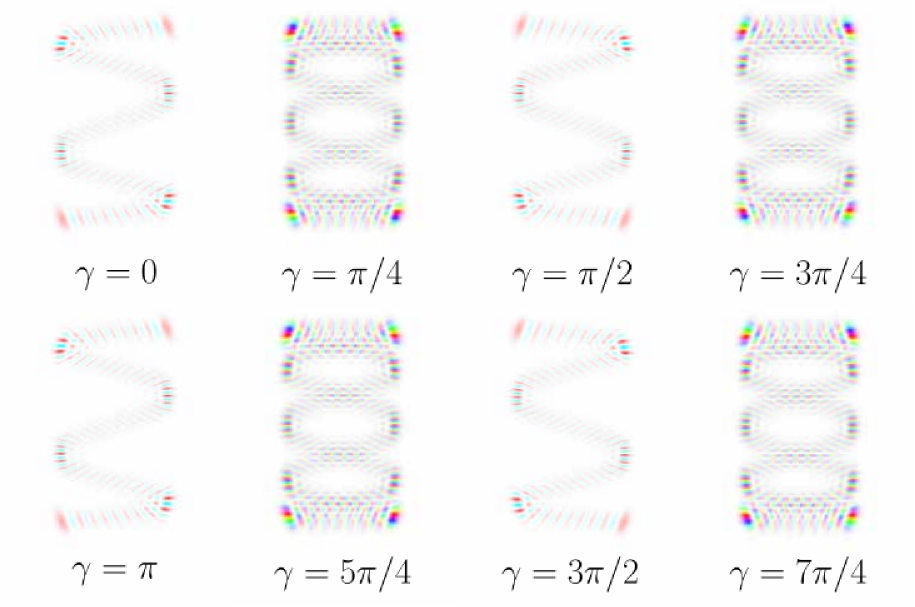}}\;\;\;\;\;
    \subfloat[\((s,t)=(5,-1)\)]{\includegraphics[width=0.4\textwidth]{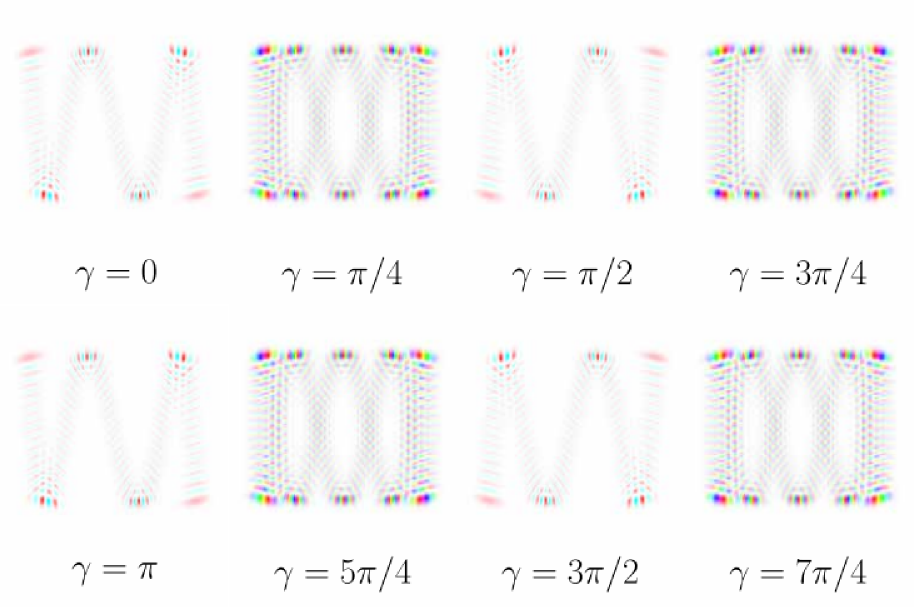}}\\
    \subfloat[\((s,t)=(-2,6)\)]{\includegraphics[width=0.4\textwidth]{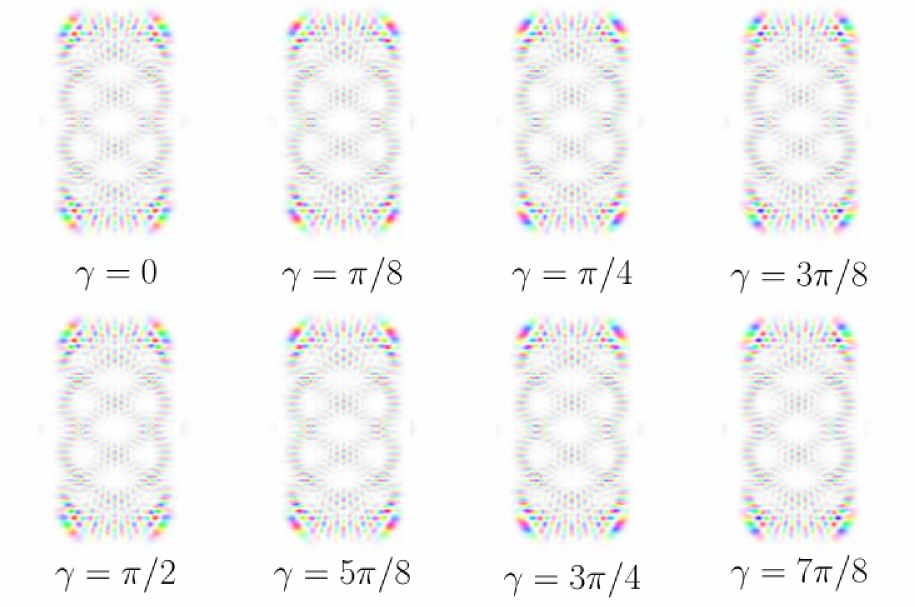}}\;\;\;\;\;
    \subfloat[\((s,t)=(6,-2)\)]{\includegraphics[width=0.4\textwidth]{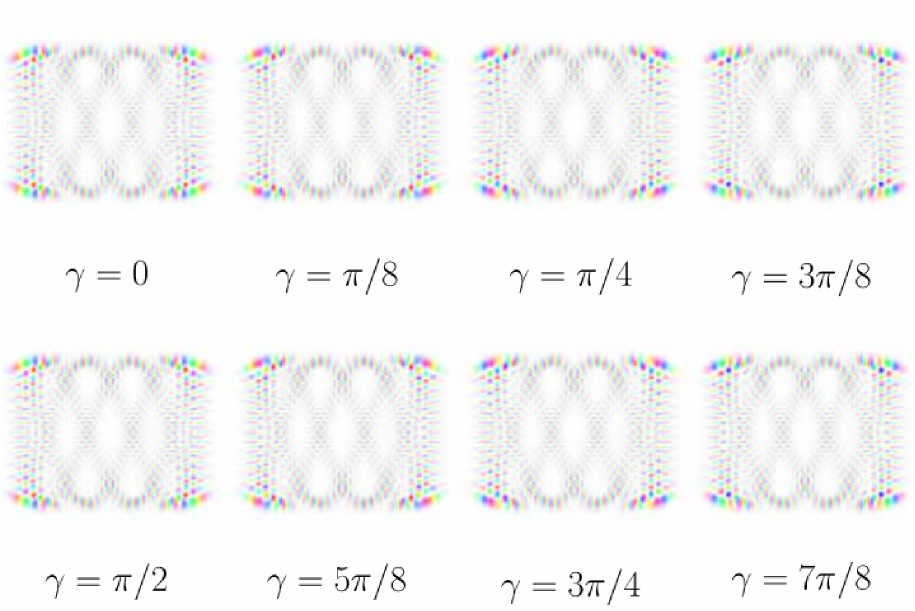}}\\
    \caption{
    Schematic of phase distribution profiles for SU(2) modes under various \((s,t)\) configurations influenced by the PB phase:
    Top row: \((s,t) = (1,3)\) and \((3,1)\) with a periodicity \(P(\gamma) = \pi\); second row: \((s,t) = (0,4)\) and \((4,0)\) with \(P(\gamma) = \frac{\pi}{2}\); third row: \((s,t) = (-1,5)\) and \((5,-1)\) with \(P(\gamma) = \pi\); bottom row: \((s,t) = (-2,6)\) and \((6,-2)\) with \(P(\gamma) = \frac{\pi}{2}\). Each column corresponds to different values of \(\gamma\), showcasing the evolution of the phase distribution as \(\gamma\) progresses.
    }
    \label{periodic_law_gamma_phase}
\end{figure*}

\newpage
\begin{figure*}[!htbp]
    \centering
    \subfloat[\((s,t)=(1,3)\)]{\includegraphics[width=0.4\textwidth]{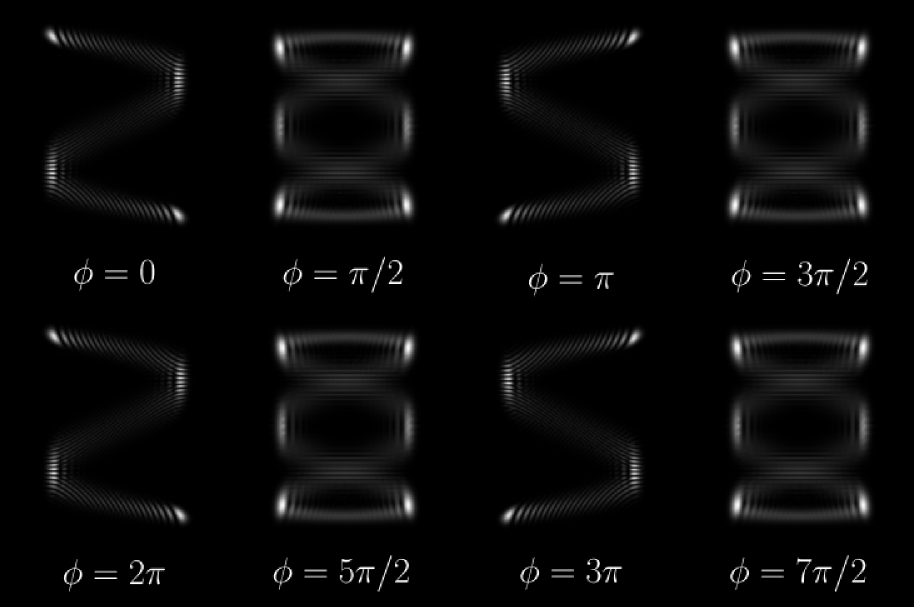}}\;\;\;\;\;
    \subfloat[\((s,t)=(3,1)\)]{\includegraphics[width=0.4\textwidth]{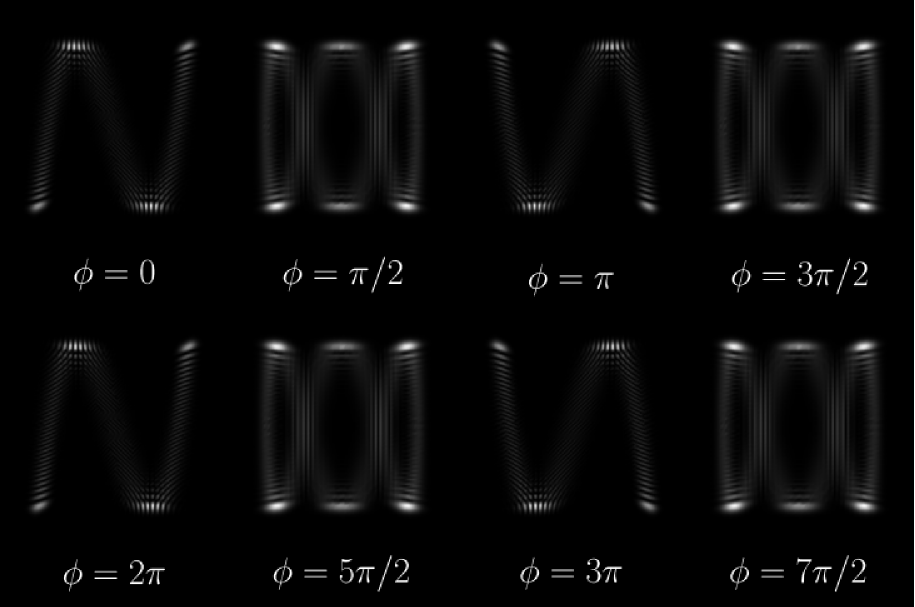}}\\
    \subfloat[\((s,t)=(0,4)\)]{\includegraphics[width=0.4\textwidth]{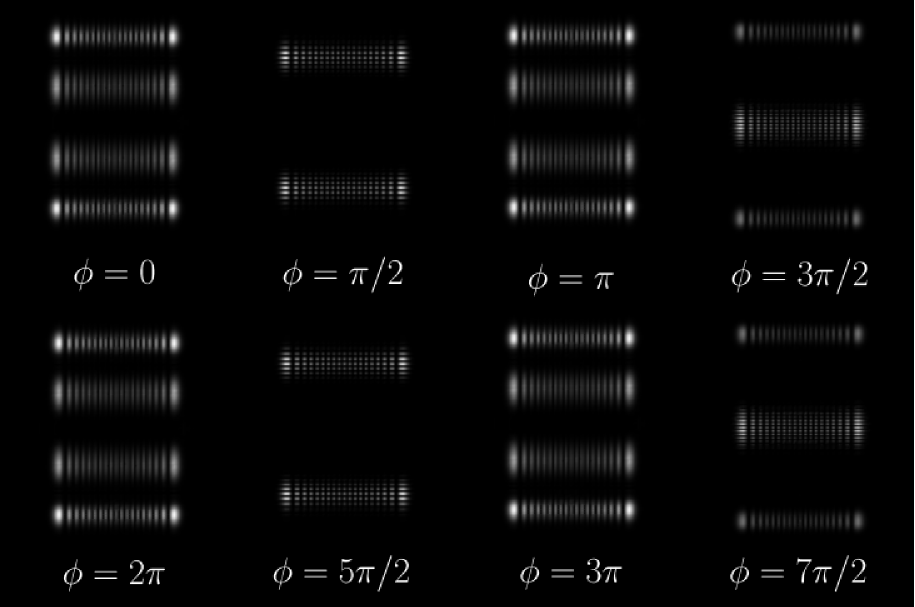}}\;\;\;\;\;
    \subfloat[\((s,t)=(4,0)\)]{\includegraphics[width=0.4\textwidth]{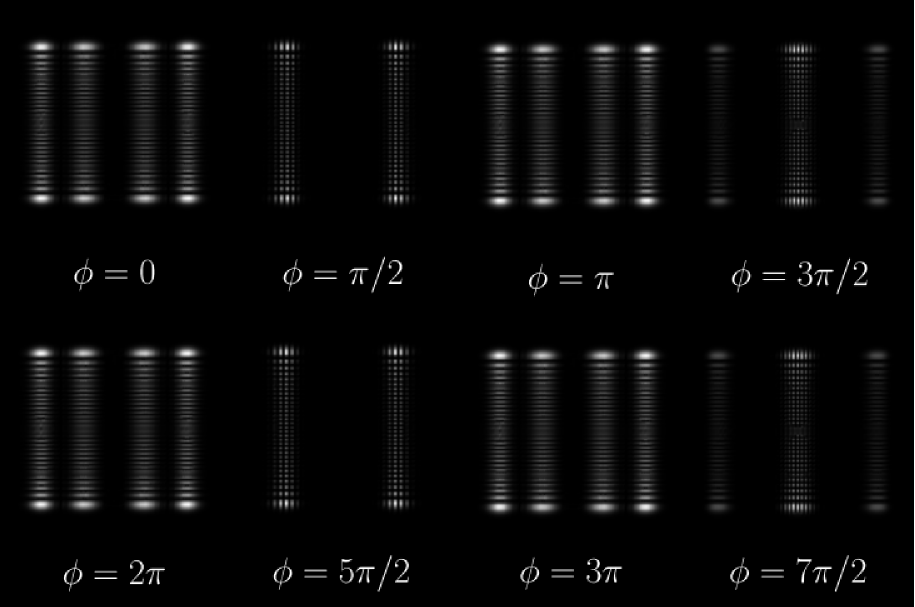}}\\
    \subfloat[\((s,t)=(-1,5)\)]{\includegraphics[width=0.4\textwidth]{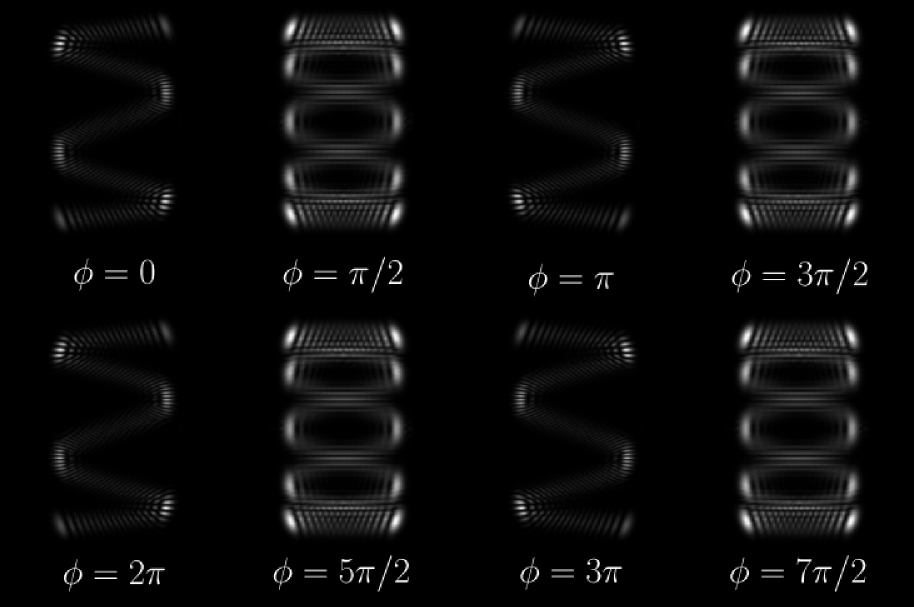}}\;\;\;\;\;
    \subfloat[\((s,t)=(5,-1)\)]{\includegraphics[width=0.4\textwidth]{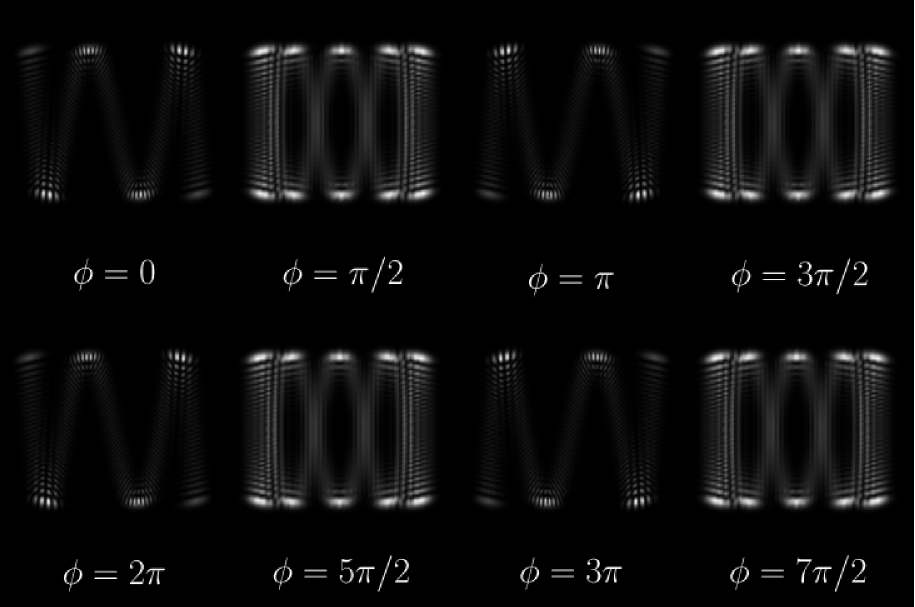}}\\
    \subfloat[\((s,t)=(-2,6)\)]{\includegraphics[width=0.4\textwidth]{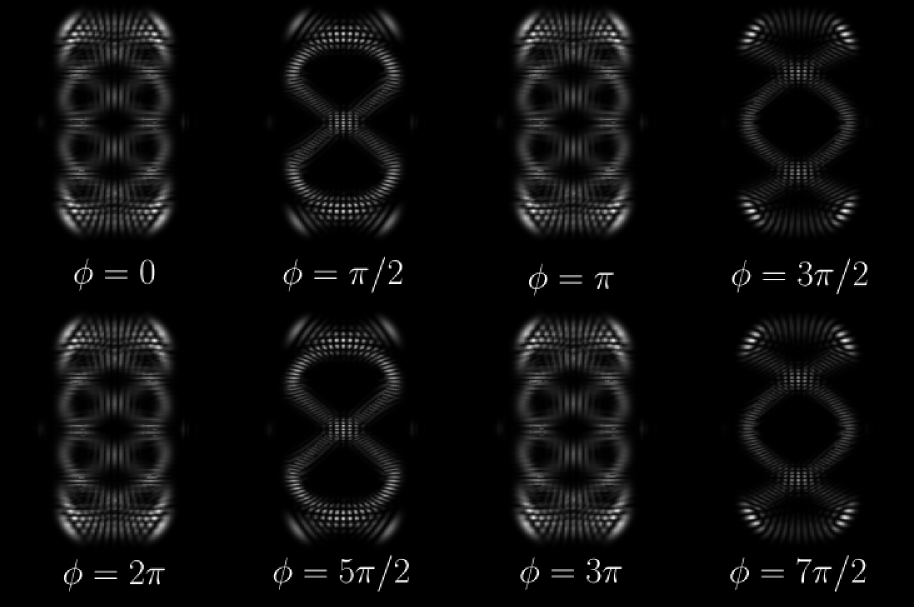}}\;\;\;\;\;
    \subfloat[\((s,t)=(6,-2)\)]{\includegraphics[width=0.4\textwidth]{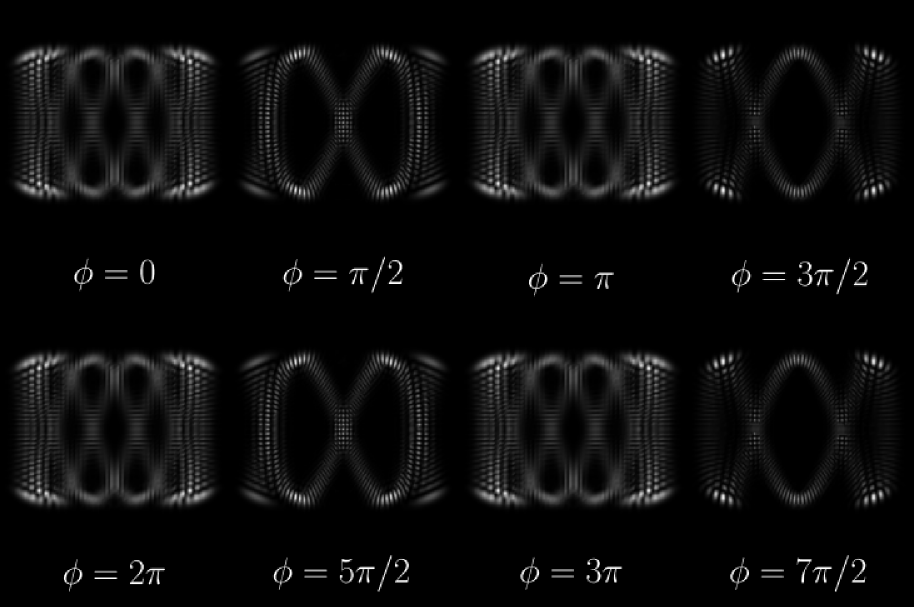}}\\
    \caption{
    Schematic of intensity distribution profiles for SU(2) modes under various \((s,t)\) configurations influenced by the coherent phase:
    Top row: \((s,t) = (1,3)\) and \((3,1)\) with a periodicity \(P(\phi) = 2\pi\); second row: \((s,t) = (0,4)\) and \((4,0)\) with \(P(\phi) = 2\pi\); third row: \((s,t) = (-1,5)\) and \((5,-1)\) with \(P(\phi) = 2\pi\); bottom row: \((s,t) = (-2,6)\) and \((6,-2)\) with \(P(\phi) = 2\pi\). Each column corresponds to different values of \(\phi\), showcasing the evolution of the intensity pattern as \(\phi\) progresses.
    }
    \label{periodic_law_phi}
\end{figure*}

\newpage
\begin{figure*}[!htbp]
    \centering
    \subfloat[\((s,t)=(1,3)\)]{\includegraphics[width=0.4\textwidth]{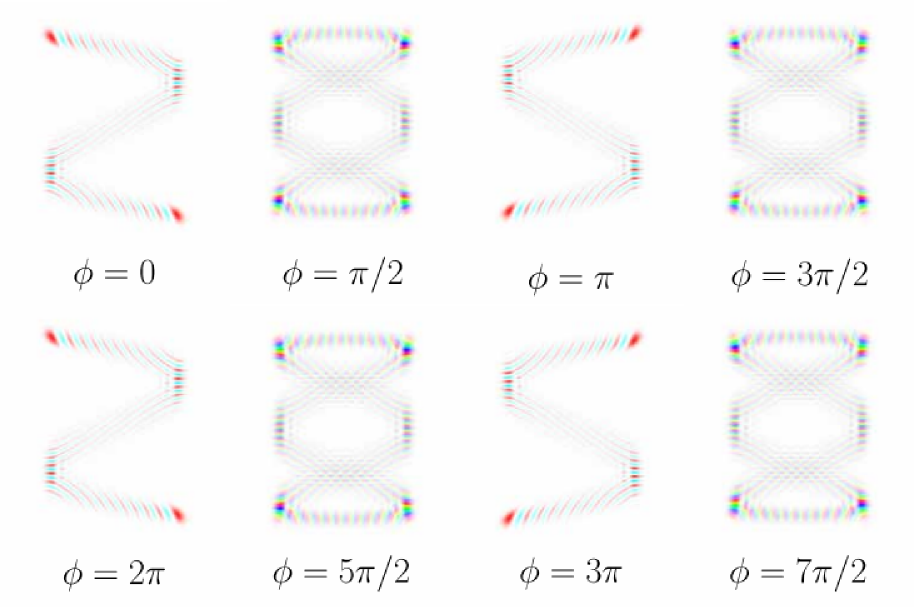}}\;\;\;\;\;
    \subfloat[\((s,t)=(3,1)\)]{\includegraphics[width=0.4\textwidth]{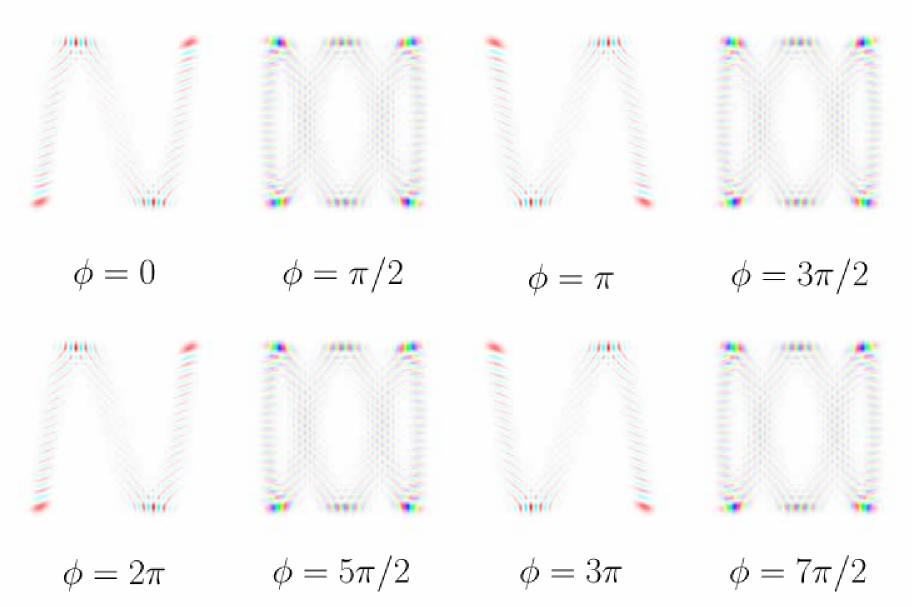}}\\
    \subfloat[\((s,t)=(0,4)\)]{\includegraphics[width=0.4\textwidth]{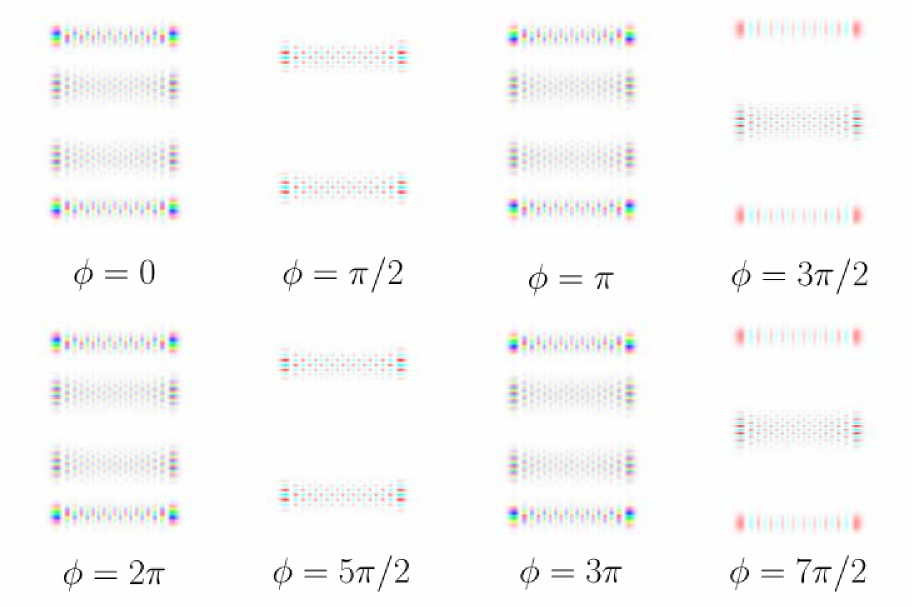}}\;\;\;\;\;
    \subfloat[\((s,t)=(4,0)\)]{\includegraphics[width=0.4\textwidth]{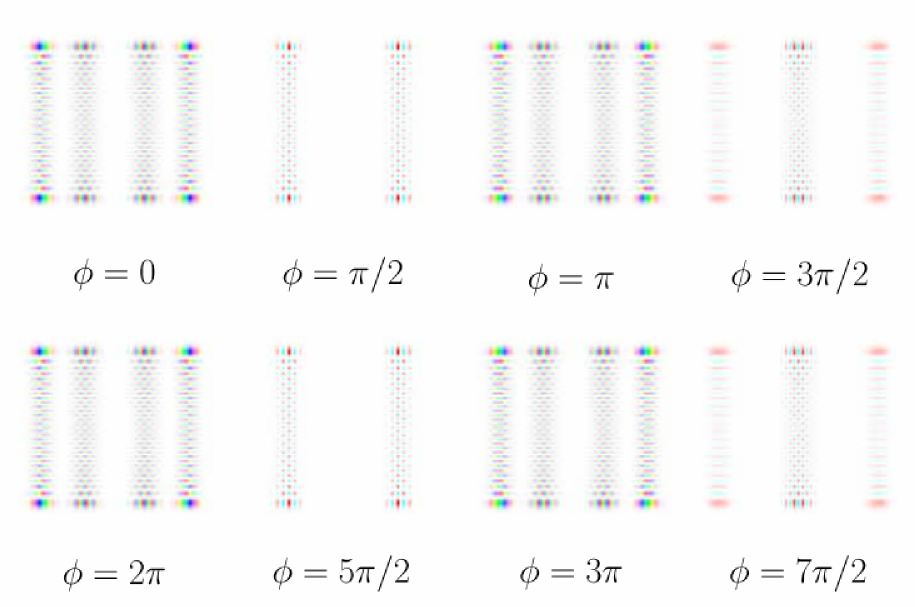}}\\
    \subfloat[\((s,t)=(-1,5)\)]{\includegraphics[width=0.4\textwidth]{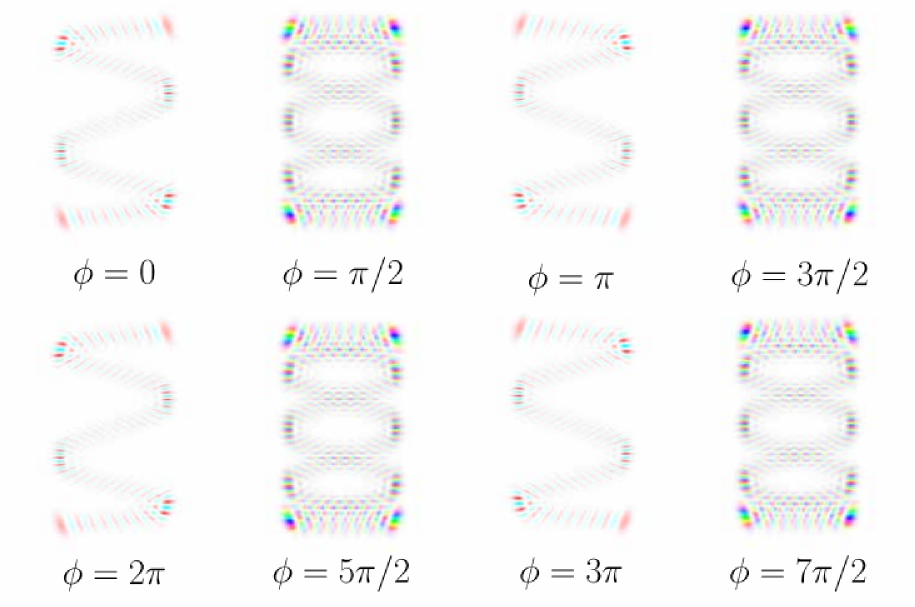}}\;\;\;\;\;
    \subfloat[\((s,t)=(5,-1)\)]{\includegraphics[width=0.4\textwidth]{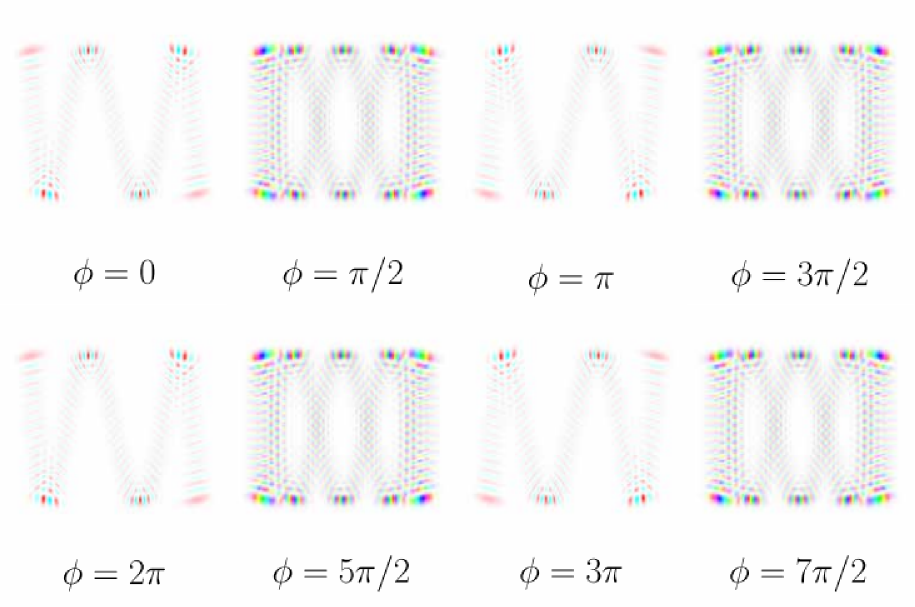}}\\
    \subfloat[\((s,t)=(-2,6)\)]{\includegraphics[width=0.4\textwidth]{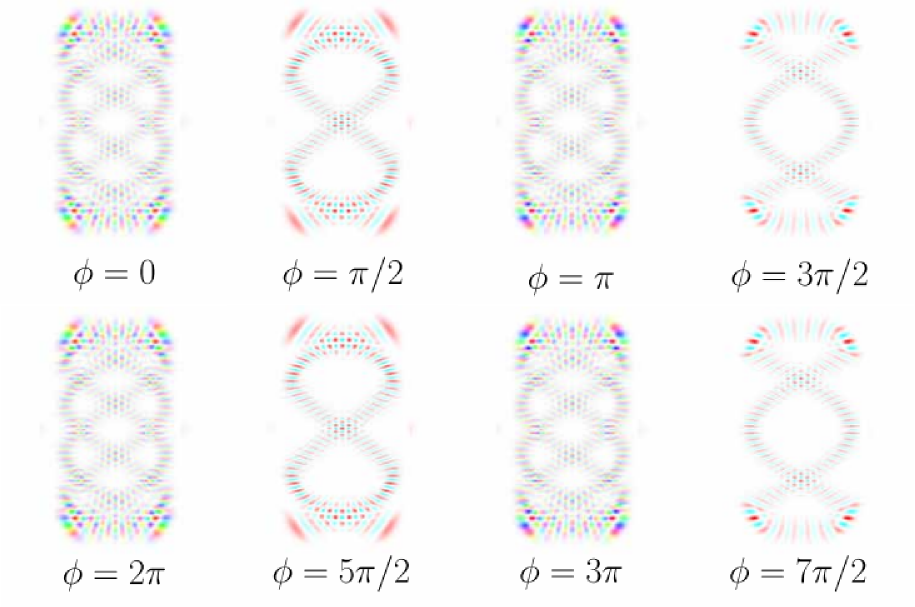}}\;\;\;\;\;
    \subfloat[\((s,t)=(6,-2)\)]{\includegraphics[width=0.4\textwidth]{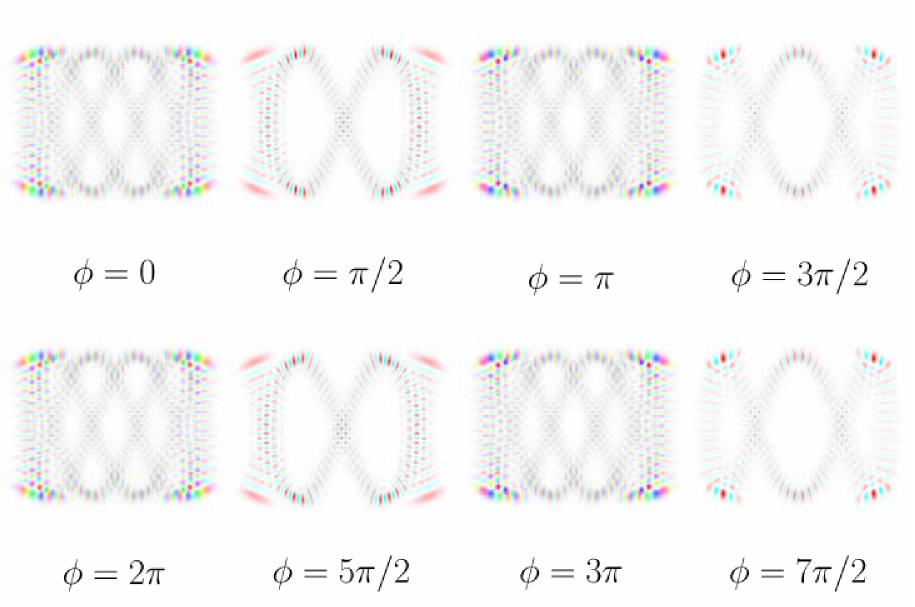}}\\
    \caption{
    Schematic of phase distribution profiles for SU(2) modes under various \((s,t)\) configurations influenced by the coherent phase:
    Top row: \((s,t) = (1,3)\) and \((3,1)\) with a periodicity \(P(\phi) = 2\pi\); second row: \((s,t) = (0,4)\) and \((4,0)\) with \(P(\phi) = 2\pi\); third row: \((s,t) = (-1,5)\) and \((5,-1)\) with \(P(\phi) = 2\pi\); bottom row: \((s,t) = (-2,6)\) and \((6,-2)\) with \(P(\phi) = 2\pi\). Each column corresponds to different values of \(\phi\), showcasing the evolution of the phase distribution as \(\phi\) progresses.
    }
    \label{periodic_law_phi_phase}
\end{figure*}

\newpage

\section{Multibeam Interference Patterns}

This section presents multibeam interference patterns to illustrate how the PB phase influences the spatial structure of SU(2) beams. These examples help to establish the physical intuition behind non-interferometric PB phase measurements.

\begin{figure*}[!htbp]
    \centering
    \subfloat[]{\includegraphics[width=0.20\textwidth]{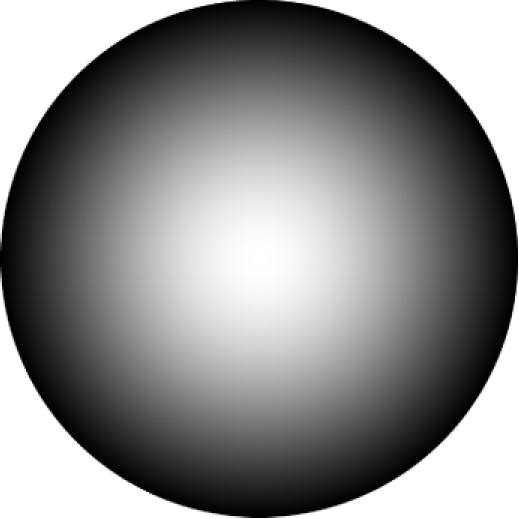}}\;\;\;\;
    \subfloat[]{\includegraphics[width=0.20\textwidth]{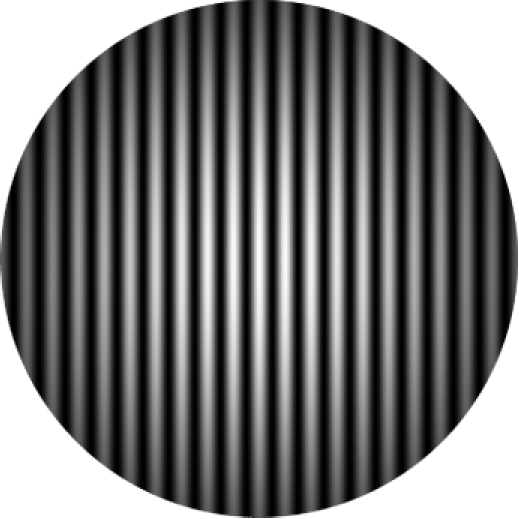}}\;\;\;\;
    \subfloat[]{\includegraphics[width=0.20\textwidth]{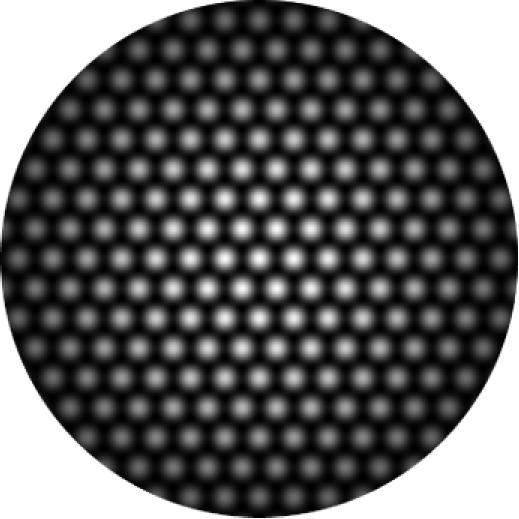}}\;\;\;\;
    \subfloat[]{\includegraphics[width=0.20\textwidth]{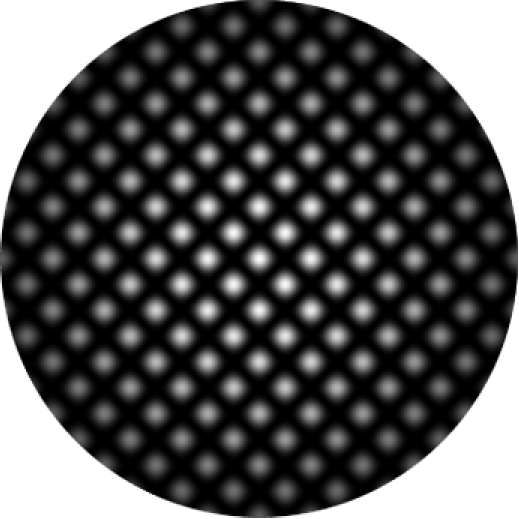}}
    \\[1ex]
    \subfloat[]{\includegraphics[width=0.20\textwidth]{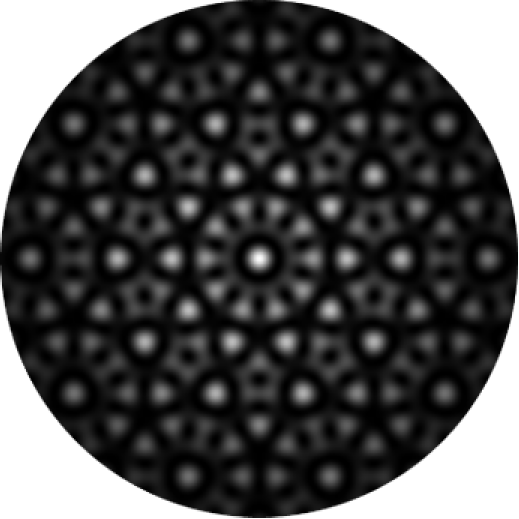}}\;\;\;\;
    \subfloat[]{\includegraphics[width=0.20\textwidth]{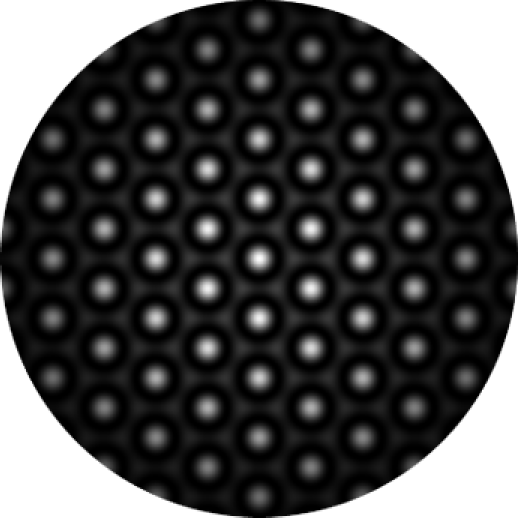}}\;\;\;\;
    \subfloat[]{\includegraphics[width=0.20\textwidth]{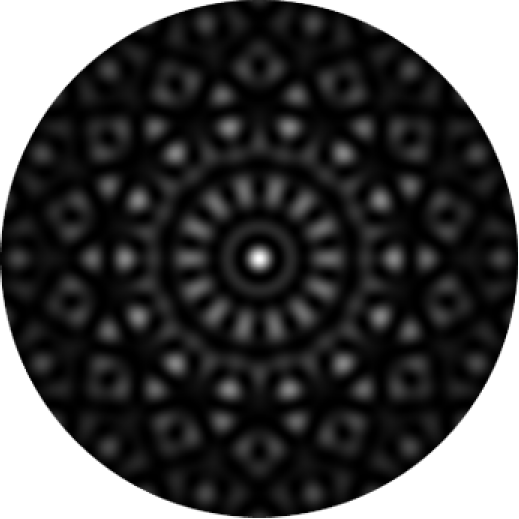}}\;\;\;\;
    \subfloat[]{\includegraphics[width=0.20\textwidth]{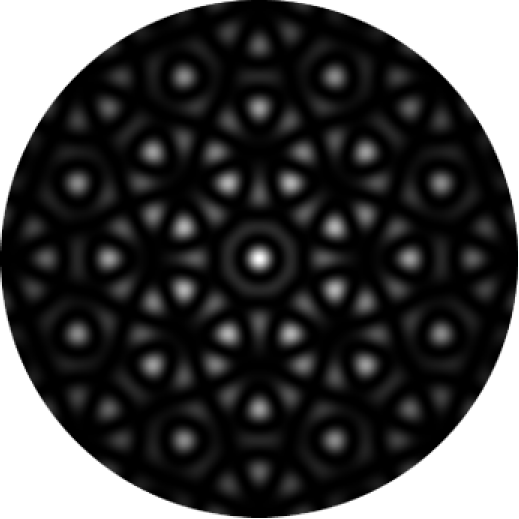}}
    \\[1ex]
    \subfloat[]{\includegraphics[width=0.20\textwidth]{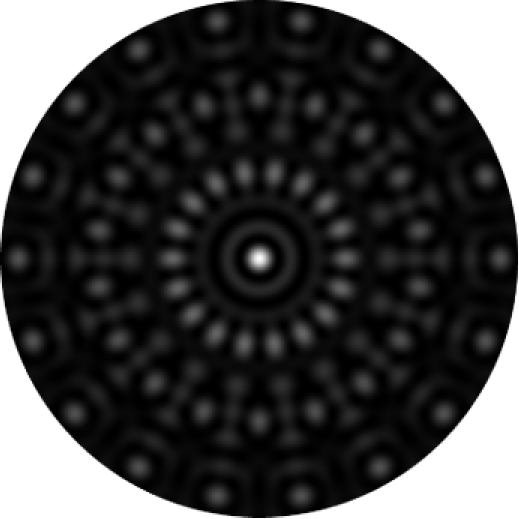}}\;\;\;\;
    \subfloat[]{\includegraphics[width=0.20\textwidth]{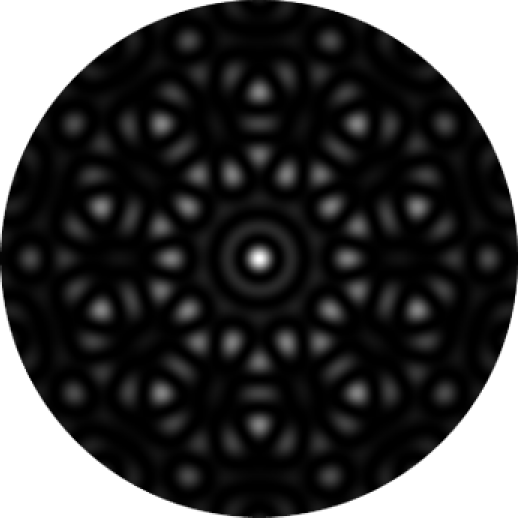}}\;\;\;\;
    \subfloat[]{\includegraphics[width=0.20\textwidth]{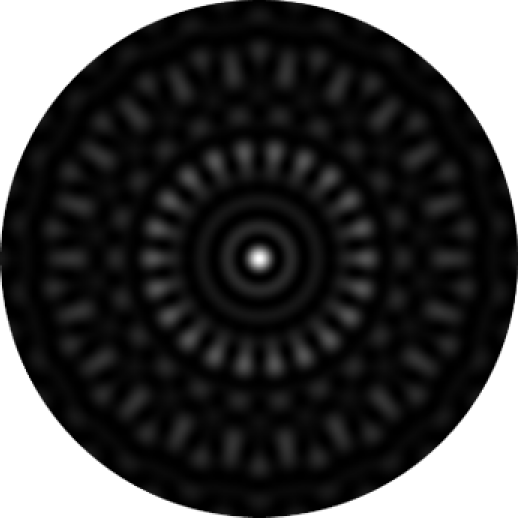}}\;\;\;\;
    \subfloat[]{\includegraphics[width=0.20\textwidth]{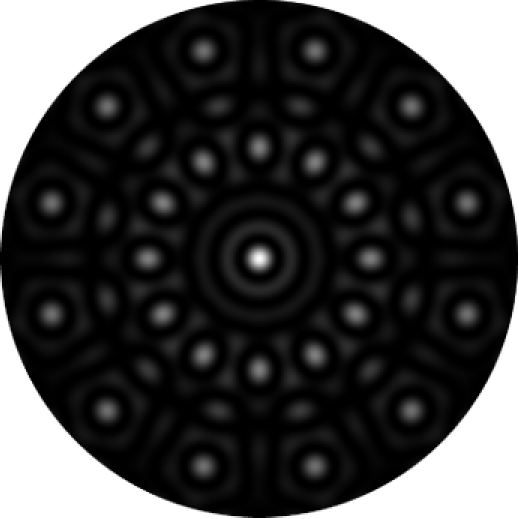}}
    \caption{
    Representative intensity patterns from one to twelve beams: (a) single Gaussian beam without interference, (b) two-beam interference, and (c--l) multibeam interference patterns for 3 to 12 beams. As the number of interfering beams increases, the spatial structures become increasingly intricate, resembling the complexity found in SU(2) coherent beam superposition.
    }
    \label{superposition_and_interference}
\end{figure*}

\section{A Heuristic Fiber-Bundle Interpretation of the PB Phase in SU(2) Beams}
\label{sec:fiber_bundle_SU2}

In this section, we outline a geometric viewpoint that may help interpret the 
Pancharatnam--Berry (PB) phase of SU(2) coherent beams using concepts from 
fiber-bundle theory.  
The purpose of this discussion is not to establish a new formal framework, 
but to offer an intuitive picture that is consistent with the notation, 
expressions, and conclusions already derived in the main text.  
Throughout this section we therefore restrict ourselves to 
interpretive statements that reproduce the PB-phase relations of the main
manuscript—for example the eigenmode contribution 
$\Theta_{K}=\gamma\,[m-n+K(s-t)]$ and the total SU(2) PB phase 
$\Theta_{\mathrm{SU(2)}}=\sum_{K=0}^{M}\Theta_{K}$.  
The fiber-bundle language is used here only as an illustrative aid to visualize 
how these factors can be regarded as arising from U(1) holonomies over closed 
paths on the relevant parameter spheres, without making claims of 
mathematical completeness or generality.

\subsection{SU(2) coherent state and eigenmode decomposition}

The SU(2) coherent state used in the main text can be written as
\begin{equation}
    \left|\Psi_{\sss{m,n,l}}^{\sss{M,s,t}}\right\rangle
    = \frac{1}{2^{\sss{M/2}}}
    \sum_{K=0}^{M}
    \binom{M}{K}^{\sss{1/2}}
    \mathrm{e}^{\sss{\mathrm{i}K(\phi-\frac{\pi}{2})}}
    \left|\psi_{\sss{m+sK,\,n+tK,\,l-K}}\right\rangle.
    \label{eq:SU2_bundle_decomposition}
\end{equation}
Here \(m\) and \(n\) are the transverse indices of the lowest-order Hermite--Gaussian (HG) mode, \(l\) is the longitudinal index, \(s\) and \(t\) specify the step sizes in the two transverse directions, \(M\) is the highest step index, and \(\phi\) is the coherent phase.  The index \(K\in\{0,\ldots,M\}\) labels the eigenmodes that participate in the coherent superposition.

For later convenience we define the transverse indices and mode order of the \(K\)-th eigenmode as
\begin{equation}
    m_{\sss K} = m + sK,\qquad
    n_{\sss K} = n + tK,\qquad
    N_{\sss K} = m_{\sss K} + n_{\sss K}.
    \label{eq:mnK_def}
\end{equation}
The difference of the transverse indices,
\begin{equation}
    \Delta_{\sss K} \equiv m_{\sss K} - n_{\sss K}
    = (m-n) + K(s-t),
    \label{eq:DeltaK_def}
\end{equation}
plays the role of an effective ``magnetic quantum number'' for the \(K\)-th eigenmode in the OAM Poincar\'e-sphere description.

Each eigenmode \(\left|\psi_{\sss{m_{\sss K},n_{\sss K},l-K}}\right\rangle\) can be mapped onto an OAM Poincar\'e sphere with mode order \(N_{\sss K}\).  Mode transformations along closed geodesic loops on this sphere are then naturally described by the geometry of a fiber bundle.

\subsection{Hopf fibration and the geometric phase of a single eigenmode}

The Hopf fibration provides a canonical example of a principal fiber bundle relevant to our problem.  The key ingredients are:
\begin{itemize}
    \item \textit{Base manifold (BM):} the Poincar\'e sphere \(S^{2}\), with spherical coordinates \((\alpha,\beta)\).
    \item \textit{Total space:} the three-sphere \(S^{3}\), representing normalized complex field states.
    \item \textit{Structure group (SG):} the unitary group \(U(1)\), whose elements \(\mathrm{e}^{\mathrm{i}\Theta}\) act as local phase rotations on the field.
    \item \textit{Fiber:} circles \(S^{1}\) attached to each point of the base manifold, representing the \(U(1)\) phase degree of freedom.
\end{itemize}
In this picture, a point on the base manifold specifies the \emph{intensity} distribution of an eigenmode, while the fiber coordinate encodes its overall phase.  A closed loop on \(S^{2}\) lifts to a path in the total space \(S^{3}\).  Because the fiber is nontrivially twisted over the base manifold, a cyclic evolution in \((\alpha,\beta)\) generally produces a non-zero geometric (Berry) phase.

For the \(K\)-th eigenmode, we denote by \(|\psi_{\sss K}(\alpha,\beta)\rangle\) the state obtained by transporting an initial eigenmode along a path on the sphere.  The Berry connection one-form is
\begin{equation}
    \mathcal{A}_{\sss K}
    = \mathrm{i}\,\big\langle \psi_{\sss K}(\alpha,\beta)\big|
    \mathrm{d}\,\psi_{\sss K}(\alpha,\beta)\big\rangle,
    \label{eq:Berry_connection_def}
\end{equation}
and the corresponding Berry curvature two-form is
\begin{equation}
    \mathcal{F}_{\sss K} = \mathrm{d}\mathcal{A}_{\sss K}.
    \label{eq:Berry_curvature_def}
\end{equation}

For SU(2)-type mode transformations on the Poincar\'e sphere, a direct calculation (see, e.g., standard derivations for SU(2) coherent states) yields the explicit form
\begin{equation}
    \mathcal{A}_{\sss K}
    = \frac{\Delta_{\sss K}}{2}\,(1-\cos\beta)\,\mathrm{d}\alpha,
    \qquad
    \mathcal{F}_{\sss K}
    = \frac{\Delta_{\sss K}}{2}\,\sin\beta\,\mathrm{d}\beta\wedge\mathrm{d}\alpha,
    \label{eq:Berry_connection_explicit}
\end{equation}
where \(\Delta_{\sss K}\) is defined in Eq.~\eqref{eq:DeltaK_def}.  The geometric (PB) phase accumulated when the eigenmode is transported adiabatically along a closed loop \(\mathcal{C}\) on the sphere is
\begin{equation}
    \Theta_{K}
    = \oint_{\mathcal{C}} \mathcal{A}_{\sss K}
    = \iint_{\mathcal{S}} \mathcal{F}_{\sss K}
    = \frac{\Delta_{\sss K}}{2}\,\Omega,
    \label{eq:Berry_phase_general}
\end{equation}
where \(\mathcal{S}\) is any oriented surface on the sphere bounded by \(\mathcal{C}\), and \(\Omega\) is the solid angle subtended by \(\mathcal{C}\) on \(S^{2}\).

In the main text we consider closed geodesic loops that enclose a solid angle \(2\gamma\), i.e.,
\begin{equation}
    \Omega = 2\gamma.
    \label{eq:Omega_gamma_relation}
\end{equation}
Substituting Eq.~\eqref{eq:Omega_gamma_relation} into Eq.~\eqref{eq:Berry_phase_general} gives the PB phase associated with the \(K\)-th eigenmode:
\begin{equation}
    \Theta_{K} = \Delta_{\sss K}\,\gamma
    = \gamma\,[m-n + K(s-t)].
    \label{eq:thetaK_final}
\end{equation}
Equivalently, the corresponding phase factor is
\begin{equation}
    \Phi_{\sss K}
    = \exp\!\left\{\mathrm{i}\,\Theta_{K}\right\}
    = \exp\!\left\{\mathrm{i}\,\gamma\,[m-n + K(s-t)]\right\}.
    \label{eq:PhiK_final}
\end{equation}
This result is precisely the phase factor obtained from the continuous-evolution analysis in the main text, showing that the fiber-bundle description is fully consistent with the mode-transformation picture on the Poincar\'e sphere.

\subsection{Total PB phase of an SU(2) coherent state}

We now lift the discussion from a single eigenmode to the full SU(2) coherent state.  When each eigenmode in the superposition of Eq.~\eqref{eq:SU2_bundle_decomposition} undergoes the same closed evolution on the Poincar\'e sphere, its state acquires the PB phase factor \(\Phi_{\sss K}\) given by Eq.~\eqref{eq:PhiK_final}.  The evolved SU(2) state is then
\begin{equation}
    \left|\Psi_{\sss{m,n,l}}^{\sss{M,s,t}}\right\rangle
    = \frac{1}{2^{\sss{M/2}}}
    \sum_{K=0}^{M}
    \binom{M}{K}^{\sss{1/2}}
    \mathrm{e}^{\sss{\mathrm{i}K(\phi-\frac{\pi}{2})}}
    \Phi_{\sss K}
    \left|\psi_{\sss{m+sK,\,n+tK,\,l-K}}\right\rangle.
    \label{eq:SU2_after_evolution}
\end{equation}
Comparing Eqs.~\eqref{eq:SU2_bundle_decomposition} and \eqref{eq:SU2_after_evolution}, we see that the geometric information is completely contained in the set of phase factors \(\Phi_{\sss K}\).

Because the Poincar\'e-sphere evolution is diagonal in the eigenmode basis, the total PB phase of the SU(2) coherent state can be regarded as the sum of the geometric phases of all constituent eigenmodes:
\begin{equation}
    \Theta_{\text{SU(2)}}
    = \sum_{K=0}^{M} \Theta_{K}
    = \sum_{K=0}^{M} \gamma\,[m-n + K(s-t)].
    \label{eq:SU2_PB_sum}
\end{equation}
Evaluating the sum explicitly gives
\begin{equation}
    \Theta_{\text{SU(2)}}
    = \gamma\Bigg[
        (m-n)\sum_{K=0}^{M} 1
        + (s-t)\sum_{K=0}^{M} K
    \Bigg]
    = \gamma\Bigg[
        (m-n)(M+1)
        + (s-t)\frac{M(M+1)}{2}
    \Bigg].
    \label{eq:SU2_PB_closed_form}
\end{equation}
Rewriting Eq.~\eqref{eq:SU2_PB_sum} without evaluating the sums, we recover exactly the expression used in the main text [Eq.~\eqref{SU(2)_PB_phase}]:
\begin{equation}
    \Theta_{\text{SU(2)}}
    = \sum_{\sss{K=0}}^{\sss M}\gamma\,[m-n+K(s-t)].
    \tag{\ref{SU(2)_PB_phase}}
\end{equation}


This shows that the PB phase of an SU(2) coherent state can be viewed as the total holonomy of a family of \(U(1)\) fibers over the base manifold \(S^{2}\), where each eigenmode contributes a geometric phase proportional to its index difference \(\Delta_{\sss K} = m_{\sss K}-n_{\sss K}\) and to the solid angle \(\Omega = 2\gamma\) enclosed by the mode-transformation path.

\begin{figure*}[!htbp]
    \centering
    \includegraphics[width=\textwidth]{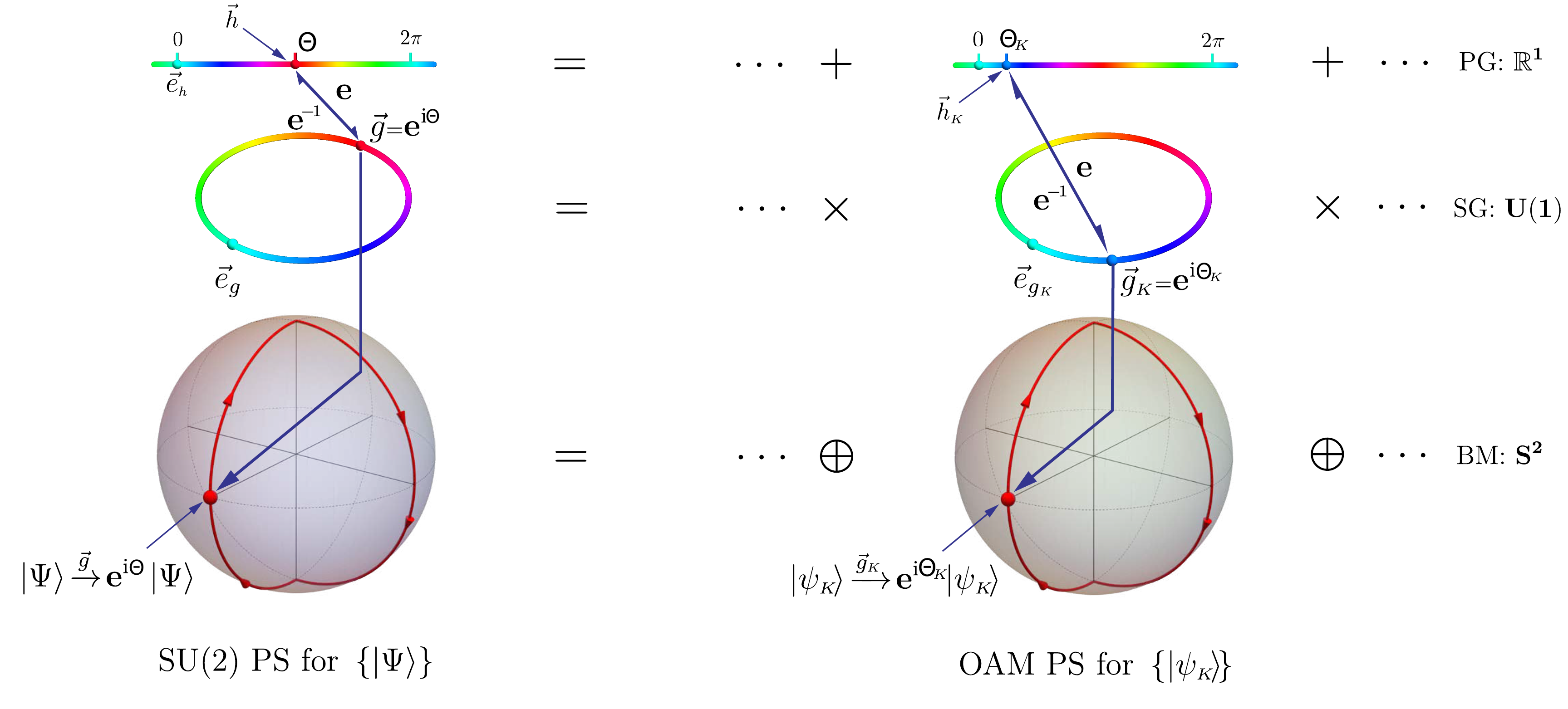}
    \caption{
    \textbf{Geometric and fiber-bundle interpretation of PB-phase generation in SU(2) coherent states.}
    The diagram is organized vertically according to the standard fiber-bundle hierarchy: the parametric group (PG, top), the structure group (SG, middle), and the base manifold (BM, bottom).  
    The left column depicts the SU(2) Poincar\'e sphere associated with the coherent-state family $\{|\Psi\rangle\}$, while the right column shows the OAM Poincar\'e spheres corresponding to the constituent eigenmodes $\{|\psi_{K}\rangle\}$.  
    Along each vertical stack, the PG traces a geodesic loop on the sphere; the SG represents the associated $U(1)$ phase rotation, and the BM records the geometric path that encloses a solid angle $2\gamma$.  
    Transporting an eigenmode around this loop produces the PB phase 
    $\Theta_{K}=\gamma\,[m-n+K(s-t)]$, determined jointly by the enclosed solid angle and the mode indices, in agreement with Eq.~\eqref{SU(2)_PB_phase}.  
    The figure illustrates how the PB phase arises as the $U(1)$ holonomy of the fiber bundle, linking the geometry of the base manifold to the algebraic structure of SU(2) coherent-state superpositions.
    }
    \label{direct-sum decomposition}
\end{figure*}

\subsection{Geometric interpretation of Fig.~\ref{direct-sum decomposition}}

The schematic in Fig.~\ref{direct-sum decomposition} summarizes this fiber-bundle picture.  The vertical structure in each panel (SU(2) PS on the left, OAM PS on the right) corresponds to three levels:
\begin{enumerate}
    \item \textit{Parameter group (PG):} a one-dimensional real line \(\mathbb{R}^{1}\), parameterized by a coordinate \(h\) or by its \(2\pi\)-periodic projection \(\Phi\).  
    In the diagram, both \(h\) and \(\Phi\) appear along the same horizontal line because the exponential map \(e:\mathbb{R}\rightarrow U(1)\) identifies them: an arbitrary real parameter \(h\in\mathbb{R}\) projects to the phase variable
    \[
        \Theta = h \!\!\!\mod 2\pi,
    \]
    which is the physically relevant coordinate on the circle.  
    Thus \(h\) labels points on the universal covering space \(\mathbb{R}\), while \(\Theta\in[0,2\pi)\) labels their images on the U(1) phase circle. The PG therefore consists of the full real line, even though the induced phase variable is periodic.
    
    \item \textit{Structure group (SG):} the unit circle \(U(1)\), whose generic elements \(\vec{g}_{\sss K}=\exp(\mathrm{i}\Theta_{\sss K})\) act as phase rotations on the eigenmodes and encode the PB phase \(\Theta_{K}\).  
    The map from PG to SG is exactly the exponential map \(h\mapsto e^{\mathrm{i}h}\), and the point labelled \(\Theta\) in the figure is the angular representative of that same element on the circle.

    \item \textit{Base manifold (BM):} the Poincar\'e sphere \(S^{2}\), on which the red closed curve represents the geodesic loop with enclosed solid angle \(\Omega=2\gamma\).
\end{enumerate}

For each eigenmode, the mapping from the PG to the SG (top to middle) is therefore given by the exponential map
\[
    e:\mathbb{R}^{1}\to U(1),\qquad h\mapsto e^{\mathrm{i}h},
\]
with \(\Theta = h\!\!\!\mod 2\pi\) used as the angular coordinate on \(U(1)\).  
The projection from the SG to the BM (middle to bottom) encodes how the phase fiber is attached over each point on the sphere.  
The left panel depicts this structure for the SU(2) coherent-state space \(\{|\Psi\rangle\}\), and the right panel for the eigenmodes \(\{|\psi_{\sss K}\rangle\}\).  
The equality signs between rows emphasize that all three descriptions—PG, SG, and BM—are mathematically equivalent ways of characterizing the same geometric phase accumulated during a closed mode transformation.

It is useful to note that, at the level of the underlying Hilbert space,
the SU(2) coherent state does not belong to a single OAM Poincar\'e sphere
but is constructed from multiple mode-order sectors.
Each sector of fixed transverse order \(N_{K}=m_{K}+n_{K}\) spans a
Hilbert subspace \(\mathcal{H}_{N_{K}}\), and the full state belongs to the
direct-sum space
\[
    \mathcal{H}_{\text{SU(2)}}=\bigoplus_{K=0}^{M}\mathcal{H}_{N_{K}}.
\]
The geometric picture shown in Fig.~\ref{direct-sum decomposition} therefore captures how the PB phase is generated on each OAM Poincar\'e sphere
individually, while the SU(2) PB phase arises from combining the contributions
from all mode-order subspaces appearing in this direct-sum decomposition.

In summary, the fiber-bundle perspective makes explicit that the PB phase in SU(2) beams arises from the nontrivial \(U(1)\) holonomy of eigenmodes transported along closed geodesics on the Poincar\'e sphere.  
The eigenmode-level phase \(\Theta_{K} = \gamma[m-n+K(s-t)]\) and the SU(2) PB phase \(\Theta_{\text{SU(2)}}\) in Eq.~\eqref{SU(2)_PB_phase} are thus direct geometric consequences of the underlying bundle structure.

{$^*$y.shen@soton.ac.uk}

\bibliography{apssamp}